\PassOptionsToPackage{usenames,dvipsnames}{xcolor}
\documentclass[onecolumn,secnumarabic,amssymb, nobibnotes, aps]{revtex4-2}

\usepackage{enumerate}
\usepackage{color}
\usepackage{cancel}
\usepackage{natbib}
\usepackage[dvipsnames,usenames]{xcolor}
\usepackage{pstricks}
\usepackage{graphicx}
\usepackage{amsmath}
\usepackage{amsfonts}
\usepackage{amssymb}
\usepackage{float}
\usepackage{xr}
\begin{document}

\title{
Characterisation of a multistable turbulent wake: application of an improved regime identification with analytical model training.
}


\author{A. Barlet$^{1}$, P. Bragan\c ca$^{2}$, C. Cuvier$^{2}$,
J. Rolland$^{2}$}
\email{joran.rolland@centralelille.fr}
\affiliation{$^{1}$ Centre INRIA de l'universit\'e de Rennes, IRMAR UMR 6625}
\affiliation{$^{2}$ Univ. Lille, CNRS, ONERA, Arts et M\'etiers Institute of Technology, Centrale Lille,
UMR 9014 - LMFL - Laboratoire de M\'ecanique des Fluides de Lille - Kamp\'e de F\'eriet, F-59000 Lille, France.}

\date{\today}

\begin{abstract}
This article presents the experimental study and the modelling of the multistable jet in the wake of two side by side square bars separated by a distance $G$ at Reynolds number $Re=U_\infty H/\nu=10000$
(with $U_\infty$ the velocity of the incoming flow, $H$ the bar side and $\nu$ the kinematic viscosity).
The velocity field downstream of the bars is measured by means of two dimensional two components Particle Image Velocimetry (PIV).
We use the weighted transverse position of the jet $Y_m$ and the jet width $w$ to characterise the regimes of multistability as the gap ratio $G/H$ is increased.
Three main regimes of multistability are possible: tristability (where the jet points toward either of the bar or flutters along the centreline),
bistability (where the jet points toward either of the bars), and monostability (with a wide jet along the centreline).
In our wind tunnel, tristability is observed  for $G/H\in [1.15,1.25]$, bistability is observed for  $G/H\in [1.5,2.65]$ and monostability is observed for  $G/H\in [3.0,3.5]$. 
Within the first two ranges of $G/H$, there exists gap ratios for which multistability is more complex.
For instance, at $G/H=1.8$, the transverse symmetry remains broken, the position of the jet is closer to the centreline,
however the jet spends very little time in one of the bistable states, which results in a monomodal histogram of $Y_m$.
Moreover, the change of regime from bistable to monostable at $G/H=2.7$ occurs through a complex temporal behaviour.
At that gap ratio, the wake alternates between a highly fluctuating jet along the centreline (which can be wide or hollow) 
and a more regular narrow jet in either of the bistable positions.
In order to analyse the simple and complex multistability regimes as well as the transition from bistable to monostable, 
we construct an analytical stochastic differential equation (SDE) modelling $Y_m$ for each gap ratio.
These SDEs are written with
polynomial drift and diffusion.
For this matter we use a data based method 
that
finds an trade--off between simplicity of the model (smaller number of monomials) and precision.
A first key advantage of the use of the data-based model fitting method is that when the flow is tristable, bistable or monostable, 
we recover the drifts expected from the theory of bifurcations, but we are now able to
correct it with the right multiplicative noise expressed by the diffusion. 
The second key advantage is that we can also fit atypical drift expressions when the multistability regime is complex that help us make sense of the jet behaviour.
Finally, using the fitted SDE, we can draw a parallel between the change of regime at $G/H=2.7$ and the transition to chaos \emph{via} temporal intermittency.
In that transition, the corresponding systems go from a simple temporal behaviour (for instance periodical) to a complex one,
 alternating between chaos and the initial simple state.
We find that the fitted SDE and dynamical systems undergoing temporal intermittency display the same type of global bifurcation. 
Finally, the information extracted from the model provides readability to the regime diagrams constructed for this system using $Y_m$ and $w$.
\end{abstract}

\maketitle

\section{Introduction}

While they display complex fluctuations in space and time, turbulent flows can nevertheless generate large scale coherent circulations.
Very often, several distinct large scale coherent circulations can exist for the same values of the control parameters of the flow.
In that case, the flow can be multistable: it spends most of its time in one configuration or another and can
rapidly and 
unpredictably transit from one configuration to another.
In this article, we will call these changes of state \emph{transitions}.
Experimental realisations of bistable turbulent flows can be found in aerodynamics.
For instance the bistable wake behind two parallel bars was described briefly by \cite{biermann1934interference}.
Since then, this flow has been studied by many authors, as latest reviewed by \cite{zhou2016wake}.
Progress in metrology has led to more precise description.
In particular, Kim \& Durbin showed some key points of this bistable flow \cite{kim1988investigation}. In aerodynamics, a lot
of attention was also devoted to the bistable wake of the Ahmed body during the last decade \citep{grandemange2013turbulent}. 
Bistability has to be taken into account when controlling this wake \cite{barros2017forcing}.
Other cases of bistability in aerodynamics can be found in pendulums in flows \citep{gayout2021rare},
and during the drag crisis of a sphere \citep{norman2011unsteady,deshpande2017intermittency}.
Bistability is also observed experimentally in geophysical fluid dynamics, for instance in two dimensional turbulence \citep{sommeria1986experimental},
and in turbulent dynamos \citep{Berhanu2007}.
While data is more scarce, bistable turbulent flows were also studied numerically,
for instance in the case of transitional wall flows \citep{rolland2011pattern},
turbulent convection \citep{podvin2017precursor},
or two dimensional geophysical flows, using rare events simulation methods \citep{bouchet2019rare}.
In the context of geophysical fluid dynamics,
bistability in atmospheric \citep{herbert2020atmospheric} or oceanic \citep{cini2024simulating,soons2024optimal}
flows can have strong implication for climate.

One important question in the study of multistable turbulent flows is the (dis)appearance of multistability as
one (or more) of the control parameters is changed.
One would like to pinpoint the value of the control parameter where this change happens,
and whether this change occurs more continuously \citep{rolland2011pattern} or discontinuously \citep{alam2011wake,bouchet2019rare,ditlevsen2023warning}.
In order to understand these observations, performing formal analogies with other
branches of physics is helpful.
The disappearance of bistability is reminiscent of both bifurcations in autonomous dynamical systems \citep{manneville2010instabilities} and
phase transitions in statistical physics \citep{landau2013statistical}.
It is sometimes termed a tipping point in the climate science context \citep{ditlevsen2023warning,cini2024simulating}.
In this article we will term it a \emph{bifurcation}, by some abuse of language.  
A first point of comparison is the fact that some symmetries or invariances (defined on average for a turbulent flow),
are often broken when the flow goes from monostable to bistable.
For instance planar  symmetries can be broken for bistable wakes \citep{grandemange2013turbulent}
or turbulent dynamos \citep{Berhanu2007}.
This means that we will often follow the change of regime using an \emph{Order Parameter} :
a scalar quantity
which measures the degree of symmetry breaking.
Thus, it goes from zero, in the disordered phase, to non zero in the ordered phase, in the case of a bifurcation or for a phase transition
(in ensemble average and in the thermodynamic limit).
In the context of bifurcations in dynamical systems, the order parameter measures
the amplitude of stable fixed points that correspond to the multistable states.
In that line of thought, the unstable fixed points that lie between the multistable
state are also important, since they are often a central part of transitions.
In that context, the unstable fixed point is termed a transition state and 
 characterises the transition mechanism \citep{bouchet2016generalisation}.
Given these measurements, we are interested in making sense of the dependence of the order parameter
on control parameters, of the shape of the bifurcation diagram,
of the type of regime change.  

Simplified readable models for bifurcations are often considered to study multistability of more complex systems.
These models consist in differential equations describing the time evolution
 of a variable whose average should represent as faithfully as possible the order parameter of the system. 
Moreover, we expect a degree of readability from the model.
Thus, from the differential equation, we would like to be able to deduce the position of the multistable states
for a given value of the control parameters.
Moreover, as the control parameters are varied, we  would like to be able to extract the change of regime
from the model.
In order to describe turbulent and/or unpredictable flows, two main classes of low degrees of freedom
models were used: deterministic chaotic models \citep{rikitake1958oscillations,ito1980chaos}
and random stochastic models \citep{rolland2011pattern,barros2017forcing,ditlevsen2023warning}.
Both types of models can generate time series of variables representing the order parameter
that spend a long duration near one state and unpredictably leave for the other state. 
A major difference between these two classes of dynamics lies in the range of time correlations
of the variables they produce and represent.
In particular, stochastic models are more apt at reproducing fast decorrelations and time spectra with flat, so called white, ranges.
Given a choice of model type, the following question is that
of model construction.
Unlike the study of a bifurcation in a laminar flow \citep{manneville2010instabilities,deng2021galerkin},
where one can perform multiscale expansions or an elimination of fast variables,
there is no systematic method to derive
the equivalent of an amplitude equation from a purely turbulent flow.  
Thus, models are built on a case by case basis.
Some models can actually be derived from the governing
equations using a Galerkin method and a truncation \citep{shukla2016statistical}.
However, a closure is often needed for these models to faithfully represent either the
flow, or at least the time dependence of relevant order parameters \citep{podvin2017precursor}.
These closures are often \emph{ad hoc} and require an input from sampled data.
This implies models are mostly  constructed using dedicated data based methods.

Constructing a chaotic model from data is a challenge.
It is possible to work on a simpler system, assuming it undergoes similar
physical effects (see \citep{rikitake1958oscillations,ito1980chaos} 
and more recently \citep{gissinger2010morphology} for models of dynamos), however, such a model hardly accounts for the flow patterns.
Moreover, setting the model parameters so that it can quantitatively reproduce the
order parameters is difficult, since there is no straightforward link between the data estimated from
the flow and the parameters of the model.
Finally, while both turbulence and chaos are deterministic (though unpredictable),
the time fluctuations of signals sampled in turbulent flows can be closer
to probabilistic, stochastic systems.
As it happens, constructing a stochastic model can be more straightforward.
This can happen provided the order parameter we wish to model decorrelates fast enough in time and amplitude.
In that case, the temporal fluctuations of the order parameter, while random, are strongly structured.
The dynamics is modelled by a Langevin equation,
while its probability density function follows the
equivalent Fokker--Planck equation \citep{gardiner2009stochastic}.
These differential equations are ruled by a pair of functions:
one controlling deterministic effects, termed \emph{drift} (very often containing the metastable states),
and one controlling the random fluctuations, termed \emph{diffusion}.
If one can estimate these functions from data, then the whole dynamics can be reproduced.
Strong simplifications can be performed, using symmetry (breaking) arguments for instance \citep{petrelis2009simple},
which reduces these functions to a few parameters that can then be estimated very simply from data \citep{rolland2011pattern}.
Following similar methods, the probability density function (PDF) of the system can be fitted  \citep{barros2017forcing}.
Such an approach cannot account for all possible symmetry breakings, especially complex ones \citep{rolland2018finite},
nor can it account simply for
state dependent noise.
Moreover, this requires distinct estimations for the amplitude of
multistable state on the one hand and time scales on the other hand.
Recently, more systematic approaches were applied to construct Langevin equations using data \citep{boujo2019stochastic,callaham2021nonlinear,callaham2022empirical}.
This type of model construction method requires a strict \citep{boujo2019stochastic}
or a more loose \citep{boujo2019stochastic,callaham2021nonlinear} assumption
as to the functional dependence of the drift and diffusion.
Nevertheless, these two functions can be computed from the estimation of a
single group of functions: empirical Kramers--Moyal coefficients \citep{gardiner2009stochastic} and histograms,
followed by a model parameter optimisation.
The resulting functions can then be used to generate new time series, complement data sets,
 discuss the properties of the system \citep{rolland2011pattern}, and propose some properties found at the crossover,
 despite the complexity of the system \citep{rolland2018finite}.
 Indeed: they are not based on an \emph{a priori} on which term should should be retained
 in the SDE and still generate sparse models. 

\begin{figure}[!ht]
\centerline{
\textbf{(a)}
\includegraphics[width=5.5cm]{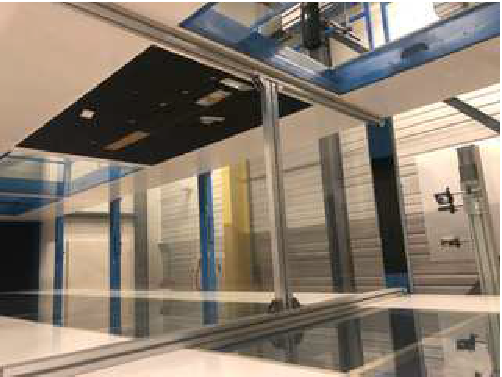}
\textbf{(b)}
\includegraphics[width=5.5cm,clip]{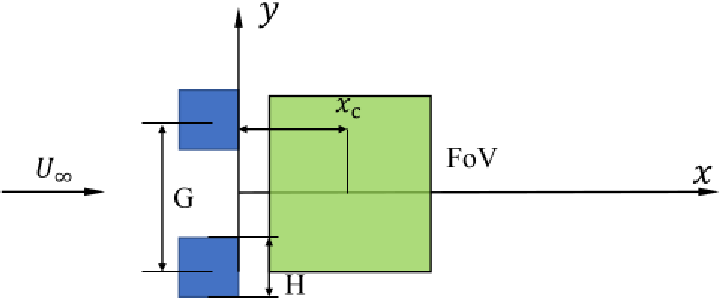}
\textbf{(c)}
\includegraphics[width=5.5cm,clip]{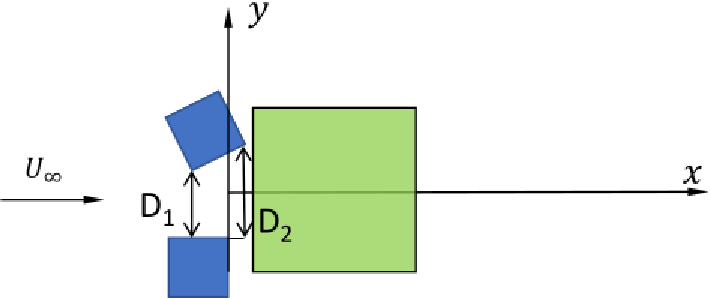}
}
\centerline{
\textbf{(d)}
\includegraphics[width=8.25cm,clip]{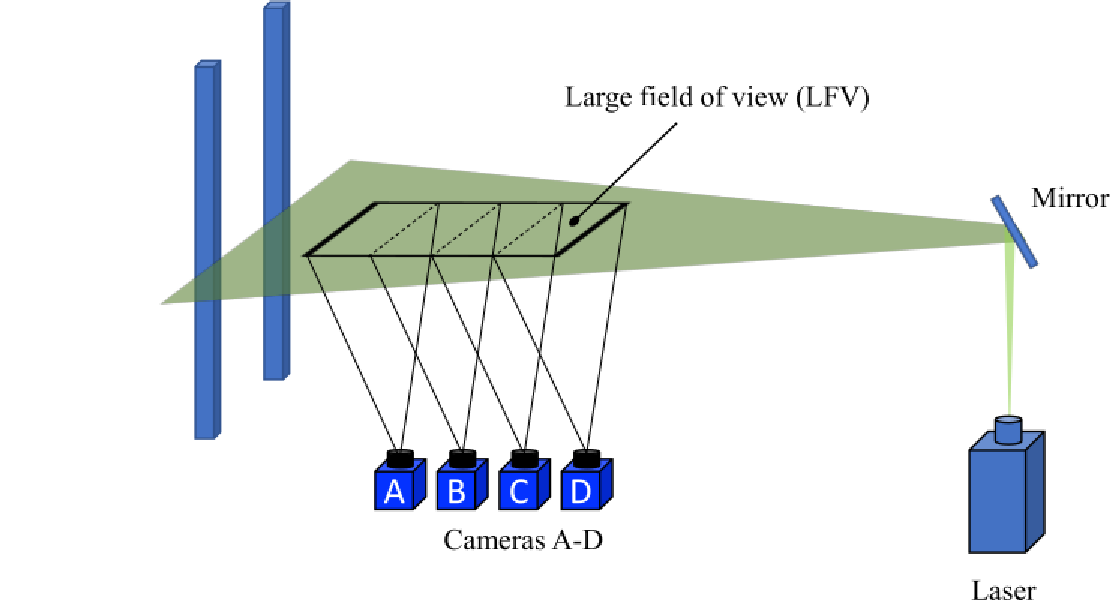}
\textbf{(e)}
\includegraphics[width=5.5cm,clip]{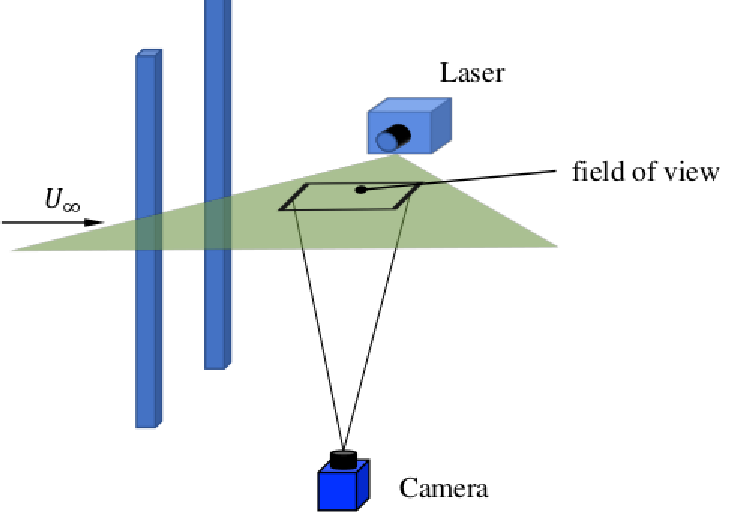}
}
\caption{(a) Photograph of the parallel bars in the LMFL boundary layer wind tunnel, the upstream direction is on the left (courtesy of J.G. Chen).
(b) Sketch of the two bars (in blue) configuration seen in a horizontal plane, indicating the distance between the bars centre $G$,
the size of the bar $H$, the upstream velocity $U_\infty$, the streamwise $\vec x$ and spanwise $\vec y$ axes, the field of view (in green)
where velocity is recorder by means of PIV and the offset $x_c$ between the bars and the centre of the field of view.
(c) Sketch of the bars in the case where one of the bars was deliberately turned on its axis.
The sketch indicates the distances $D_1$ and $D_2$ between the corners of bar two and the side of bar one
used to determine the rotation angle.
(d) Sketch of the experiment in perspective showing the PIV set up used to record the Large Field of View by \cite{chen2021turbulence}.
(e) Sketch of the experiment in perspective showing the PIV set up used in high frequency measurements and in the parametric study.
(Figures (b,c,e) are adapted from \citep{chen2021turbulence}, figure (d) is reprinted from \citep{chen2021turbulence}).}
\label{skint}
\end{figure}

We wish to test how these automatically constructed models render 
the various regimes of multistability in a wide interval of control parameter from experimental measurements.
We firstly would like to recover a simple Langevin equation when the behaviour of the system is symmetric and typical.
We secondly and more importantly would like to describe all the changes of multistability regime and
all the observed behaviour in that range of control parameters, in particular if they are complex and atypical.
One of the key points of this text is the demonstration that automated model construction can generate a SDE that can be used to:
\begin{enumerate}
\item  identify metastable points and their relative stability,
\item  confirm 
the existence of transition points,
\item identify precisely the diffusion, without any \emph{a priori} as to its shape, when the noise is multiplicative,
\item  more importantly, detect the positions
where the system can spend a long time without being metastable.
In that case, the model could provide tools to characterise this almost metastable state.
\end{enumerate}
Such an approach  would provide a large improvement in stochastic model construction, compared to
earlier studies where the form of the model is prescribed.
Therefore, we will work on multistability in a configuration
which is convenient to construct in a laboratory \citep{ishigai1972experimental,kim1988investigation,sumner1999fluid,alam2011wake,chen2021turbulence}
or simulate (with various degrees of modelling of turbulence) \citep{afgan2011large}:
the multistable jet in the wake behind two side by side bars (Fig.~\ref{skint}).
We sketch this configuration in figure~\ref{skint} (b):
this corresponds to an incoming homogeneous flow of velocity $U_\infty$
towards two parallel bars of square section with side $H$, whose centres are separated by a distance $G$
arranged perpendicularly to the incoming flow direction.
One can refer to successive reviews \citep{zdravkovich1977review,zdravkovich1987effects,sumner1999fluid,zhou2016wake} and references within
for an overview of the literature on the configuration.
Provided the Reynolds number $Re=HU_\infty/\nu$ (where $\nu$ is the kinematic viscosity of the fluid)
is large enough, the flow always detaches at the upstream corners of the  square bars \citep{alam2011wake},
which have a turbulent wake \citep{zhou2019extreme,chen2021turbulence}.
When the bars are brought close enough to one another (as measured by the gap ratio $G/H$),
the two wakes can interact.
This leads to coupled vortex streets \citep{alam2011wake}.
As the bars are brought closer to one another, the flow coming out the bars arranges  into a thin jet that
can reconnect to the wake further downstream.
This jet can spend a relatively long time (compared to the periods of the wakes)
either pointing toward one bar or the other.
While most works report bistability, it has been noted that when two circular cylinders are brought close enough,
but not so close that they generate the wake of a single object, the flow can
be tristable \citep{sumner1999fluid} and the jet can reside at the central position.
This phenomenon can occur in the wake of bars of square \citep{alam2003aerodynamic,alam2011wake}
or circular section \citep{kim1988investigation,sumner1999fluid} (and possibly other regular shapes).
The symmetry breaking was also reported when the wake is laminar \cite{kang2003characteristics}.
Note that a comparable symmetry breaking can also occur in the fluidic pinball, a similar configuration with a third upstream cylinder, again
at Reynolds numbers low enough for the flow to be laminar \citep{deng2021galerkin}.
This multistability was described, at least qualitatively,
for almost a century \citep{biermann1934interference},
and the first bifurcation diagram following the Strouhal numbers of the flow was constructed a short time after \citep{spivack1946vortex}
(although the bistable character of the flow was not fully understood at the time).
Point measurements, often of base pressure on the bars and/or hot wire anemometry
 \citep{hori1959experiments,ishigai1972experimental,bearman1973interaction,zdravkovich1977interference,zdravkovich1987effects,kim1988investigation},
were performed in the following years, leading to an increasingly precise description of
the multistable flow.
A first full characterisation of bistability in the case of cylindrical bars was performed by \cite{kim1988investigation}.
Among other things, they computed the distribution of waiting times and the parametric dependence
of the mean waiting time.
In the last 25 years, it has become possible to measure the planar velocity field
in the wake of the bars by means of PIV \citep{sumner1999fluid,alam2011wake,chen2021turbulence}.
Meanwhile the flow has become amenable to Large Eddy Simulations \citep{afgan2011large}. 
The different types of bifurcations present and the variety of available means to
investigate this flow encourages to use it as a test case for methods of processing and modelling of multistable flows.
In particular, the possibility to access velocity fields quantitatively will give
us the possibility to compute dedicated scalars from fields obtained numerically or experimentally,
and construct stochastic models for their time evolution.

In this work, we intend to investigate the bistable wake between two square bars experimentally, at the level of the velocity fields.
This will be the occasion to test the ability of model training methods to characterise the change of regime between multistable and monostable flows.
For this matter we will use the dataset sampled by \cite{chen2021turbulence}, and we will extend this dataset to
have a time resolved view of the change of direction of the jet and perform a parametric study in gap ratio.
The properties of these datasets and their experimental sampling are presented in section~\ref{sexp}.
In section~\ref{sproc}, we present how we process the resulting datasets to obtain scalars used to follow quantitatively the
changes of direction of the jet and changes of regime.
Using these time series, we will train parametric stochastic models.
The principles of the models and their construction is presented in section~\ref{smodc}.
In section~\ref{sres}, we present the flow states as the gap ratio is
increased, and the flow goes from tristable to bistable to monostable (\S~\ref{sover}),
using visualisations and our scalars.
We then present how the main features of the model can capture the properties
of multistability as well as more complex situations (\S~\ref{srmod}).
The scenarii of regime changes are then presented in a
synthetic manner using bifurcation diagrams constructed from
the various processing techniques and the models (\S~\ref{sdiag}).
Finally, the results are discussed in the conclusion (\S~\ref{sconc}).

\section{Experimental setup and measurement method}\label{sexp}

In this section, we present the experimental facility and the procedure
used to sample the data analysed in this article.
We firstly present the wind tunnel in which the measurements have been carried out, our notations and
the setup of the two parallel bars (\S~\ref{tun}).
We secondly present  the main properties of the three measurement campaigns (\S~\ref{dat_expc}).   
Further technical details on PIV set-ups and experimental conditions will be given in appendix~\ref{app_exp}.

\subsection{Properties common to all experimental campaigns: LMFL wind tunnel, bar configurations and PIV measurements}\label{tun}

All the data presented in this article were sampled in the Laboratoire de M\'ecanique des Fluides de Lille (LMFL) closed loop wind tunnel (visible in photograph in Fig.~\ref{skint} (a)).
The detailed properties of the tunnel are presented by \cite{carlier2005experimental}. The flow in the tunnel was further characterised by \cite{cuvier2017extensive}.
The tunnel cross section has a height of $1\text{m}$ and a width of $2$m.
The temperature of the air in the tunnel is kept within $0.15\text{K}$ of the command value.
The atmospheric pressure is free and is systematically measured each measurement day (pressure variations of several hPa over a day are also noted).
The double bar setup is visible in the photograph in figure~\ref{skint} (a), seen from upstream.
The bars are placed at  $5.6$m from the entrance of the test section of the tunnel, at an equal distance of less than one meter from the walls of the tunnel.
The setup is sketched, seen from above, in figure~\ref{skint} (b).
We indicate the square bar width $H=3\cdot 10^{-2}$m and $G$, the variable gap between the bars, measured from one bar centre to the next.
We will denominate the streamwise direction of the incoming flow by $x$ and the spanwise direction of the bar alignment by $y$.
By an abuse of language, the $y>0$ positions are termed the left side and $y<0$ positions the right side.
The origin of the coordinate system is set on the downstream face of the bars in $x$, at middistance of their inner corners in $y$.
In all experimental campaigns, in the core of the cross section of the tunnel, a steady streamwise flow of incoming velocity $U_\infty=5.0\text{m}\cdot\text{s}^{-1}$ arrives on the bars.
The incoming rate of turbulence was  tabulated by \cite{carlier2005experimental} and measured again during the third experimental campaign (\S~\ref{app_exp}.\ref{sup_param}).
It is defined as the standard deviation of the incoming velocity divided by $U_{\infty}$.
It is measured by means of Hot Wire Anemometry and it is below $0.3\%$.
A statistically steady turbulent boundary layer is tripped
at the entrance of the test section of the tunnel
on the top and bottom walls of the tunnel \citep{cuvier2017extensive}.
It has been verified that it displayed the tabulated properties of such a boundary layer all along the wind tunnel.
At the streamwise position of the bars, the turbulent boundary layer on the top and bottom walls is approximately $0.1$m thick
and can thus engulf the rails (of height $4.5\cdot 10 {-2}\text{m}$, \S~\ref{app_exp}.\ref{compiv}) and the top and bottom of the bars.
Furthermore, a boundary layer of thickness up to  $0.1$m can be found on the side walls of the tunnel \citep{cuvier2017extensive}.
We thus expect that even the outermost part of the flow around the bars, that detaches from the downstream outer
corners of the bars, is not impacted by these side walls.
Due to the friction on the walls of the tunnel in the boundary layers,
a moderate head loss is found in the test section that leads to
a pressure gradient of $-0.2\text{kg}\cdot\text{m}^{-2}\cdot\text{s}^{-2}$.

In figure figure~\ref{skint} (b), we sketch in green the Field of View.
It is the rectangle of side $\Delta x$ and $\Delta y$ downstream of the bars, where the two components of horizontal velocities
$u_x$ and $u_y$ are measured by means of two dimensional two components Particle Image Velocimetry (PIV).
Its centre is placed at $y=0$ and $x=x_c$.
It will systematically be placed at mid vertical height (approximately 0.5m)
in the tunnel, well outside the top, bottom and sides boundary layers.
In PIV measurements, velocity fields will be sampled at frequency $f$, on a grid of step ${\rm d}x$.
Continuous runs contain $N_t$ time frames. The processing of images for PIV ensures a smallest resolved length of about 
the interrogation window size $\delta x$.

Using the control parameters of the flow, it is possible perform the classical non-dimensionalisation of high Reynolds number flows.
Most lengths will be of order $H$, while most velocities will be of order $U_\infty$. 
As a consequence, the kinetic energy of the flow is of order $U_\infty^2/2$.
Similarly, we expect pressure gradients of order $\rho U_\infty^2/2H$
(which are three orders of magnitude larger than
the average pressure gradient along the tunnel).
The natural time scale of wake flows is $H/U_\infty=6\cdot 10^{-3}$s. It is commensurate with the observed
periods $T_{vs}$ of vortex shedding in that wake. This can for instance be checked from the measured
Strouhal numbers of the flow $H/U_\infty T_{vs}$, which are contained in the interval $[0.1, 0.4]$ \citep{spivack1946vortex,kim1988investigation}.
As it has been reported in the literature and as we will see in this article, 
all time scales associated with multistability are several orders of magnitude larger than $H/U_\infty$.
The flow will be controlled by two dimensionless control parameters: the gap ratio $G/H$ and the Reynolds number $Re=HU_\infty \rho/\mu$.
Since the temperature is controlled, the dynamic viscosity of the fluid is constant. 
The small variations of atmospheric pressures will lead to small variations of density $\rho$.
The control parameters are such that the Reynolds number will always be at most $5$\% below $10^4$.
As a consequence, we will present data where length are rescaled by $H$, velocities by $U_\infty$, kinetic energy by $U_\infty^2$
as a function of physical time and the dimensionless control parameter $G/H$, similarly to \citep{kim1988investigation,afgan2011large}.
This choice leads to moderate values of durations, while all other quantities are of order $1$.

\subsection{Experimental campaigns and datasets}\label{dat_expc}

In this article, we use data measured in three experimental campaigns, each with different measurement control parameters.
The data sampled in these three successive campaigns are respectively termed the first dataset, the second dataset and the third dataset.
Inside each dataset, records are performed in what we term a \emph{group of runs}, where $N$ runs
of velocity fields recorded in a given duration (the same for each run) are recorded one
after the other, with a short time of approximately $10$s between the end of a run and
the beginning of the next.
In this section we present the rationale behind each campaign and dataset, the
values of the control parameters directly impacting our study (summarised in table~\ref{expparam}) and the main properties of PIV measurements.
The numerical values of all control parameters are given in the appendices (\S~\ref{app_exp}.\ref{app1std},~\ref{app_exp}.\ref{100hz} and~\ref{app_exp}.\ref{sup_param}), along with
the details of the optics mounted on the cameras and properties of laser sheets used for PIV measurements. 
They will depend on the size of
the field of view, the sampling frequency and the purpose of each specific campaign.

The first dataset studied in this article was sampled by \cite{chen2021turbulence} (see also \S~\ref{app_exp}.\ref{app1std}). 
It was originally designed to investigate out of equilibrium turbulence in a wake. 
The presence of the bistable jet led us to include this dataset in our study.
The velocity fields were measured using a setup sketched in figure~\ref{skint} (d),
in a Field of View larger than what will be used in the next two datasets, 
at three gap ratios where three distinct regimes of multistability were seen. 

The second dataset was sampled at a higher frequency of $f=100$Hz to study transitions in themselves.
Indeed, it had been observed in the first dataset sampled at $f=4$Hz that
the jet changes direction over a duration which is less than half a second.
This is visible in time series following
the position of the jet (Fig.~\ref{posyw} (d)). This had also been observed by \citep{kim1988investigation} (see their figure~4). 
The comparison is relevant because the time scales in their experiments are of the same order of magnitude as ours.
As a consequence, this direction change was underresolved in time in the first dataset, while it
is time resolved in this second dataset.
While the properties of transitions is beyond the scope of this article, 
the sampled data proved an enlightening complement for the study of multistability regime changes. 
Firstly, this dataset will help us validate some hypotheses required to model our process by a stochastic differential equation. 
Secondly, we use it to test the sensitivity to controlled experimental imperfections, mostly obtained by turning either of the bars by a positive
or a negative angle (Fig.~\ref{skint} (c)).
In this dataset, we focused on a single gap ratio of $\frac{G}{H}=2.4$,
given the amount of data required to quantitatively characterise transitions.
For this measurement campaign, the field of view for PIV measurements was also focused on the jet (see figure~\ref{skint} (e)).
It was observed (see Fig.~5 of \citep{chen2021turbulence}, and Fig.~\ref{posyw}, (a,b), Fig.~\ref{vis1-25} (b,c,d), Fig.~\ref{vis3-5} (b,c) of this article)
that the jet extends at most to four $H$ downstream of the bars (an approximate distance of $1.2\cdot 10^{-1}$m),
and approximately $1.5H$ on the sides (an approximate distance of $ 4.5\cdot 10^{-2}$m).
This led us to choose a field of view with a size of approximately $\Delta x\times\Delta y\simeq 10^{-1}\text{m}\times 10^{-1}\text{m}$ for this
campaign and the next.
We recorded 130 runs in this high frequency dataset. 
We give details on the selection of the corresponding runs
in the appendix (\S~\ref{app_exp}.\ref{100hz}).

The third dataset was also specifically sampled for this work.
It had been observed in the velocity fields from the first dataset, sampled at three values of $G/H=1.25$, $2.4$ and $3.5$,
that the flow had three distinct regimes (tristability, bistability and monostability, respectively) \citep{chen2021turbulence}.
The purpose of this dataset was to perform a parametric study for $G/H$ in the range $[1.15, 3.5]$,
in order to investigate more finely the changes of regime in the flow as the bars are moved apart from one another.
Similarly to the second dataset, we chose a field of view focused on the jet. Similarly to the first dataset, we
chose a lower sampling rate so as to be able to record long enough continuous datasets containing several direction changes of the jet.
The parameters of our PIV measurements are nevertheless slightly different from what we had used in the first and second dataset, as they were
specifically tailored for the need of the parametric study.
Note that for some values of $G/H$, several groups of runs were recorded on different days (with distinct numbers of runs, atmospheric pressure
and details of the bars assembly). They will be distinguished with v followed by the number of the group of runs.
For instance, we will present the data processing using the velocity measured in the group of runs $G/H=2.4$v2,
the second out of the four groups of runs obtained at $G/H=2.4$.
Similarly,  we will examine the difference between the properties of multistability in the two groups of runs 
obtained at $G/H=1.8$, denominated $G/H=1.8$v1 and $G/H=1.8$v2.
The list of gap ratios, the list of groups of runs for these gap ratios, the number of runs per group of runs, and the corresponding values of Reynolds numbers
for this experimental campaigns are given in the appendix (\S~\ref{app_exp}.\ref{sup_param}).
Details on some experimental imperfections are also given in this appendix.  

\begin{table}
\begin{center}
\begin{tabular}{|c|c|c|c|c|c|c|c|c|}
\hline Dataset & $f$ (Hz) & $\frac{G}{H}$ &$dx$ &$\delta x$  &$\Delta x$ & $\Delta y$ & $x_c$ & $N_t$ \\ \hline
First dataset &4 & 1.25, 2.4, 3.5  &$0.761$&$1.6$ &741& 158&387 & 2000 \\ \hline
Second dataset  &100 &  2.4 &$0.629$&$1.5$  &96.9&102 &52 & 1550 \\ \hline
Third dataset &5 & $[1.15, 3.5]$  &$0.45$& $1.2$ &95.1&113 & 51.1 & 10000 \\ \hline
 \end{tabular}
\end{center}
\caption{Table giving the main properties of the velocity fields sampled in each PIV measurement campaign.
We indicate the corresponding dataset,
the frequency of acquisition $f$ (in Hz), the gap ratios $\frac{G}{H}$,
the grid step $dx$  on which the velocity fields are sampled,
the smallest scale  $\delta x$  resolved by PIV measurements (see \cite{soria1996investigation} and \S~\ref{app_exp}.\ref{compiv} for details), 
the streamwise and transverse size of the fields of view $\Delta x$ and $\Delta y$,
 the position
of the centre of the field of view $x_c$ and the number of time frames in a run $N_t$.
All lengths are given in $(10^{-3})$m.
For a full description, see \citep{chen2021turbulence} and appendix~\ref{app_exp}.}
\label{expparam}
\end{table}

\section{Data processing}\label{sproc}

In order to characterise the multistable flow quantitatively and determine how
its properties change with $G/H$, several diagnostics have been computed from the
velocity fields measured by means of PIV.
We have estimated the spanwise position of the jet (\S~\ref{jpos}), from
which order parameters can be derived, as well as the jet width (\S~\ref{jwid}), so as to further characterise the disappearance of
multistability. These scalar quantities will come in the form of time series, from which we will extract the time
intervals where the jet points in one direction or another (\S~\ref{stated}).
They will be used to derive the properties of the flow when the jet points in a specific direction.

\subsection{Computation of jet position and width}


\subsubsection{Jet position}\label{jpos}

It had been noted in the literature \citep{kim1988investigation,alam2011wake} and in the study by Chen \emph{et al.}
that the jet could point toward the left (positive values of $y$),
toward the right (negative values of $y$), or in some cases along the centreline (values of $y$ close to zero).
In order to follow time series of the direction of the jet, we systematically compute $Y_m(t)$,
the $y$ position of the jet spatially averaged in a window,
using a weighting by the in-plane kinetic energy $E_k=\frac{u_x^2+u_y^2}{2}$.
We will systematically use $E_k$ to visualise and characterise the jet: \cite{chen2021turbulence} showed using the numerical data of \cite{zhou2019extreme},
that it contained more than 90\% of the total kinetic energy in the area we focus on.
The streamwise component of velocity will be the main contributor to the in-plane kinetic energy, owing to the moderate angle of the jet relatively to the plane $y=0$.

\begin{figure}[!ht]
\centerline{
\includegraphics[width=5.5cm]{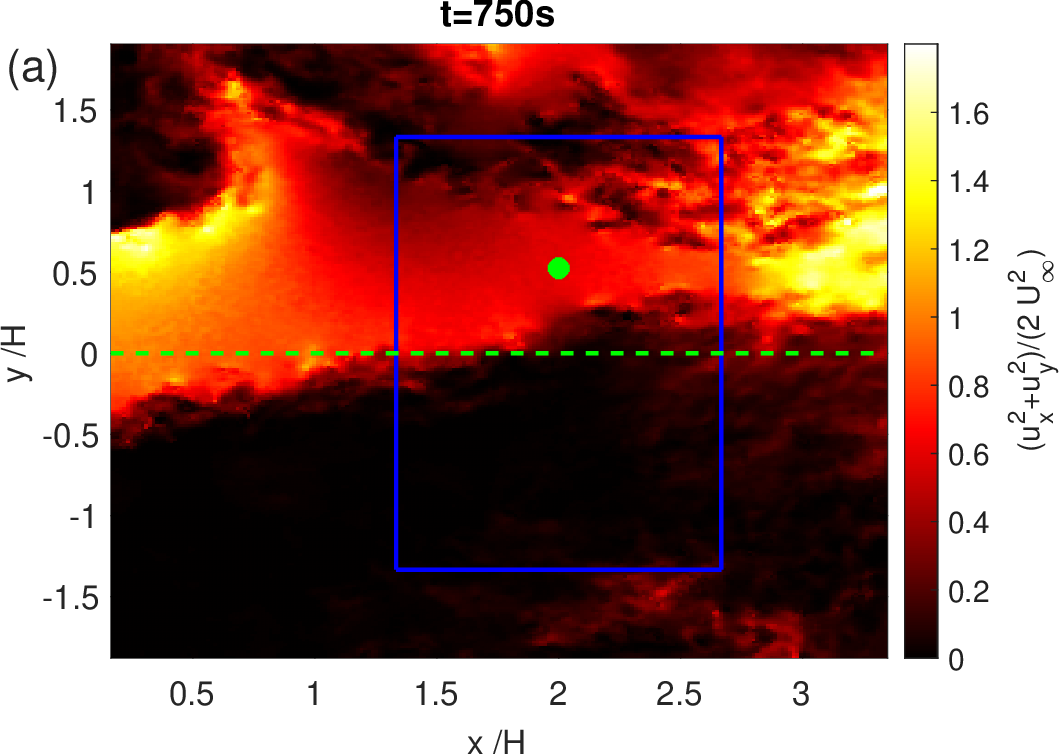}
\includegraphics[width=5.5cm]{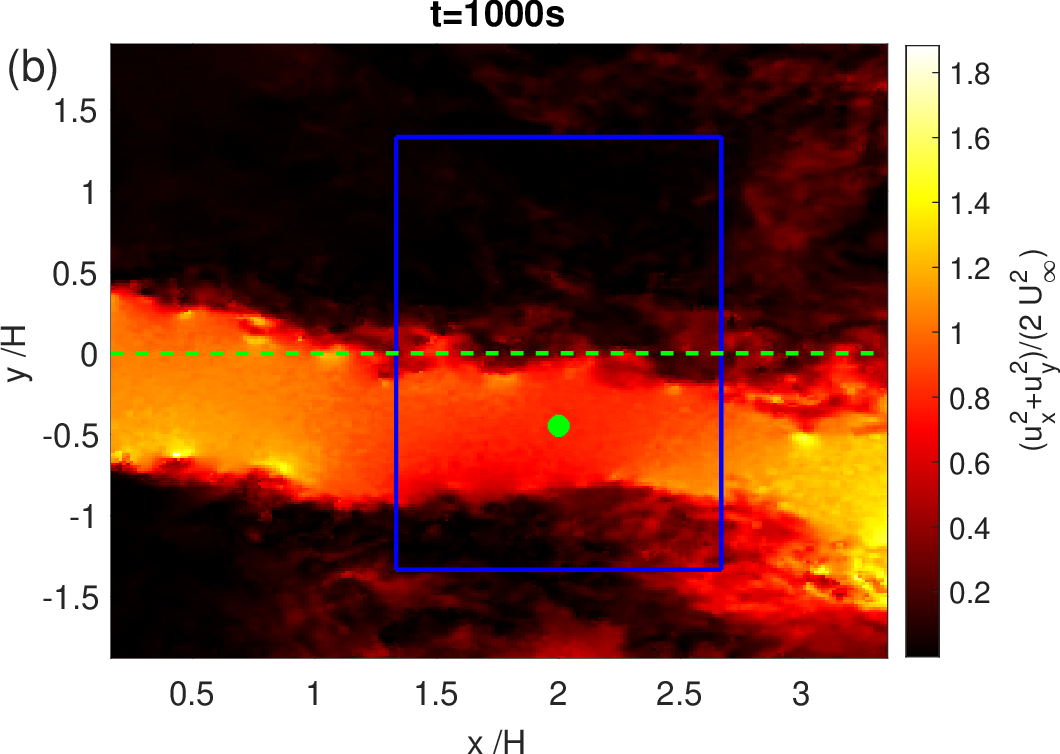}
\includegraphics[width=5.5cm]{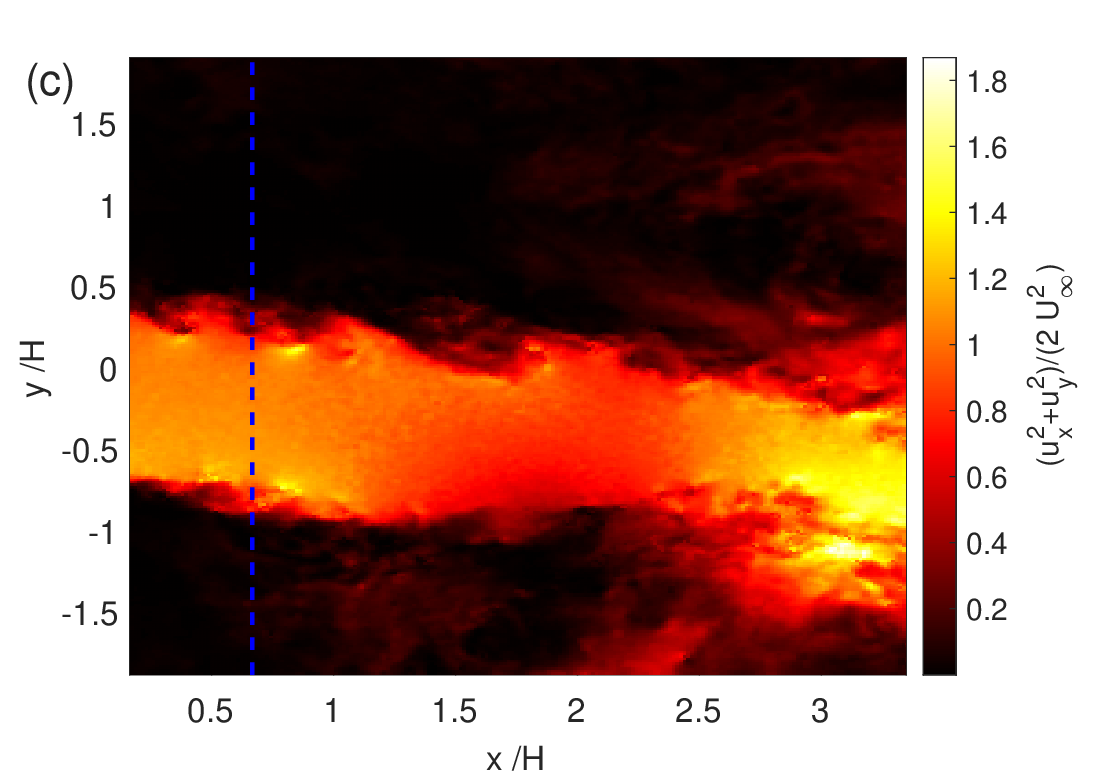}
}
\centerline{
\includegraphics[width=14cm,clip]{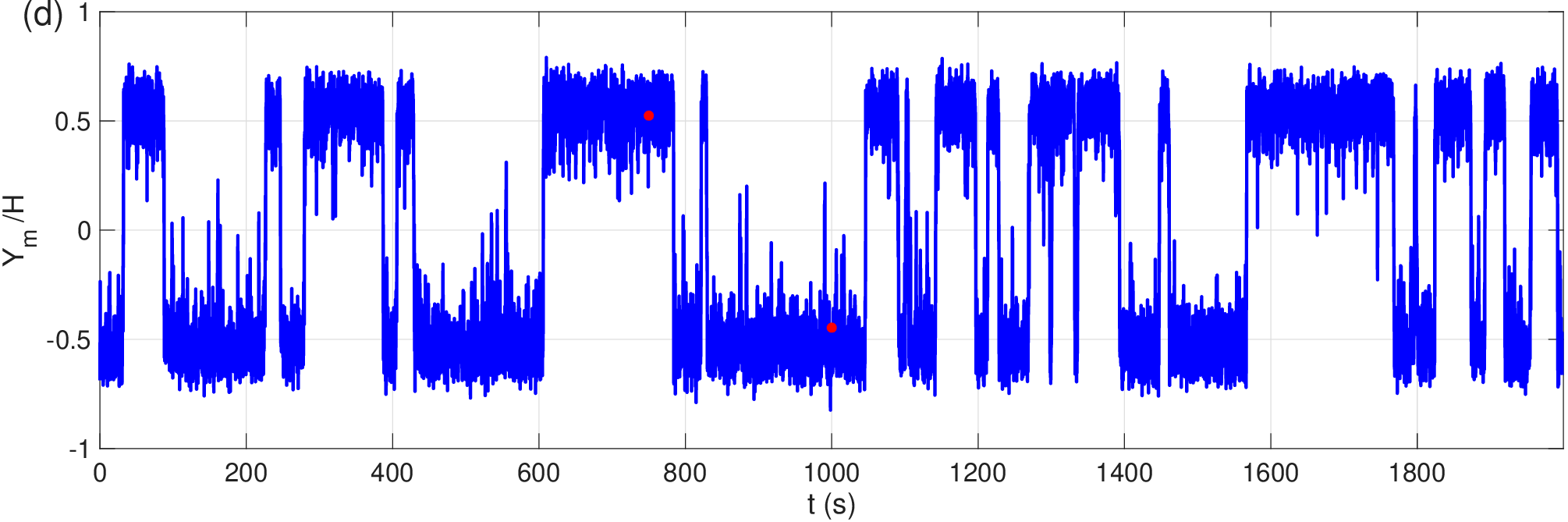}
\includegraphics[width=6.5cm]{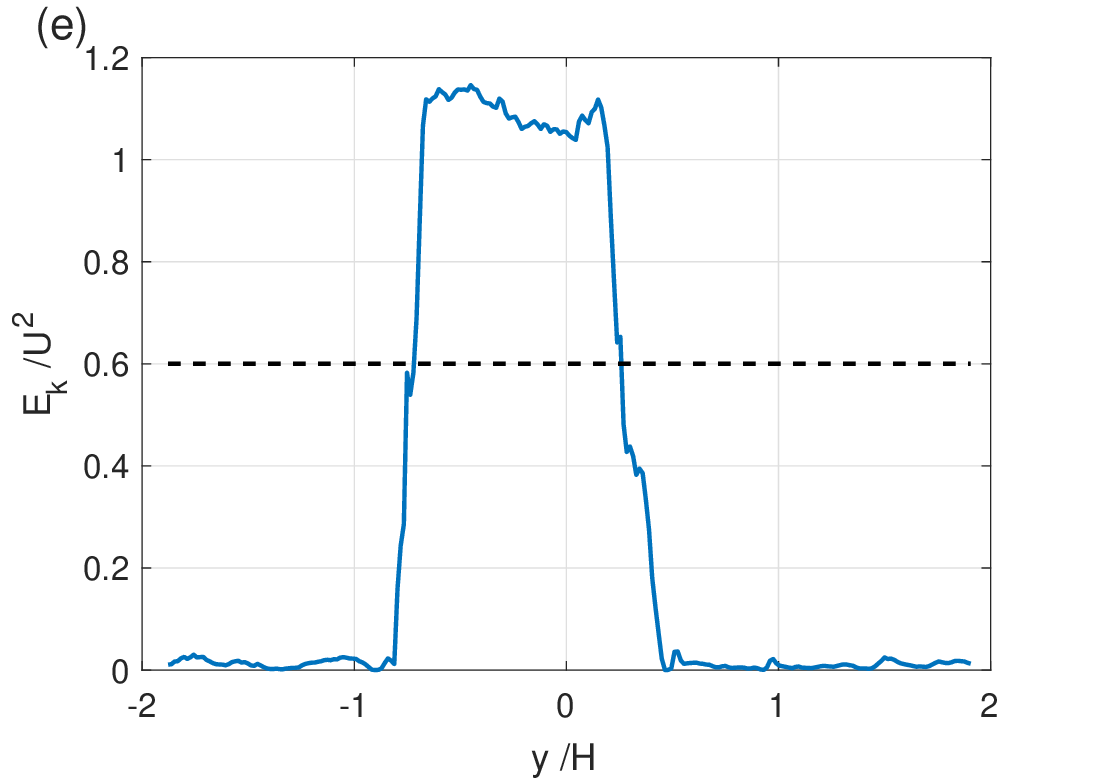}
}
\caption{Colour levels of the in-plane kinetic energy $(u_x^2+u_y^2)/(2U_\infty^2)$ computed using the velocity field measured by means of PIV in the horizontal field
of view in the third experimental campaign with gap ratio $G/H=2.4$ v2
(a) at time $t=750$s,
(b) at time $t=1000$s.
The blue frames indicate the region in space $4H/3<x<8H/3$, $-4H/3<y<4H/3$ 
used to compute $Y_m$ and the dashed green line indicates  the line $y=0$.
For both panels, the green dots are placed at $x=2H$ and $y=Y_m(t)$ so as to indicate the weighted position of the jet.
(c) In plane kinetic energy from the same run at time $t=1500$s.
 The blue dashed line indicates the position $x=2H/3=0.02$m used for the
measurement of the jet width.
(d) Time series of the position of the jet $Y_m$ computed from the same dataset as the kinetic energy shown in (a,b) (using (Eq.~(\ref{defym}))).
The two red dots indicate the instants in time where colour levels of panels (a,b) are computed.
(e) Profile of kinetic energy as a function of $y$ obtained at $x=2H/3$ 
from the field of $E_k/U_\infty^2$ displayed in (c). The black dashed line indicates
the threshold used for the computation of the width of the jet.}
\label{posyw}
\end{figure}

Quantitatively, $Y_m$ results from the weighted average of $y$ in the window downstream of the bars defined as $x\in [4H/3, 8H/3]=[0.04,0.08]$m,
and $y\in[-4H/3,4H/3]=[-0.04,0.04]$m.
This window is shown on top of colour levels of the rescaled in-plane kinetic energy $(u_x^2+u_y^2)/(2U_\infty^2)$
in figure~\ref{posyw} (a,b), at two successive instants.
The $x$ interval starts far enough downstream of the bars so that the variation of spanwise position of the jet there is large enough,
and ends at a position which is contained in all fields of views.
The $y$ interval is large enough so as to contain the most intense values of in-plane
kinetic energy corresponding solely to the jet, for all jet positions and
for all values of the control parameters. In particular, in the case of the first dataset,
this excludes the range of $y$ where the flow that has detached from the bar corners.
These values of $y$ where ``parasitic'' values of $E_k$ are found 
can fall to as low as $|y|>5H/3$ for smaller gap ratios (see for instance Fig.~\ref{vis1-25} (b,c,d) and \citep{alam2011wake,chen2021turbulence}).

Our practical computation of $Y_m$ takes into account the fact that
the velocity fields measured by means of PIV are discretised on a rectangular grid.
We choose to write the spatial dependence of $u_x$, $u_y$ using the discrete position of grid points $x$ and $y$ 
(related to grid indices $n_x$ and $n_y$, such that $x=x_c-\frac{\Delta x}{2}+n_x{\rm d}x$ and $y=-\frac{\Delta y}{2}+n_y {\rm d}x$).
For each field of view, we determine $[N_{x,1},N_{x,2}]$, the range of streamwise indices $n_x$
such that the corresponding $x$  
position is in the required interval  $4H/3\le x\le 8H/3$. 
Similarly, we determine $[N_{y,1},N_{y,2}]$ the range of spanwise indices  $n_y$
such that the corresponding $y$  
position is in the required interval $-4H/3 \le y\le4H/3$.  
The indices $N_{x,1}$, $N_{x,2}$, $N_{y,1}$ and $N_{y,2}$ are such that $x(N_{x,1/2})=X_{1/2}$ and
$y(N_{y,1/2})=Y_{1/2}$ are the first grid positions greater than or equal to the corresponding bounds.
We then define $Y_m$, the weighted $y$ position of the jet
\begin{equation}
Y_m(t)=
\frac{\sum_{x=X_{1}}^{X_{2}}\sum_{y=Y_{1}}^{Y_{2}}y(u_x^2(x,y,t)+u_y^2(x,y,t))}
{\sum_{x=X_{1}}^{X_{2}}\sum_{y=Y_{1}}^{Y_{2}}(u_x^2(x,y,t)+u_y^2(x,y,t))}
\,,
\label{defym}
\end{equation}
This ratio of sums over the discrete positions is
 an approximation of an integrated weighted average 
$Y_m \simeq \frac{ \int_{x=4H/3}^{8H/3}\int_{y=-4H/3}^{4H/3} y \frac{u_x^2+u_y^2}{2}\,{\rm d}x{\rm d}y}{\int_{x=4H/3}^{8H/3}\int_{y=-4H/3}^{4H/3} \frac{u_x^2+u_y^2}{2}\,{\rm d}x{\rm d}y }$.


We indicate how our definition of $Y_m$ leads to a precise estimation of the jet position.
Colour levels of in-plane kinetic energy computed using the velocity fields
measured by means of PIV in the third dataset at $G/H=2.4$ v2 (first run)
are presented at $t=750$s and $t=1000$s in figure~\ref{posyw} (a) and (b) respectively. We note that
the in-plane kinetic energy takes non negligible values mostly for $y>0$ in the first case and
mostly for $y<0$ in the second case.
This is stressed by the position of the jet relatively to the green dashed line.
We compare these two visualisations to the time series of $Y_m(t)$ computed from the same group of runs in figure~\ref{posyw} (d).
The corresponding values of $Y_m$ at $t=750$s and  $t=1000$s are indicated by the two red dots in the time series.
additionally, they are indicated as a green dot on top of the colour levels of the in-plane kinetic energy in figure~\ref{posyw} (a,b).
This shows us that $Y_m$ is larger than $0$ when the jet is on the left and smaller than zero when the jet is on the right.

This scalar is directly related to measurements of the jet angle with respect to the centreline \citep{kim1988investigation}
performed using a cross hot wire anemometer (see \citep{bruun1996hot} \S~5.4 for the principle of the technique).
That angle can also be measured in LES of the flow \citep{afgan2011large}.
Moreover, since we find a lower pressure on the bar toward which the jet is pointing,
there is also a monotonous relation between $Y_m$ and the difference of the base pressure coefficients measured on the bars \citep{kim1988investigation,alam2011wake}.
This position is also related to another measure of the symmetry breaking: the fact that the two recirculations
bubbles downstream of each square bar have two different lengths and widths \citep{alam2011wake}.
Note finally that similar wake position estimation methods, using a weighting obtained from the flow fields, have also been used.
For instance, in the case of the wake of the Ahmed body, the two dimensional centroid of the wake have been computed \citep{haffner2020mechanics}.
Again, that observable gives an information similar to components of the pressure gradient \citep{barros2017forcing}.

\subsubsection{Jet width}\label{jwid}

We use a second diagnostic to follow the state of the jet in the form of the width of the
jet at its base $w$.
It is fairly straightforward to access this information from velocity fields measured by means of PIV, in contrast to measurements of base pressure
on the bars.
 This quantity will help us characterise the disappearance of bistability
when the gap ratio $G/H$ is changed.
For this matter, for any given dataset, we set a common base at $x\simeq 2H/3$  
and we determine $x_{w}=\arg\min |x-2H/3|$, 
the closest corresponding index on the grid.
This position is indicated by the blue dashed line on top of the colour levels of in-plane kinetic energy in figure~\ref{posyw} (c).
We determine the location of the jet using the in-plane kinetic
energy as a function of $y$ at this streamwise index
\begin{equation}
E_k(x_{w},y,t)=\frac{u_x(x_{w},y,t)^2+u_y(x_{w},y,t)^2}{2}\,.
\end{equation}
An example of the in-plane kinetic energy on the line $x\simeq 2H/3= 0.02$m is displayed in figure~\ref{posyw} (e).
It has a step-like dependence in $y$: we will use this to measure the width.
We consider that the jet is present if this kinetic energy is larger than $E_k\ge E_{k,w}= 0.6U_\infty^2=15\text{m}^2\cdot\text{s}^{-2}$
 (as indicated by the black dashed line).
This value is approximately half of the maximum of the kinetic energy inside the whole jet
in most cases (which goes from approximately  $6U_\infty^2/5$ 
at $G/H=1.15$
to approximately  $9U_\infty^2/5$ 
at $G/H=3.5$),
so as not to include secondary fluctuations of velocity (on the sides)
nor miss parts of the jet.
 The case displayed here is typical  of what can be observed in the flow:
the in-plane kinetic energy decreases very fast away from the jet
(where $E_k$ is of order $\mathcal{O}(0.04U_\infty^2)$)  
so that the jet detection is rather insensitive
on the precise value of the threshold $E_{k,w}$.
The jet width is then $w=n_y^t{\rm d}y$, where $n_y^t$ the number of gridpoints where $E_k(x_{w},y,t)\ge E_{k,w}$.
This width will increase with the width of the jet if it is continuous, while it will be small if the jet is broken down.


\subsection{State detection and computation of exit times}\label{stated}

From the observation of the time series of figure~\ref{posyw} (d),
we visually note that $Y_m$ fluctuates either around a positive local average value (of order $\mathcal{O}(0.5H)$) 
or a negative average value (of order $\mathcal{O}(-0.5H  )$)) 
with a standard deviation (of order $\mathcal{O}(H/6)$)  
that appears notably smaller than the average
when the jet is on the left or on the right (the difference between average positions being five to six times the standard deviation).
We also note that $Y_m$ changes sign rapidly as the jet goes from left to right or right to left.
This type of time signal is typical of bistability \citep{kim1988investigation,alam2003aerodynamic,rolland2011pattern}.
We would like to determine systematically the jet direction (left or right) and measure the time spent on
one side or the other as well as the properties of the jet when it is on either side (average position and standard deviation).
For this matter, we use a method to detect on which side (left or right) the jet is
that we had already used to study bistability in a different turbulent flow \citep{rolland2011pattern}.

\begin{figure}
\centerline{
\includegraphics[width=16cm,clip]{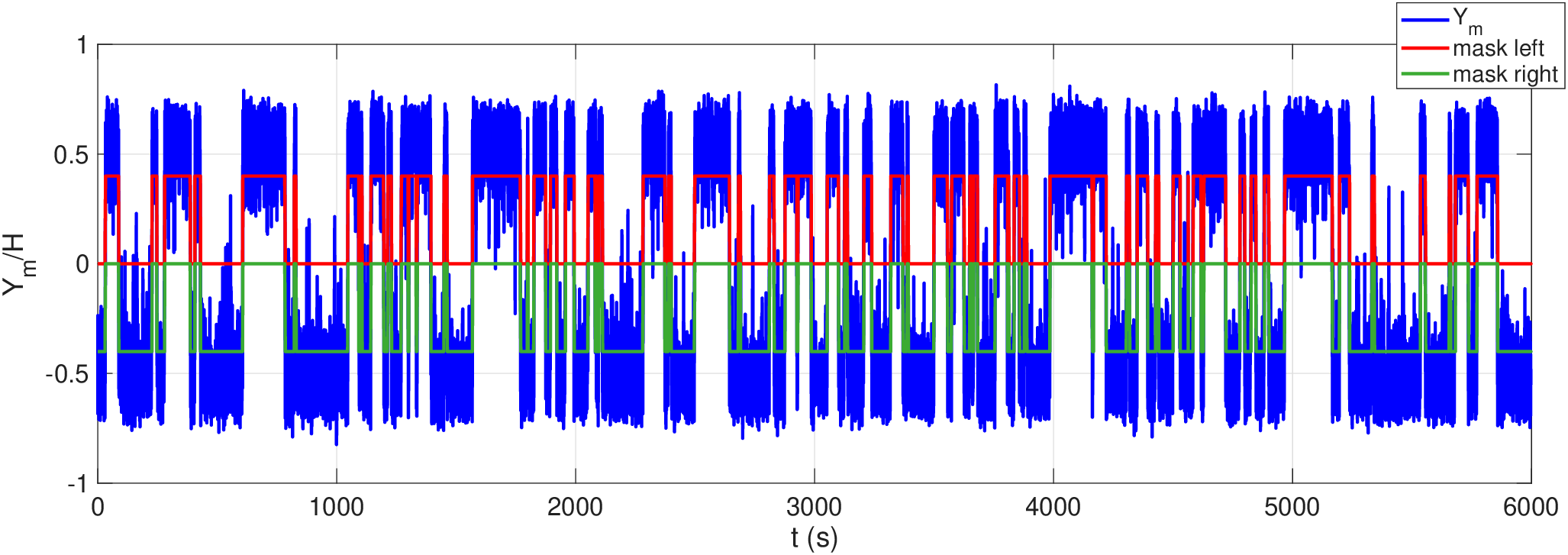}}
\caption{
 Time series of $Y_m$ computed from velocity fields sampled at $G/H=2.4$ v2 during the third experimental campaign,
the masks highlighting the presence in states $Y_m>0$ and $Y_m<0$ are superimposed.
}
\label{maskcdf_a}
\end{figure}

We proceed in the following manner:
\begin{itemize}
\item We state that the jet enters the left state if the jet was not in it (either on the right or neither) and $Y_m(t)>0.4H$. 
The jet is then in the left state until it enters the right state.
\item Conversely,  we state that the jet enters the right state if the jet was not in it (either on the left or neither) and $Y_m(t)<-0.4H$  
The jet is then in the right state until it enters the left state.
\end{itemize}
At the beginning of each group of runs, we assume that the jet is in neither states.
If the jet is detected to be in the same state at the end of a run and at the beginning of the next in a given group,
we assume that it has not changed direction in the short time interval separating the two runs.
We changed the exit condition when processing $Y_m(t)$ computed at $G/H=1.25$, where the jet is tristable (Fig.~\ref{vis1-25}).
The jet is considered to exit the left state when it reaches the centreline if $Y_m(t)<-H/30$  
and exits the right state when it similarly reaches the centreline if $Y_m(t)>H/30$  

We present the resulting state detection in figure~\ref{maskcdf_a}, where
we display the time series of $Y_m$ computed from velocity fields measured in the third dataset in the group of runs $G/H=2.4$ v2.
On top of this curve, we add a mask which is $0.4H$  
when the jet is detected to be on the left and $0$ otherwise
and a mask which is $-0.4H$ 
when the jet is detected to be on the right and $0$ otherwise.
Comparison of the curves indicates that the jet indeed fluctuates on the left or on the right when it is detected to be as such.
This procedure does not miss changes of direction nor proposes changes of direction when only a short excursion occurred.
The procedure works because the value of the threshold chosen here is not arbitrary.
It should not be so large that $Y_m(t)$ never goes above it (in absolute value) and
not so small that it detects irrelevant changes of direction while the jet is only fluttering
near the centreline \citep{lestang2018computing}.
This can happen when a distinction based on a single limit in between the two multistable state  is used \citep{alam2003aerodynamic}.
Practice indicates that choosing a threshold given by the conditional average of $Y_m$ when it is on the left
minus the corresponding condition standard deviation to detect the left position (and vice versa for the right one),
meets our criteria. We note that the value chosen here is self-coherent for all values of $G/H$
for which bistability of the jet direction is observed.

Using these state detections, we compute first passage times and their distribution (see appendix  \S~\ref{mmfpt}) 
which is used to discuss the Gaussian white noise hypothesis invoked for modelling. 
We also compute averages and standard deviations of $y$,
conditioned to being in the left or right states,
comparable to what has been done by \cite{alam2003aerodynamic} for force coefficients.
From the time series displayed in figure~\ref{maskcdf_a},
we find a conditional average of $Y_m$ on the left position of $0.53H$  
with a conditional standard deviation of $0.123H$. 
Similarly, we find a conditional average of $Y_m$ on the right position of $0.52H$ 
with a conditional standard deviation of $0.133H$.  
This indicates that the procedure is self-consistent:
We find that the value of the conditional average for $Y_m$ on the left (resp. on the right) minus (resp. plus)
the conditional standard deviation is approximately
equal to the threshold $0.4H$ 
(resp. $0.4H$ ). 
We will see in section~\ref{sbdiag} that for all values of $G/H$ considered here,
this criterion will give similar consistent measurements.

\section{Stochastic model identification}\label{smodc}

In this section, we first present the type of stochastic models
used to represent the time series of $Y_m$: Langevin equations (\S~\ref{slang}).
We then present how we compute empirical finite time Kramers--Moyal coefficients and histograms 
from the time series of $Y_m$ so as to extract the information used for modelling  (\S~\ref{ssemp}).
Finally, we present the generic polynomial expression of our models and the outline of  the method used to construct the models (\S~\ref{slreg}).
Further details on the models and the procedure are presented in appendix~\ref{app_sdei}.

\subsection{The target models: Langevin equation and finite time Kramers--Moyal coefficients}   
\label{slang}

In this subsection, we first present the features of the Langevin equations we will construct in this article (\S~\ref{slfpeq}).
In that framework, we then present how we can access the coefficients of the stochastic differential equations
from time series using the finite time Kramers--Moyal coefficients (\S~\ref{sKM}).

\subsubsection{The Langevin equation} 
\label{slfpeq}

We choose to model the sampled time series of $Y_m(t)$ by a stochastic process $Y(t)$, governed by
a one dimensional Langevin equation (see \citep{gardiner2009stochastic} \S~4.3).
We drop the subscript ${}_m$ to denote the time series originating from such models.
We write the Langevin equation in a discretised form as
\begin{equation}
{\rm d}Y(t)=f(Y(t)){\rm d}t+g(Y(t)){\rm d}W(t)\,,\label{leq}
\end{equation}
where ${\rm d}Y(t)=Y(t+{\rm d}t)-Y(t)$ is the increment of $Y$ between two time steps separated by ${\rm d}t>0$.
The right hand side of the Langevin equation comprises two terms that govern two distinct parts of the dynamics.
The drift $f$ depends solely on $Y$ and is thus deterministic.
Meanwhile $g$ is termed the diffusion and modulates the randomness in the process:
it can be viewed as the standard deviation of the noise.
Said noise is driven by the discrete Wiener process ${\rm d}W(t)$ which 
is a sequence of independent Gaussian random variables of average $\langle {\rm d}W\rangle=0$ and
covariance $\langle {\rm d}W(t){\rm d}W(t')\rangle=\delta(t-t'){\rm d}t$.
Note that $g$ can depend on $Y(t)$, in which case the noise is multiplicative.
As a consequence, in this Langevin equation, we choose an It\^o convention, so that $g(Y(t))$ and ${\rm d}W(t)$ are uncorrelated: $\langle g{\rm d}W\rangle=0$.
We choose to model our data by a stationary process, implying that both the drift $f$ and the diffusion $g$ have no explicit time dependence.
Finally, the increment of $Y$ between $t$ and $t+{\rm d}t$ solely depends on $Y(t)$, which is equivalent to stating that the process $Y(t)$ is Markovian.
Since the Langevin equation generates a stationary process, the probability for $Y(t)$ to have a value $y$ is given by $\rho(y)$ the steady PDF of
the system
\begin{equation}
\mathbb{P}(Y(t)=y)=\rho(y)\,.
\end{equation}
In the rest of the article, the lower case $y$ will denote the random variable.
We describe how this PDF is controlled by the drift and diffusion as the solution of the steady Fokker--Planck equation of the system in appendix~\ref{app_sdei}.\ref{mod_det}.\ref{app_FP}.

\subsubsection{The proxy: finite time Kramers--Moyal coefficients}\label{sKM}

When training a model using data, constructing the Fokker--Planck equation is more straightforward than directly constructing the Langevin equation.  
The proxies for this training are the finite time Kramers--Moyal coefficients (see \citep{gardiner2009stochastic} \S~11.2). 
We present them in the core of the article as we will discuss them in our analysis.
Given the stochastic process $Y(t)$,
they are obtained from the ensemble average of time increments of $Y$ with step $\tau$ in the
realisations of the process as
\begin{align}
f_\tau(y)=\frac{1}{\tau}\left\langle \left(Y(t+\tau)-Y(t)\right) \right\rangle_{Y(t)=y}\,,\,  
a_\tau(y)=\frac{1}{2\tau}\left\langle \left(Y(t+\tau)-Y(t)\right)^2 \right\rangle_{Y(t)=y}\,.\label{atau}
\end{align}
A first property of the Kramers--Moyal coefficients is that the limit of coefficients of higher order $n$ is zero provided the noise is Gaussian in the Langevin equation: 
\emph{i.e.} $\lim_{\tau\rightarrow 0}\frac{1}{n!\tau}\left\langle \left(Y(t+\tau)-Y(t)\right)^n \right\rangle_{Y(t)=y}=0\,,\, n\ge 3$. 
We have checked this hypothesis both using computation of higher order $n\ge 3$ Kramers--Moyal 
coefficients and with additional properties of the time series of $Y_m$ (\S~\ref{app_sdei}.\ref{mod_det}.\ref{stmark}, \S~\ref{mmfpt}).  
Another fundamental property of these coefficients is that we have
the limits
\begin{equation}
\lim_{\tau\rightarrow 0}f_\tau(y)=f(y)\,,\lim_{\tau\rightarrow 0}a_\tau(y)=a(y)=\frac{g^2(y)}{2}\,,\label{fint}
\end{equation}
where we introduced $a(y)$, the covariance of the noise. 
This tells us that using an information from the process discretised in time (Eq.~(\ref{atau})), 
we can recover the properties of the underlying continuous stochastic differential equation (Eq.~(\ref{fint})).   
Conversely, the finite time Kramers--Moyal coefficients can be obtained from the properties of the underlying model.
They can be computed from well chosen solutions of the backward Fokker--Planck equation.
We describe the computation of these solutions and their use to reconstruct the finite time Kramers--Moyal coefficients in appendix~\ref{app_sdei}.\ref{mod_det}.\ref{app_FP}.

To conclude, in figure~\ref{compaf}, we illustrate the successive functions   mentioned in this section,  that are used in a model construction.
In that figure, we present the drift and first finite time Kramers--Moyal coefficient (Fig.~\ref{compaf} (a)),
the diffusion and second Kramers--Moyal coefficient (Fig.~\ref{compaf} (b)) and finally
the histograms and PDFs of the models (Fig.~\ref{compaf} (c)).

\subsection{Binned empirical estimates: Kramers--Moyal coefficients, histograms}\label{ssemp} 

In this section, we present quantities that are computed using data: 
the empirical finite time Kramers--Moyal coefficients  
and the histograms. 
They will be estimated using our sampled time series and will thus be used as objectives for the model construction.


We firstly present how we compute empirical finite time Kramers--Moyal coefficients from the time series of $Y_m$.
The first coefficient is denoted by $f_\tau^e$ and the second is denoted by $a_\tau^e$.
In practice, they are computed by binning the moments of the time increments on a regular grid of $N_y+1$ points between the minimum and maximum of $Y_m$
over its time series within a group of runs.
The grid points are given by
\begin{equation}
Y_i=\min_{t}(Y_m)+\frac{\max_t(Y_m)-\min_t (Y_m)}{N_y}(i-1)\,.\label{gridcoef}
\end{equation}
For each bin $i$, defined as the interval $[Y_i,Y_{i+1}]$, we have $N_i$ instants in time $t^i$ such that $Y_m(t^i)$ is
in the bin.
We then average over all these instants as
\begin{equation}
f_{\tau}^e(Y_i)=\frac{1}{N_i}\sum_{t^i}\frac{1}{\tau}(Y_m(t^i+\tau)-Y_m(t^i))\,,\,
a_{\tau}^e(Y_i)=\frac{1}{N_i}\sum_{t^i}\frac{1}{2\tau}(Y_m(t^i+\tau)-Y_m(t^i))^2\,.
\end{equation}
If the $Y_m$ are generated by an ergodic stochastic process and the duration of the time series
goes to infinity, these empirical Kramers--Moyal coefficient converge to the average
in the bin of the ensemble averaged Kramers--Moyal coefficients (Eq.~(\ref{atau})).   
The finite time $\tau$ is chosen such that the
process describing $Y_m$ can be viewed as Markovian at time scale $\tau$.
We verify this in practice in appendix~\ref{app_sdei}.\ref{mod_det}.\ref{stmark}. 
In our study we will use $\tau=0.2$s and $N_y=32$. In our case, a time series of $Y_m$ comprise all the values
computed from a group of runs.


We give an example of a first empirical Kramers--Moyal coefficient  $f_\tau^e$ in figure~\ref{compaf} (a),
and second empirical Kramers--Moyal coefficient $a_\tau^e$ in figure~\ref{compaf} (b).
These coefficients are computed from rescaled $Y_m$
obtained from the second group of runs measured for $G/H=2.4$v2 from the third dataset.


Histograms will be computed from the time series of $Y_m$ by binning on the same grid (Eq.~(\ref{gridcoef})) as the empirical Kramers--Moyal coefficients.
We denote them by $\mu(Y_m)$.
Their modes will be systematically computed, if present, in the three distinct intervals $Y_m<-0.005\text{m}=-5H/30$,
$Y_m\in [-0.005\text{m},0.005\text{m}]$ and $Y_m>0.005\text{m}=5H/30$.
We chose these intervals so as to be able to distinguish all types of modes, regardless of the regime of multistability (bistability, tristability and monostability).
The value of the mode is then that of the bin where the local maximum is found.
We present an example of histogram obtained from the second group of runs measured for $G/H=2.4$v2 from the third dataset in figure~\ref{compaf} (c).

\subsection{Langevin regression}\label{slreg}

We finally present the method for model construction.
We first present the polynomial expression chosen for the drift and the diffusion.
We explain how this choice of function
 is rooted in the study of bifurcations in autonomous dynamical systems (\S~\ref{ssratio}).
We then explain how a given function with a fixed structure is fitted (\S~\ref{fitfunc}).
Technical details on the cost function computation and minimisation, model structure selection 
and an illustration of a model computation are given in appendix~\ref{app_sdei}.\ref{appmodcomp}.

\subsubsection{Rationale for the regression}\label{ssratio}

\begin{figure}
\centerline{
\includegraphics[width=5.5cm]{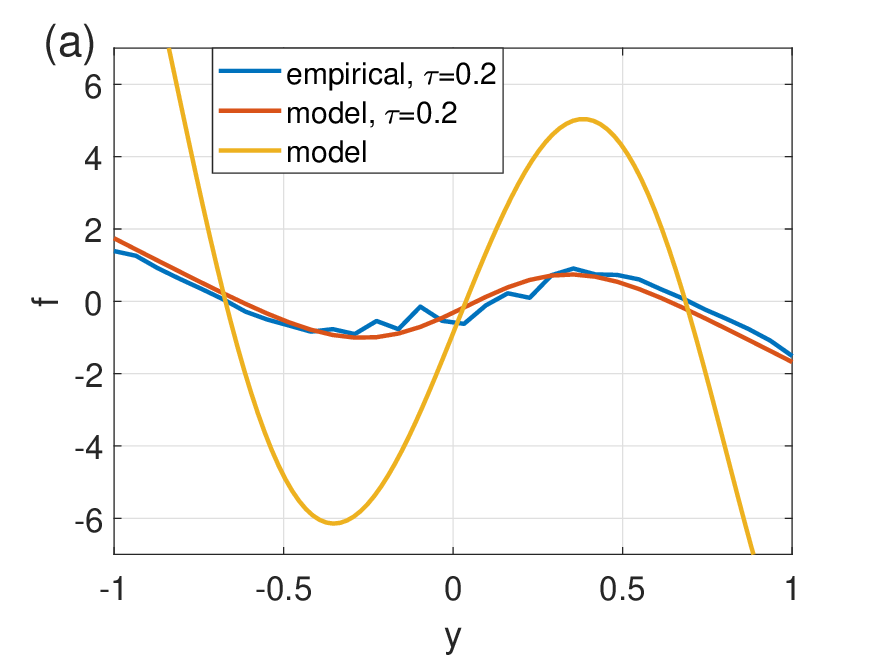}
\includegraphics[width=5.5cm]{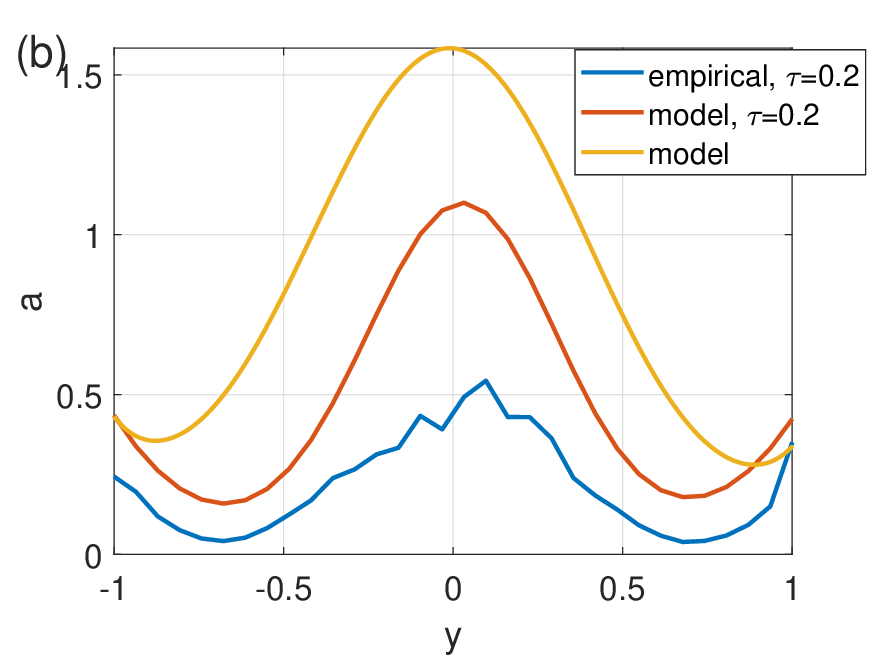}
\includegraphics[width=5.5cm]{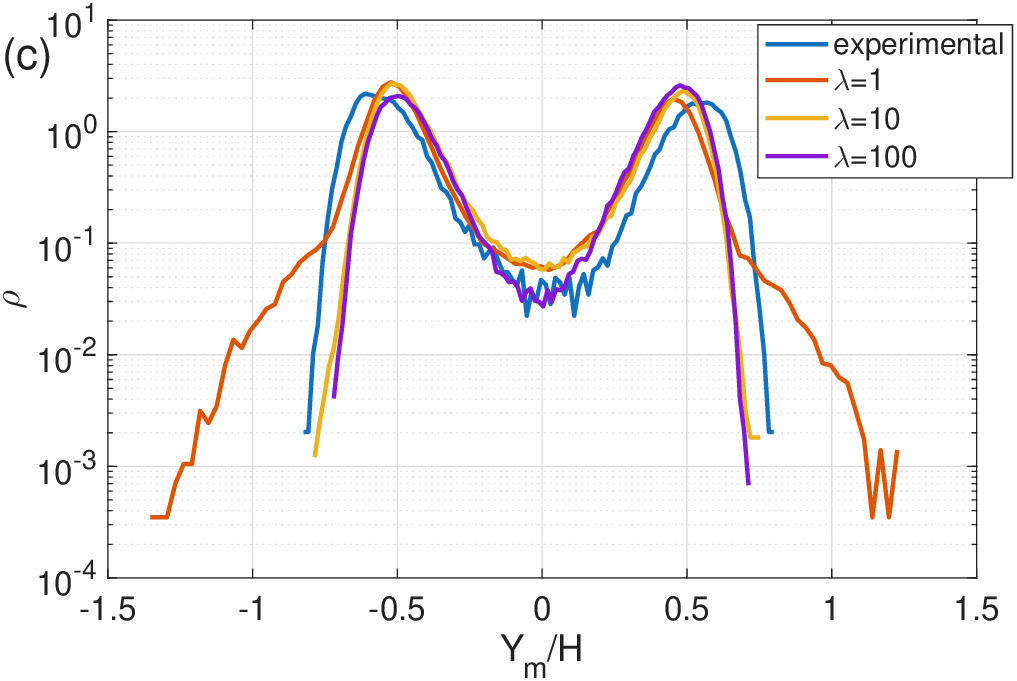}
}
\caption{
Models and binned quantities computed using the time series of $Y_m$ from the group of runs $G/H=2.4$v2.
(a,b) Finite time Kramers--Moyal coefficient at $\tau=0.2$, (empirical and derived from the model)
and fitted model after rescaling from $Y_m$ to $Y_r$.
The Lagrange multiplier used to include the match between the histogram of $Y_m$ and the PDF of the model in the training is set to $\lambda=10$.
(a) First Kramers--Moyal coefficient $f_\tau$ and drift $f$.
(b) Second Kramers--Moyal coefficient $a_\tau$ and diffusion $a$.
(c) Histogram of $Y_m$ computed from time series compared to the histograms of the model computed using three values of Lagrange multiplier $\lambda=1$, $10$ and $100$.
}
\label{compaf}
\end{figure}

Given a system described by a Langevin equation~(\ref{leq}) (or equivalently a Fokker--Planck equation~(\ref{fpeq}) \S~\ref{app_sdei}.\ref{mod_det}.\ref{app_FP})
the relation between the drift and diffusion on the one hand and the finite time Kramers--Moyal coefficients on the other hand  
is invertible: one can be obtained from the other and \emph{vice versa}.
However, given an empirical estimate of the finite time Kramers--Moyal coefficients from a finite duration time series (Fig.~\ref{compaf}),
performing an inversion will lead to several problems:
\begin{itemize}
\item Errors on the estimate of the Kramers--Moyal coefficients, causing them to be irregular, could lead to a poor estimate of the drift and diffusion.
Indeed the computation would have to be performed numerically, given the discretised coefficients and the integration could be unstable.
\item Even if a smooth estimate of the drift and diffusion were obtained, those estimates would lack the interpretability
of analytical formulae.
\end{itemize}

For these reasons we turn to a method that can adjust an analytical model for the drift and diffusion
from empirically sampled Kramers--Moyal coefficients and select an optimum between the precision of the model and its simplicity.
We follow a procedure initiated by \citep{lade2009finite,honisch2011estimation} and
further developed by \citep{callaham2021nonlinear,callaham2022empirical} and adapt their code \citep{callaham2021nonlinear}.
We will describe the procedure in two steps: firstly the adjustment of a given function (\S~\ref{app_sdei}.\ref{appmodcomp}.\ref{fitfunc},~\ref{app_sdei}.\ref{appmodcomp}.\ref{costmin})
and secondly the inclusion of that fit into an iterative procedure to 
select the best functional form (\S~\ref{app_sdei}.\ref{appmodcomp}.\ref{sselmod},~\ref{app_sdei}\ref{appmodcomp}.\ref{spar}).

For reasons related to the theory of bifurcations and phase transitions \citep{manneville2010instabilities,landau2013statistical},
which is invoked to describe changes of regime of multistability,
we choose to fit and select models that will be polynomial in $y$.
We will write our drift $f$ and diffusion $g$ as
\begin{align}
f(y)=\sum_{k=0}^{N_f}f_{k}y^k\,,\label{edriftf}\\
 g(y)=\sum_{k=0}^{N_g}g_{k}y^k\,,\label{ediffg}
\end{align}
We will use $N_f=5$ and $N_g=4$ for most cases, except at $G/H=1.15$ and $G/H=1.25$ where we will use $N_f=7$ and $N_g=6$.
These numbers are chosen large enough so that the resulting polynomial can represent the variations of the drift and diffusion and can
lead to a converging fit, but not so large so as to limit the computational cost.
Selecting and fitting the model will then consist in selecting the right monomials in $f$ and $g$, that is to say selecting which model parameters $f_k$ and $g_k$ are non zero,
and adjusting the value of non zero $f_k$ and $g_k$.
The selection and adjustment will be based on our sampled data.

As we have stated, the choice of polynomial functions for $f$ and $g$ is not arbitrary.
The study of bifurcations or phase transitions has often highlighted ordinary differential equations
or stochastic differential equations governing the dynamics of the order parameter.
The deterministic part of these differential equations is polynomial in the large majority of cases.
 A lot of experience has thus been gained in the analysis of such models.
In that case, the metastable states are given by the value of $y$
at the stable fixed points, while the transition states will be the unstable fixed points.
As soon as the diffusion becomes smaller in amplitude compared to the drift (which is often the case), 
these stable fixed points coincide with the modes of the PDF and conditional averages of the measured variable $Y_m$.  
This means that there is a direct relation between the symmetries of the system (present and broken at bifurcations),
the position of stable and unstable fixed points (or modes and transition states), as well as the type of regime changes, on the one hand
and the leading monomials in the drift (Eq.~(\ref{edriftf}))
as well as the sign and amplitude of the model parameters in front of those monomials, on the other hand. 
This relation comes from the fact that the fixed points of the deterministic part of the Langevin equation correspond to $f(y)=0$.
In turn, they are
stable (resp. unstable) if $df/dy<0$ (resp. $df/dy>0$).
A bifurcation occurs if a fixed point appears or disappears and/or it changes stability.
All this is necessarily governed by the dependence of $f$ on $y$.

In particular, in the case where we have a reflection symmetry in the flow,
with two possible metastable positions for which the order parameter has opposite signs,
we have a common type of polynomial for the drift (Eq.~(\ref{edriftf})).
Indeed, in that case, the reflection symmetry imposes that the dynamics should be invariant under the
operation $y\rightarrow -y$.
This means that any stable and unstable fixed points of the drift should have
an exact counterpart with opposite sign.
This imposes that the drift is odd  $f(-y)=-f(y)$.
Among odd polynomial drifts, the recurring ones are actually at most quintic.
This leads to the ordinary differential equation (ODE)
\begin{equation}
\frac{d y}{dt}=f(y)=f_1 y+f_3 y^3+f_5 y^5\,.\label{oddODE}
\end{equation}
Where 
the $f(-y)=-f(y)$ symmetry imposes $f_0=f_2=f_4=0$.
This ODE is for instance found in Landau normal forms \citep{manneville2010instabilities,landau2013statistical}.
The stable and unstable fixed points of this ODE are respectively describing the multistable positions and transition states in between.
For tristable, bistable and monostable, we find three distinct values and signs for $f_1$, $f_3$ and $f_5$.
We describe these three typical cases of normal form in appendix~\ref{app_sdei}.\ref{mod_det}.\ref{app_normform}.
We complement this description with a discussion of the expression of the corresponding fixed points.
The discussion of the model we will trained will be done in view of these archetypes of ordinary differential equations.

In the simplest cases, we expect the diffusion (Eq.~(\ref{ediffg})) to keep a constant sign, that we arbitrarily choose positive.
In common model systems, the amplitude of the diffusion is moderate with respect
to that of the drift and thus its  variations mostly modulate the effect of the drift.
If we have the reflection symmetry $y\leftrightarrow -y$, then the diffusion is even $g(y)=g(-y)$.
This would cancel out the odd model parameters $g_1=g_3=0$ and retain only the even ones $g_0$, $g_2$ and $g_4$.

In this article, we will not restrict the powers of the drift and diffusion based on the observed state of multistability, nor will we forbid
even powers in the drift and odd powers in the diffusion.
Firstly, we will allow the model to account for any asymmetry contained in the experimental data (for instance, caused by imperfections of the setup).
Secondly, we will allow scenarii more complex than the three clear cut cases of monostability, 
bistability and tristability described in appendix~\ref{app_sdei}.\ref{mod_det}.\ref{app_normform}.
Therefore, while we will closely monitor $f_1$, $f_3$, $f_5$, $g_0$, $g_2$ and $g_4$.
Studying possible even model parameters in the drift and odd ones in the diffusions
may help us highlight disymmetries between the left and right states.

\subsubsection{Outline of the procedure selecting a sparse model matching the data}\label{fitfunc}

The aim of the procedure that we use in this article is to fit a polynomial expression for the drift and diffusion that is in agreement with
the data while having as few monomials as possible.
This is done using an outer loop over the number of monomial contained in the drift and diffusion,
starting from the full sum to a single monomial in both the drift and diffusion.
At each stage of the outer loop, the monomial that has the least weight in either the drift or the diffusion is removed.
This is done by having an inner loop where each monomial removal is tested, the corresponding model is fitted to the data
and the best fit is retained for the next step of the outer loop.
Once the outer loop is done, a criterion on fit quality is used to select the 
step of the outer loop (and the resulting fitted model) that results in a balance between sparsity and precision.

In this section, we give an overview on how we fit the polynomials expressing the drift (Eq.~(\ref{edriftf})) 
and diffusion (Eq.~(\ref{ediffg})), within the aforementioned loops, when said polynomials each have a fixed set of monomials.
The resulting model parameters $\{f_k\}_{1\le k \le N_f}$ and $\{ g_k\}_{1\le k \le N_g}$ will minimise a cost function
that expresses the objectives the fitted models have to meet.
The model should result in finite time Kramers--Moyal coefficients which are close to the ones estimated empirically.
Moreover, we force the model to result in a PDF which is close to the empirical histogram computed from data.
The importance of this constraint will be adjusted using a Lagrange multiplier $\lambda$.
The technical details of the cost function computation and minimisation are given in appendix~\ref{app_sdei}.\ref{appmodcomp}.\ref{costmin},
while the loops over polynomial degree are presented in appendix~\ref{app_sdei}.\ref{appmodcomp}.\ref{optimshape} and illustrated in appendix~\ref{app_sdei}.\ref{appmodcomp}.\ref{sselmod}.
We also present a parallelisation procedure we have used in appendix~\ref{app_sdei}.\ref{appmodcomp}.\ref{spar}.

The optimisation and the discussion of results will be done on time series of $Y_m$ rescaled by their maximum over the group of runs.
All subsequent quantities (drift, diffusion, Kramers--Moyal coefficients, histograms and PDF) will also be rescaled as described in appendix~\ref{app_sdei}.\ref{appmodcomp}.\ref{app_resc}. 
The rescaled quantities are denoted with a subscript ${}_r$.
The model parameters $f_{r,k}$ and $g_{r,k}$ will be computed by minimising the cost function
\begin{equation}
V=\frac{\sum_{i=1}^{N_y}(f_{\tau,i}^e-f_{\tau,i})^2}{\sum_{i=1}^{N_y}{f_{\tau,i}^e}^2}
+\frac{\sum_{i=1}^{N_y}(a_{\tau,i}^e-a_{\tau,i})^2}{\sum_{i=1}^{N_y}{a_{\tau,i}^e}^2}
+\lambda D(\mu_r\parallel \rho_r)\label{eqcost}
\end{equation}
The first two contributions 
consist in the square norm of the difference between the empirical Kramers--Moyal coefficients and model Kramers--Moyal coefficients, summed over the gridpoints.
They measure the mismatch between the data and the dynamical properties of a given model, with
given values of the coefficients $f_{r,k}$ and $g_{r,k}$.
They are normalised by the square norm of the corresponding empirical Kramers--Moyal coefficient to ensure they are both of order $\mathcal{O}(1)$. 
This has improved the computational cost of the optimisation 
as well as the precision of the computation of both coefficients the drift and diffusion, regardless of
their order of magnitude. The finite time Kramers--Moyal coefficients of 
a model are obtained through a numerical resolution of the backward Fokker--Planck equation (Eq.~(\ref{bfp}), \S~\ref{app_sdei}.\ref{appmodcomp}.\ref{costmin}) 
at each step of minimisation of $V$.
We also constrain our model to have PDF close to the histogram of $Y_m^r$.
We use the Lagrange multiplier $\lambda>0$ to tune the relative importance given to the match between
the PDF and the histogram (Fig.~\ref{compaf} (c)), with respect to the
match between empirical and model Kramers--Moyal coefficients.
The PDF of a given model is obtained through a numerical resolution of the Fokker--Planck equation (Eq.~(\ref{fpeq}), \S~\ref{app_sdei}.\ref{appmodcomp}.\ref{costmin}), 
again at each step of minimisation of $V$.
This second match between model and data is measured by the Kullback--Leibler divergence $D(\mu_r\parallel \rho_r)$ between the rescaled histogram
and the PDF of the rescaled model, defined as
\begin{equation}
D(\mu_r\parallel \rho_r)=\int \mu_r\log\left( \frac{\mu_r}{\rho_r}\right)\,{\rm d}Y_r\,.
\end{equation}
This function is always positive and cancels out when $\mu_r$ and $\rho_r$ are equal almost everywhere.
As a consequence, it is often used as a measure of the distance between
two PDFs in machine learning \citep{goodfellow2016deep}.
See also \citep{cover2012elements} for details on the properties of the Kullback--Leibler divergence.

\section{Results}\label{sres}

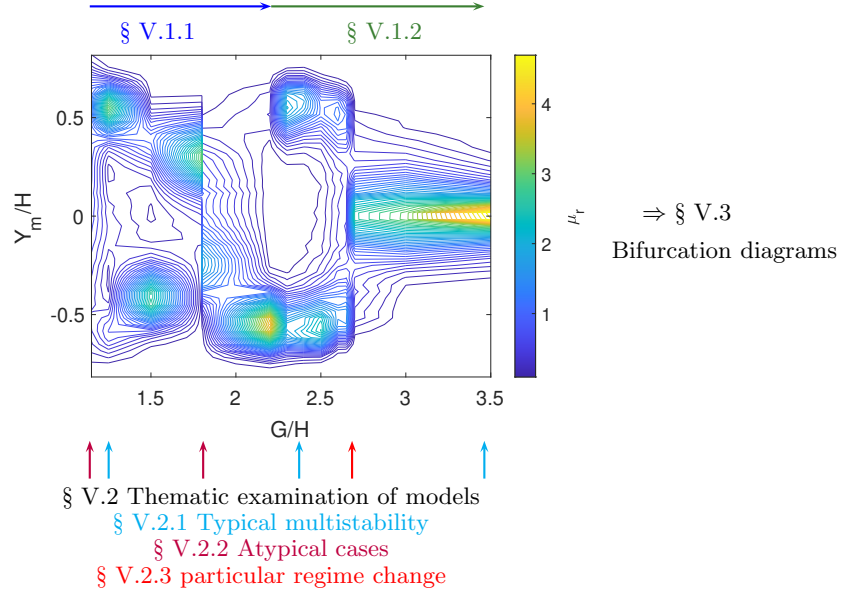
\begin{figure}[!ht]
\begin{center}
\begin{pspicture}(13,9)
\rput(5,1.5){\S~\ref{srmod} Thematic examination of models}
\rput(5,1.15){\textcolor{cyan}{\S~\ref{modtyp} Typical multistability}}
\rput(5,0.8){\textcolor{purple}{\S~\ref{modpec}  Atypical cases}}
\rput(5,0.45){\textcolor{red}{\S~\ref{modtrans} particular regime change}}
\rput(5.5,5){\includegraphics[width=8cm]{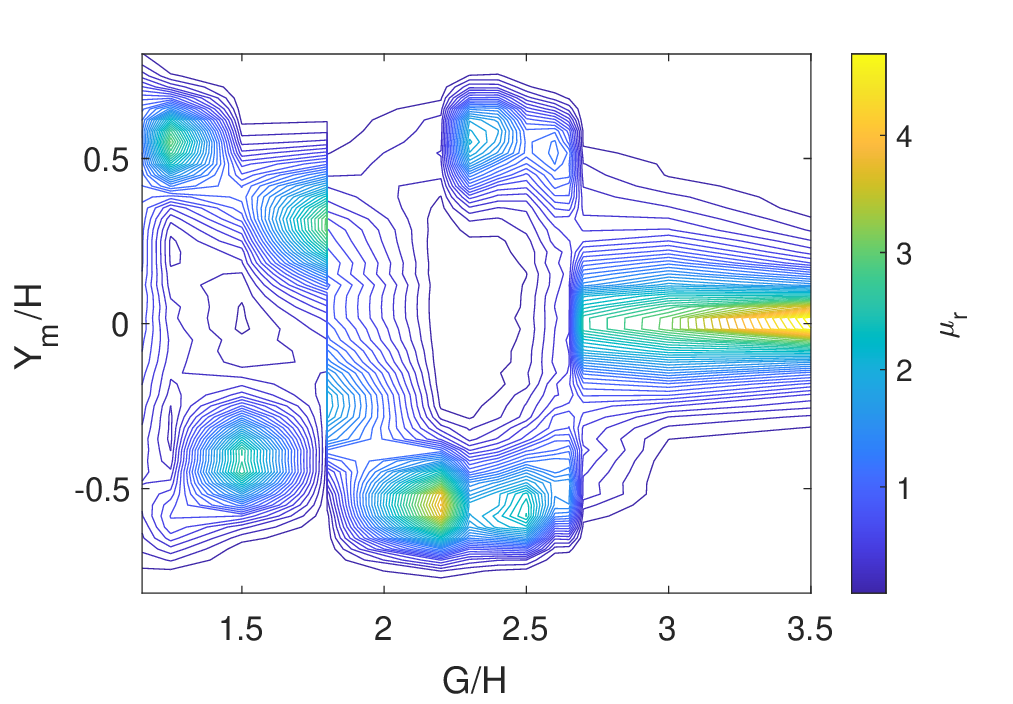}}
\psline[linecolor=cyan]{->}(2.85,1.75)(2.85,2.25)
\psline[linecolor=cyan]{->}(5.37,1.75)(5.37,2.25)
\psline[linecolor=red]{->}(6.07,1.75)(6.07,2.25)
\psline[linecolor=cyan]{->}(7.82,1.75)(7.82,2.25)
\psline[linecolor=purple]{->}(2.6,1.75)(2.6,2.25)
\psline[linecolor=purple]{->}(4.1,1.75)(4.1,2.25)
\rput(5,8.75){\S~\ref{sover} : Detailed description of dynamics}
\psline[linecolor=blue]{->}(2.6,8)(5,8)
\psline[linecolor=OliveGreen]{->}(5,8)(7.82,8)
\rput(3.5,7.65){\textcolor{blue}{\S~\ref{slgh}}}
\rput(6.5,7.65){\textcolor{OliveGreen}{\S~\ref{strns}}}
\rput(10.5,5.25){$\Rightarrow$ \S~\ref{sdiag}}
\rput(11,4.75){Bifurcation diagrams}
\end{pspicture}
\end{center}
\caption{Graphical summary of the result section~\ref{sres}.
At the centre of this figure, we place the contours of the histograms $\mu_r$ 
as a function of $Y_m/H$, on the ordinate axis, concatenated 
for all values of $G/H$ available in third group of runs, along the abscissa axis. 
We overlapped two histograms at $G/H=1.8$, arising from the two available groups of runs.
The overview indicates that different regimes of multistability are found in the range $[1.15,3.5]$, 
visible in the different numbers of maxima of the histogram at each value of $G/H$.
In section~\ref{sover}, we will present the succession of different types of dynamics in details using time series of $Y_m/H$
and visualisations of in plane kinetic energy
as $G/H$ is increased.
This progressive investigation of the flow will reveal that at some $G/H$, the temporal evolution of the jet 
is not necessarily similar to that of a classic bistable flow.
Thus, in section~\ref{srmod} we use the empirical finite time Kramers--Moyal coefficients and the trained models
to understand the underlying feature of these different types of dynamics.
For this matter, we perform a thematic presentation, regrouping cases at different values of $G/H$
that share a similar level of complexity:
in section~\ref{modtyp}, we present typical cases of multistability at $G/H=1.25$, $2.4$ and $3.5$,
atypical cases at $G/H=1.15$ and $1.8$ in section~\ref{modpec} and the change of regime toward monostability at $G/H=2.7$ in section~\ref{modtrans}.
The information gathered in section~\ref{sover} and~\ref{srmod} will finally be synthesised in the form of bifurcation diagrams in 
section~\ref{sdiag} that will pinpoint all the properties of this complex multistable flow that can only be glimpsed in this first figure.
}
\label{RES_disp}
\end{figure}

In this section, we will present the results of our processing and modelling of the velocity fields
measured by means of PIV over the three experimental campaigns.
The ultimate goal of this section will be to construct bifurcation diagrams presented in section~\ref{sdiag}.
These diagrams will give a quantitative view of the successive regimes of multistability found in the flow and of
the particular transition to monostability.
For their construction, we will firstly use visualisations and time series, which we 
present in a detailed description of the dynamics of the jet as $G/H$ is increased (\S~\ref{sover}).
We will secondly use the analysis of the empirical finite time Kramers--Moyal coefficients and the fitted models.
Their presentation will be centred on three types of behaviour: classical multistability (\S~\ref{modtyp}, $G/H=1.25$, $2.4$ and $3.5$), 
complex regimes (\S~\ref{modpec}, $G/H=1.15$ and $1.8$) and around the transition (\S~\ref{modtrans}, $2.6 \le G/H \le 2.7$).
In order to give an overview of the organisation of this section, as well as 
a first view of the richness of multistability regimes found in the flow, that does not require any analysis yet,
we display a graphic summary of the section in figure~\ref{RES_disp}.
We centre this figure around contours of the histograms $\mu_r$ of $Y_m/H$ concatenated for all $G/H\in [1.15, 3.5]$.
Thus, this result section will gradually explain why one or several maxima are found for each value of $G/H$.

\subsection{Overview of flow states}\label{sover}

We start this result section by an overview of the flow states and time series for increasing values of  $G/H$. We decompose this presentation in two parts:
we first describe the flow in the range $G/H \in[1.15,2.2]$ (\S~\ref{slgh}),
where a variety of regimes of multistability is found,
then in the range $G/H\in [2.3,3.5]$ (\S~\ref{strns}) where the flow transits from bistable to monostable.

\subsubsection{Various states of multistability at $G/H=1.25$, $G/H=1.8$ and $G/H=2.2$}\label{slgh}

\paragraph{$G/H=1.15$}

\begin{figure}
\centerline{
\includegraphics[width=14cm,clip]{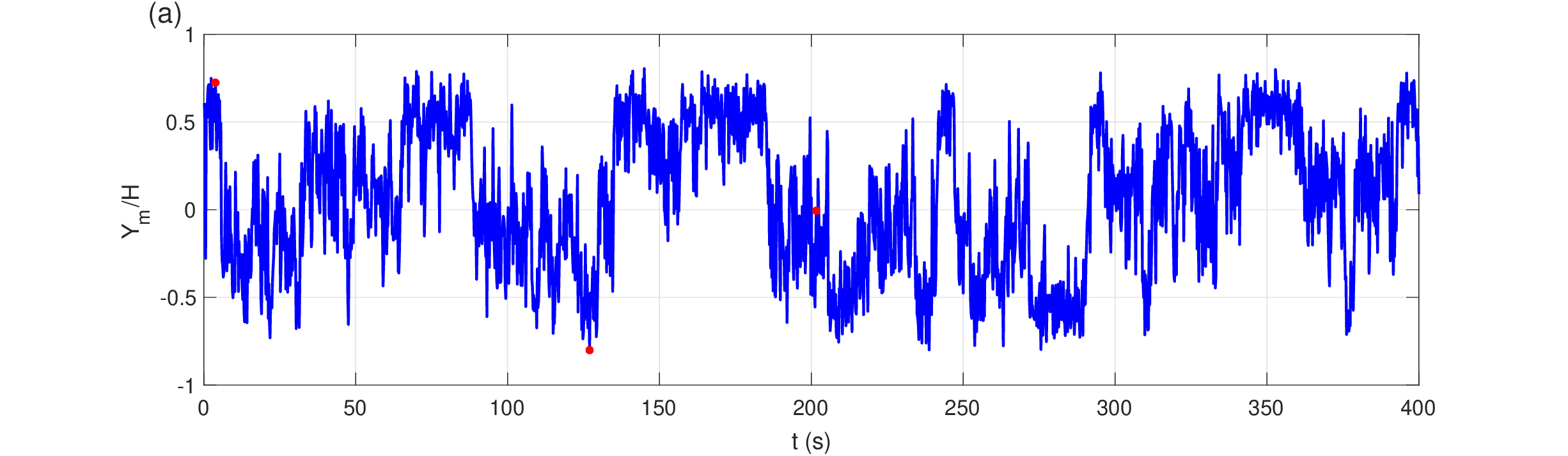}
\includegraphics[width=5.5cm,clip]{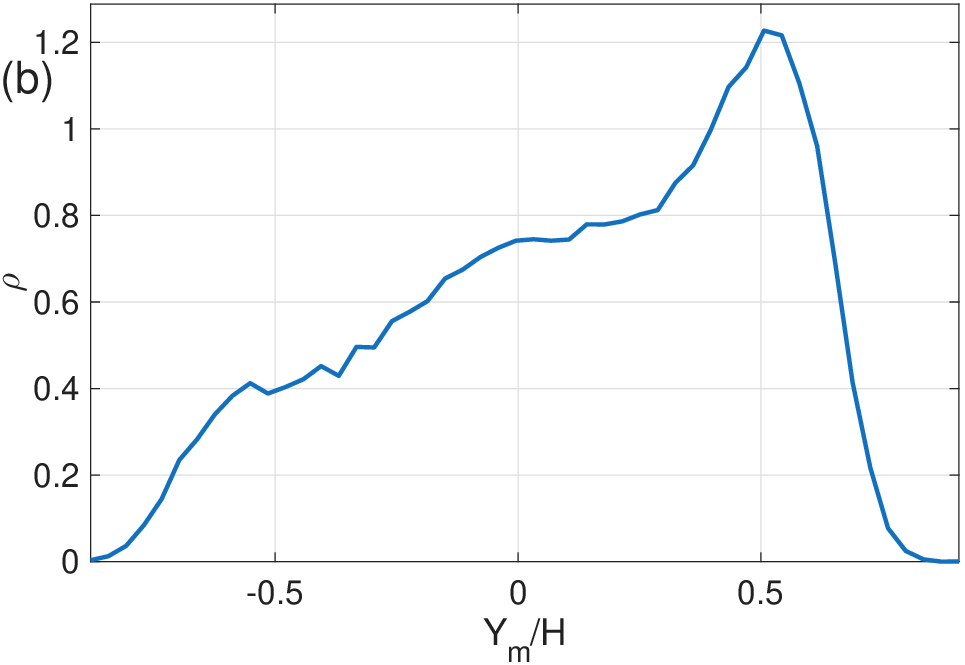}
}
\vspace{1cm}
\centerline{
\includegraphics[width=5.5cm]{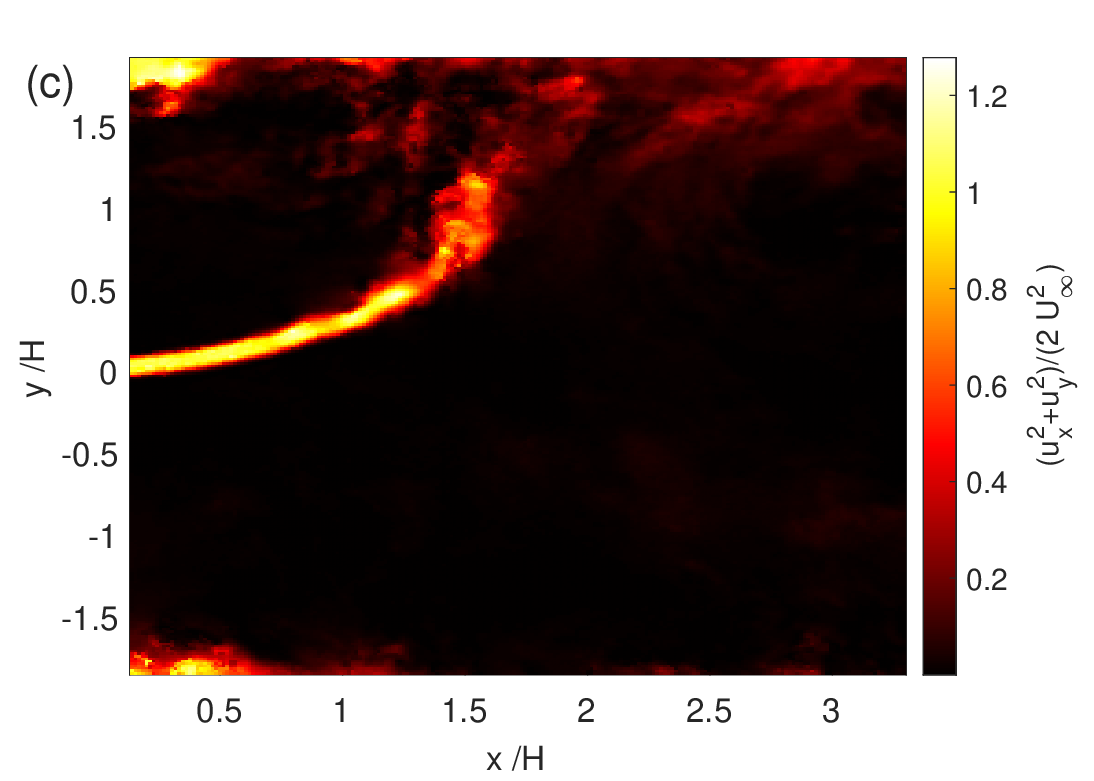}
\includegraphics[width=5.5cm]{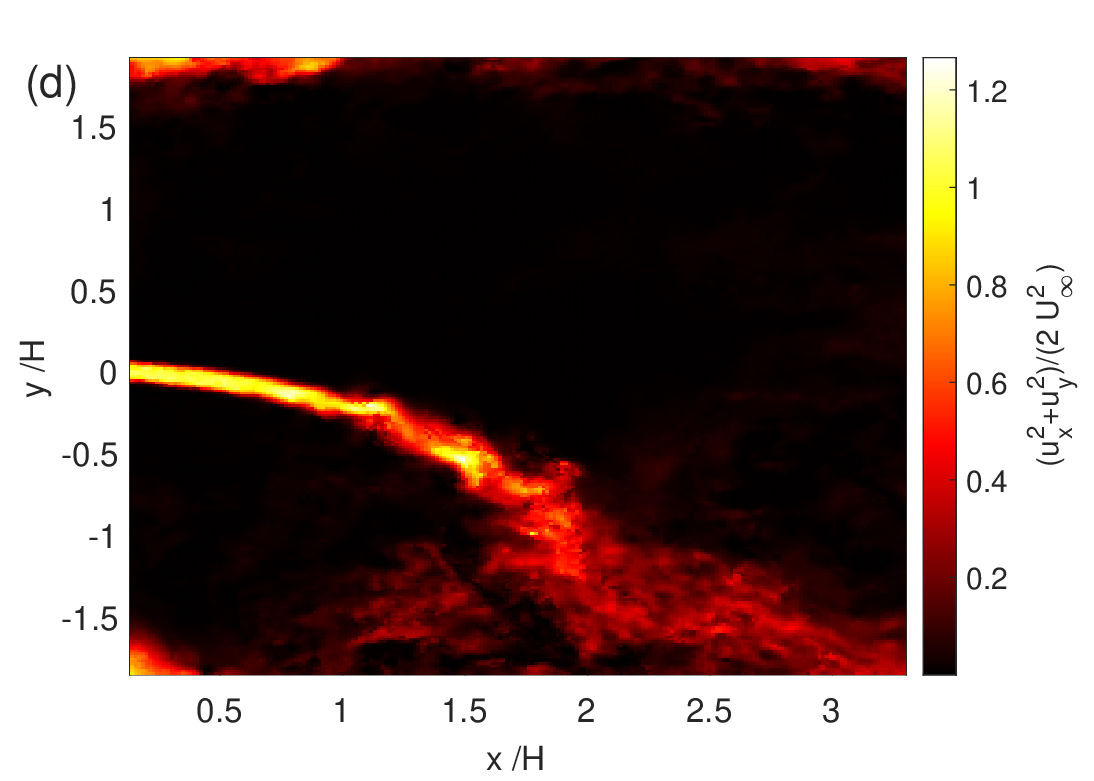}
\includegraphics[width=5.5cm]{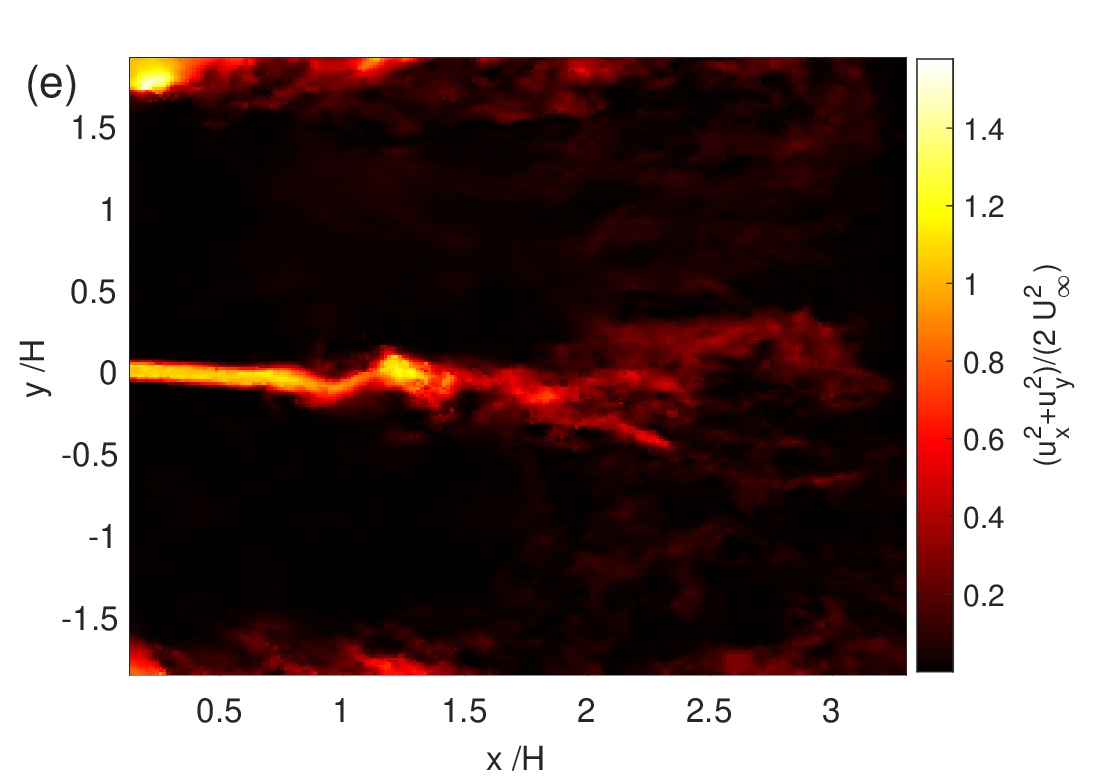}}
\caption{From the velocity fields measured by means of PIV with a gap ratio of $G/H=1.15$ during the third experimental campaign,
zoomed on the time interval $[0,400]s$:
(a) time series of the weighted position $Y_m$ of the jet (the three red dots indicate the instants in time corresponding to figures (c,d,e)),
(b) Histogram of $Y_m/H$,
colour levels of the kinetic energy in the horizontal measurement plane obtained from the same measured velocity fields at times
(c) $t=3.8$s,
(d) $t=127$s,
(e) $t=201.6$.
}
\label{vis1-15}
\end{figure}

We display the time series of $Y_m$ in figure~\ref{vis1-15} (a),
as well as visualisations of the in-plane kinetic energy when the jet is 
on the left (Fig.~\ref{vis1-15} (c)), on the right (Fig.~\ref{vis1-15} (d))
and along the centreline (Fig.~\ref{vis1-15} (e)).
The jet position $Y_m$ spans the whole range of $[-5H/6, 5H/6]$.  
We do not observe a clear cut case of multistability, where $Y_m$ would fluctuate in distinct intervals,
and change interval from time to time.
We find intervals of time where $Y_m/H$ fluctuates around $+0.5$ (for instance $t \in [0s,5.5s]$), $-0.5$ (for instance $t\in [123s, 130s]$) and around $0$ (for instance $t\in [185s, 205s]]$),
intervals of time where it apparently drifts from range to the next (for instance $t\in [206s,235s]$), 
and intervals of time where it changes rapidly (around $t\simeq 5.5s$).
In the range $Y_m\in[-5H/6, 5H/6]$, the histogram of $Y_m$ takes non negligible values and its curvature varies. 
However, only a single mode near $Y_m\simeq 0.56H$ is observed   
(Fig.~\ref{vis1-15} (b)).
We will thus investigate the broken left right symmetry 
in more details at that gap ratio using the fitted models.

\paragraph{$G/H=1.25$}

\begin{figure}[!ht]
\centerline{
\includegraphics[width=16cm,clip]{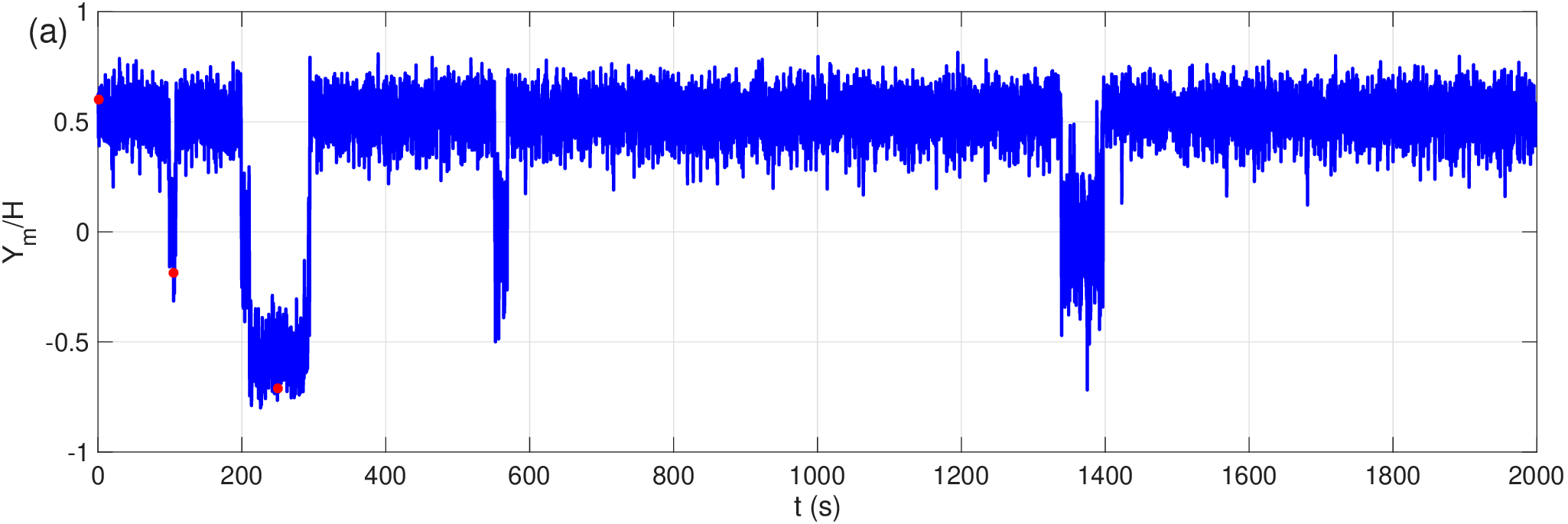}
}
\vspace{1cm}
\centerline{
\includegraphics[width=5.5cm]{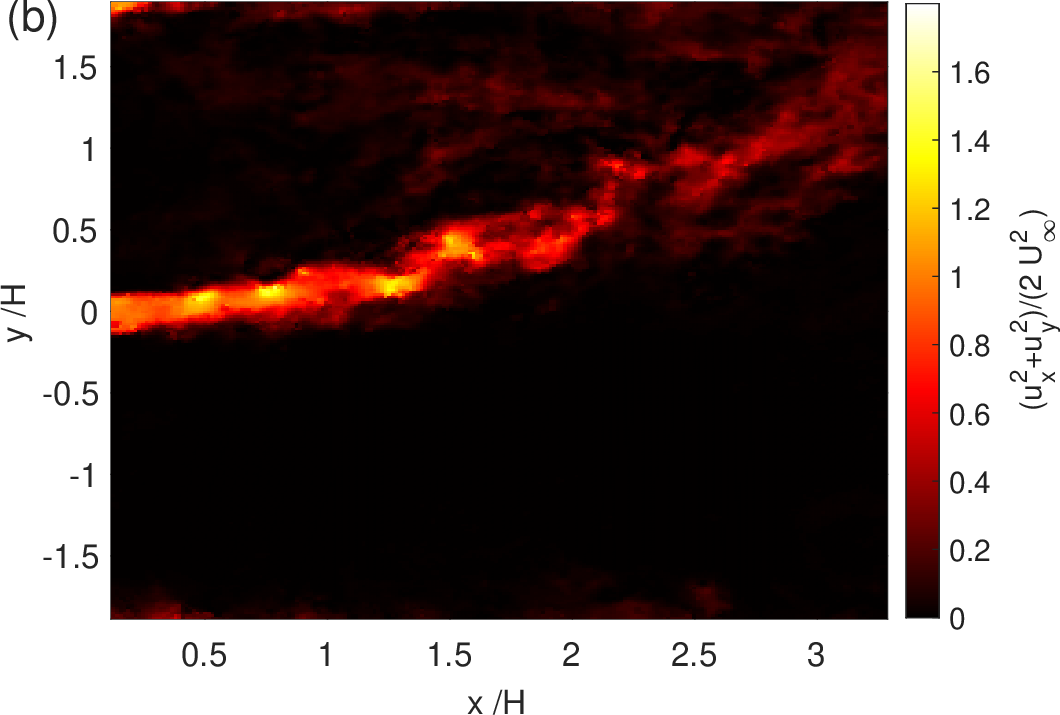}
\includegraphics[width=5.5cm]{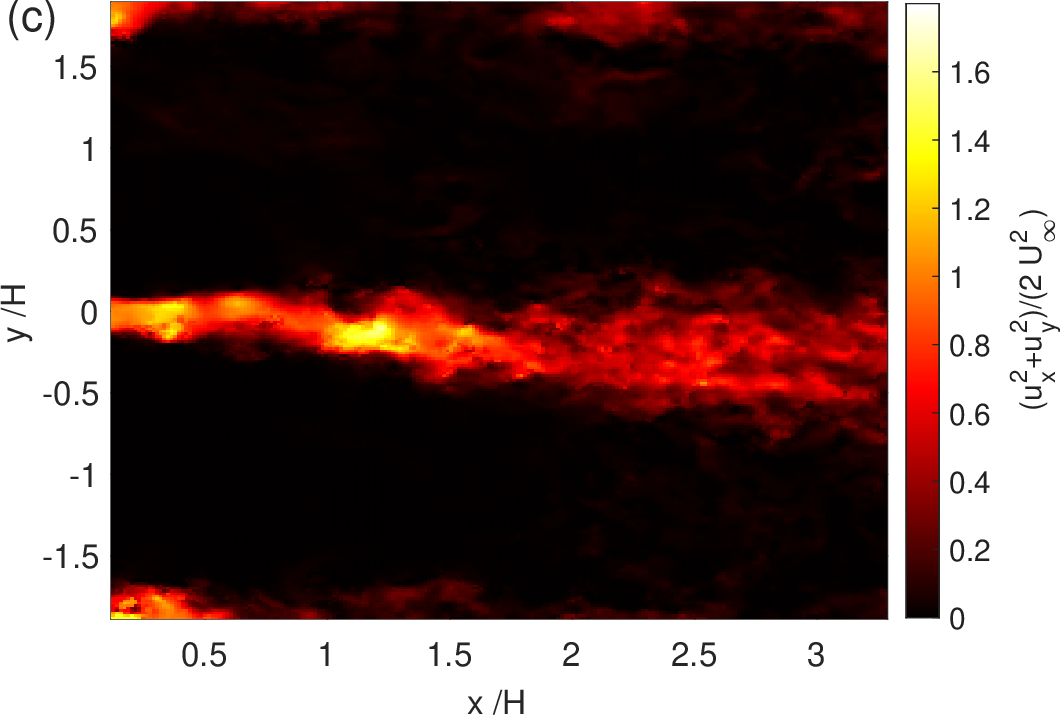}
\includegraphics[width=5.5cm]{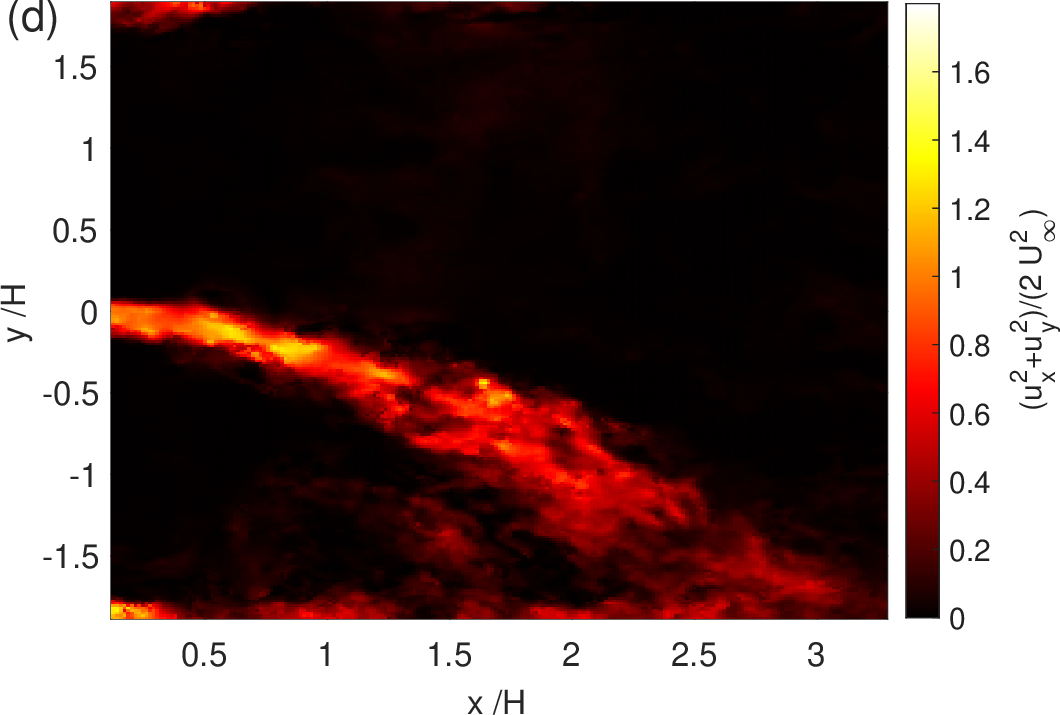}}
\caption{From the velocity fields measured by means of PIV with a gap ratio of $G/H=1.25$ during the third experimental campaign,
zoomed on the time interval $[0,2000]s$:
(a) time series of the weighted position $Y_m$ of the jet (the three red dots indicate the instants in time corresponding to figures (b,c,d)),
colour levels of the kinetic energy in the horizontal measurement plane obtained from the same measured velocity fields at times
(b) $t=0.2$s,
(c) $t=105$s,
(d) $t=250$.
}
\label{vis1-25}
\end{figure}

The flow becomes tristable.
We describe this tristability using the time series of $Y_m$ sampled during the third experimental campaign (Fig.~\ref{vis1-25} (a)).
We complement this using three visualisations of the in-plane kinetic energy $(u_x^2+u_y^2)/(2U_\infty^2)$ at times $t=0$s (Fig.~\ref{vis1-25} (b)),
$t=105$s (Fig.~\ref{vis1-25} (c)) and $t=250$s (Fig.~\ref{vis1-25} (d)) marked by red dots in the time series (Fig.~\ref{vis1-25} (a)).
The data sampled in the first experimental campaign yields 
very similar visualisations and similar time series. 
We will see in 			  
section~\ref{sbdiag} that the properties of the width of the jet and the multistable positions
are very similar from one dataset to another at that gap ratio. 
We note that the weighted position $Y_m$ not only spends time fluctuating around the value $Y_m\simeq- 0.015$m ($t=0$s Fig~\ref{vis1-25} (a,b)),
and $Y_m\simeq 0.015$m ($t=250$s Fig~\ref{vis1-25} (a,d))
but that in some time intervals (\emph{e.g.} $t\in [98\text{s},108\text{s}]$, $[198\text{s},212\text{s}]$,
$[550\text{s}, 570\text{s}]$ and $[1340\text{s},1395\text{s}]$), it is fluctuating around $0$,
as illustrated in figures~\ref{vis1-25} (a,c).
We also note that the jet never goes from the left to the right positions (or \emph{vice versa}) without spending some time fluctuating
around the central position.
As a result, the histogram of  $Y_m$ has three modes at the corresponding positions (Fig.~\ref{RES_disp}).  
This tristability at narrow gap ratio is similar to the one reported by \cite{sumner1999fluid}.
Because of shorter residency times near the right position, there is a factor of five between the value of the maxima of $\mu$
for the left and right metastable positions.
Such disymmetries are observed at all gap ratios. Disentangling the effect of finite sampling time and experimental imperfections
in those imbalances is difficult. The subject is under investigation and will be presented in an article currently in preparation. 

\paragraph{$G/H=1.5$}

The jet is bistable, with time series and visualisations of in-plane kinetic energy
comparable to what had been shown in section~\ref{jpos}.
This also results in a bimodal histogram for $Y_m$ (Fig.~\ref{RES_disp} (a)).

\paragraph{$G/H=1.8$}

\begin{figure}[!ht]
\centerline{
\includegraphics[width=14cm,clip]{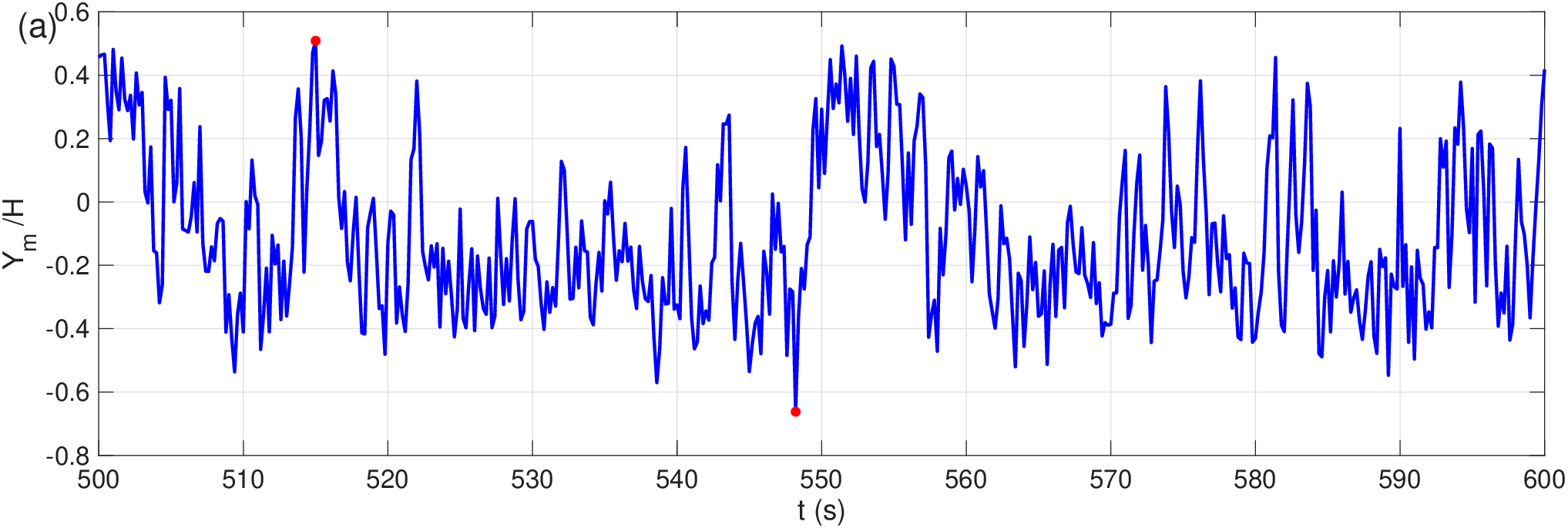}
\includegraphics[width=5.5cm,clip]{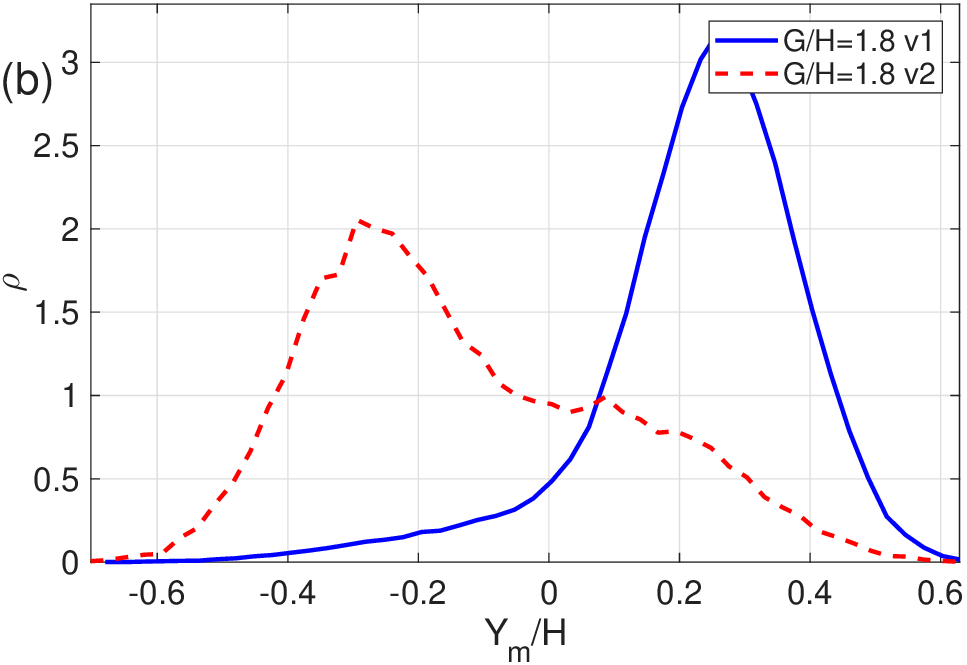}}
\vspace{1cm}
\centerline{
\includegraphics[width=5.5cm]{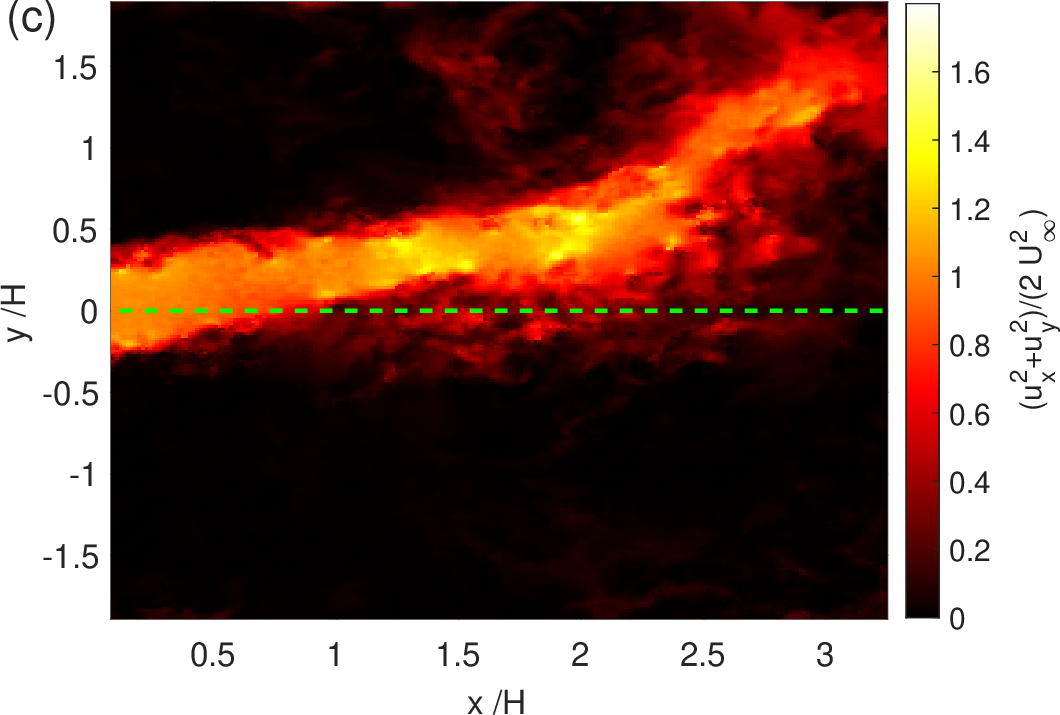}
\includegraphics[width=5.5cm]{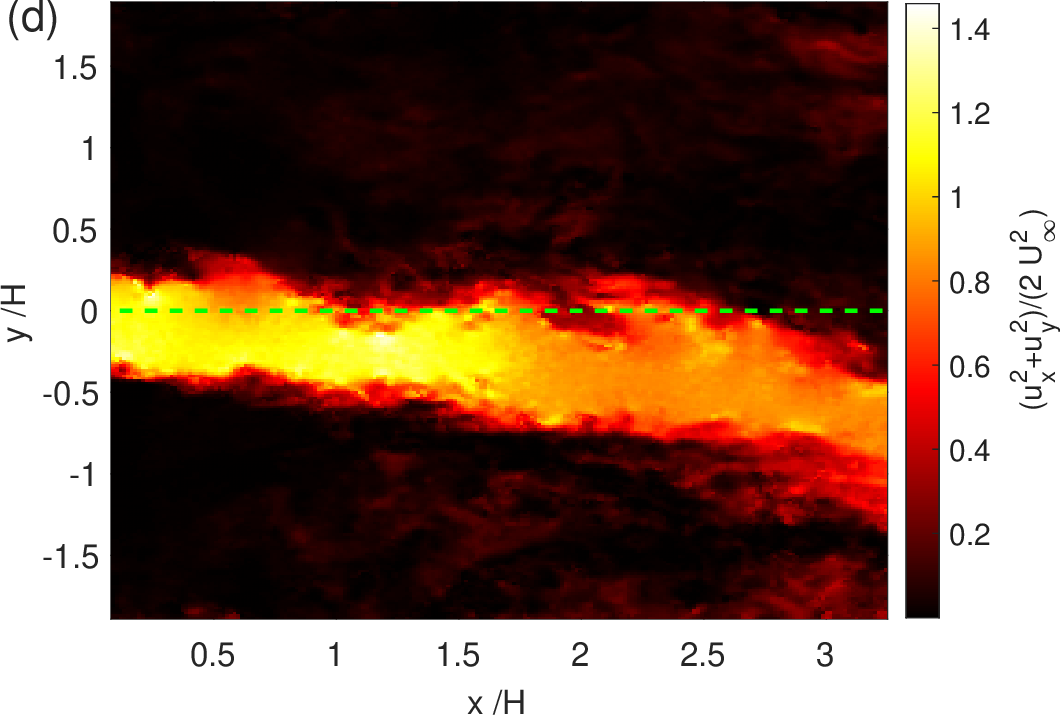}
}
\caption{From the velocity fields measured by means of PIV with a gap ratio of $G/H=1.8$ (second group of runs)
during the third experimental campaign:
(a) time series of the weighted position $Y_m$ of the jet, zoomed in the time interval $[500\text{s},600\text{s}]$
(the two red dots indicate the instants in time
corresponding to figures (c,d)),
(b)
Histogram of $Y_m/H$ from the two groups of runs $G/H=1.8$v1 and v2,
colour levels of the kinetic energy in the horizontal measurement plane obtained from the same measures velocity fields
at the time indicated by the red dots in (a), (c) $t=515$s, (d) $t=548.2$s
(the green dashed line indicates the line $y=0$m).
}
\label{vis1-8}
\end{figure}

We then present the times series of $Y_m$ and visualisations of the in-plane kinetic energy 
(Fig.~\ref{vis1-8}).
For that gap ratio, for the two groups of runs, the time series of $Y_m$
indicate that the position of the jet spans the interval $[-2H/3,2H/3]$.
We illustrate this for the second group of runs in figure~\ref{vis1-8} (a).
Excursions to $|Y_m|=2H/3$ are now much rarer than for smaller gap ratios and the weighted position of the jet
is mostly found in the range $-H/3 \le Y_m\le H/3$.
Compared to what was observed at $G/H=1.25$ (and G/H=1.5), 
the time intervals for which $Y_m$ is strictly positive (for which the jet is on the left Fig.~\ref{vis1-8} (c))
or those for which $Y_m$ is strictly negative (for which the jet is on the left Fig.~\ref{vis1-8} (d)) are much shorter.
Note that the time interval presented here has a length of $100$s, which is shorter than what is chosen for most other gap ratios.
In figure~\ref{vis1-8} (b),
we include the histograms computed from the time series of $Y_m$ obtained from both group of runs $G/H=1.8$v1 and $G/H=1.8$v2.  
The histograms constructed from the two group of runs only
present one mode, whose position, at $|Y_m|\simeq 0.26H$, is smaller in absolute value than at other gap ratios for which multistability is observed.
In line with this, the histograms take notable values in a narrower range. 
Note that while monomodal, these histograms are certainly not
centred on the $y=0$ line: the left-right symmetry is still broken.
This synthesises the observation of figure~\ref{vis1-8} (a,c,d) that the jet can be found
both on the left and right sides for short time intervals and that the transitions from
one side to the other can be rapid, owing to a lower relative stability. 
We finally note that the jet mostly favours the left side in the group of runs 1.8v1 (spending less short durations in that position),
while the jet slightly favours the right side in the group of runs 1.8v2.
Indeed, we note that the histogram is almost flat near $Y_m\simeq +0.26H$, indicating that this
position is almost metastable. The examination of the dynamics (finite time Kramers--Moyal coefficients and fitted model),
will help us make more precise assertions as to that.
The smaller value of $Y_m$ at the modes and rapid transition time may make the jet at this value
of the gap ratio more sensitive to experimental imperfections. 

\paragraph{$G/H=2.2$} 
 
\begin{figure}[!ht]
\centerline{
\includegraphics[width=16cm,clip]{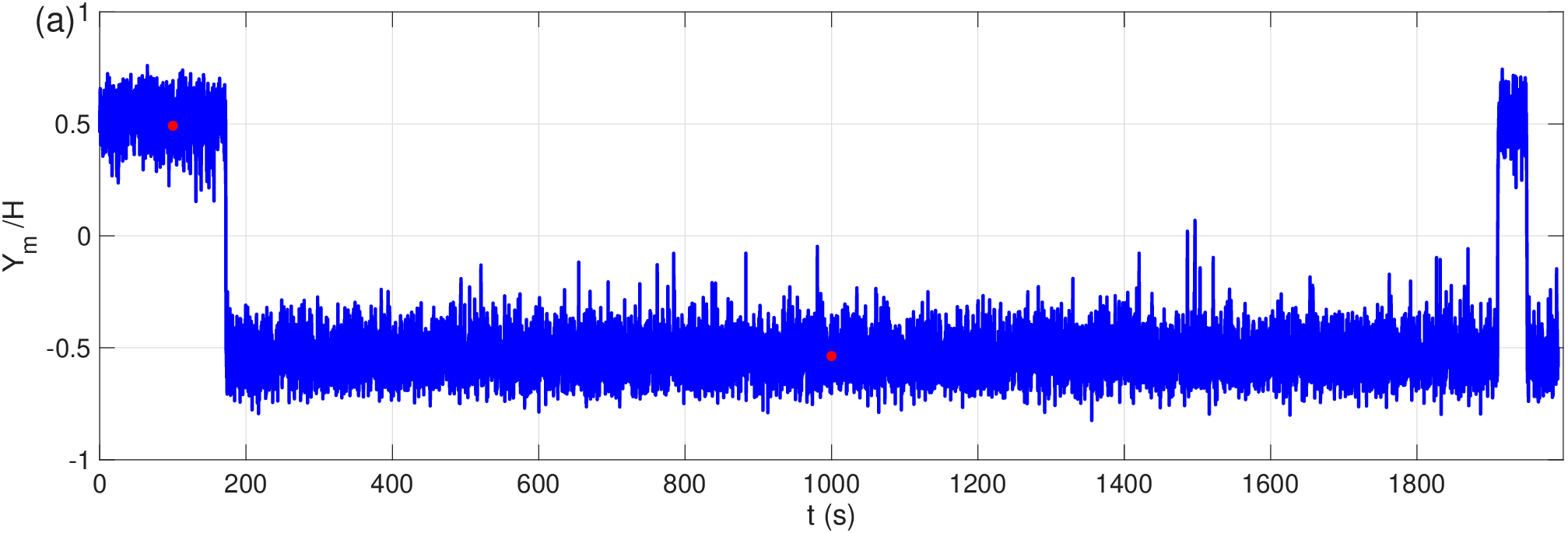}}
\vspace{1cm}
\centerline{
\includegraphics[width=5.5cm]{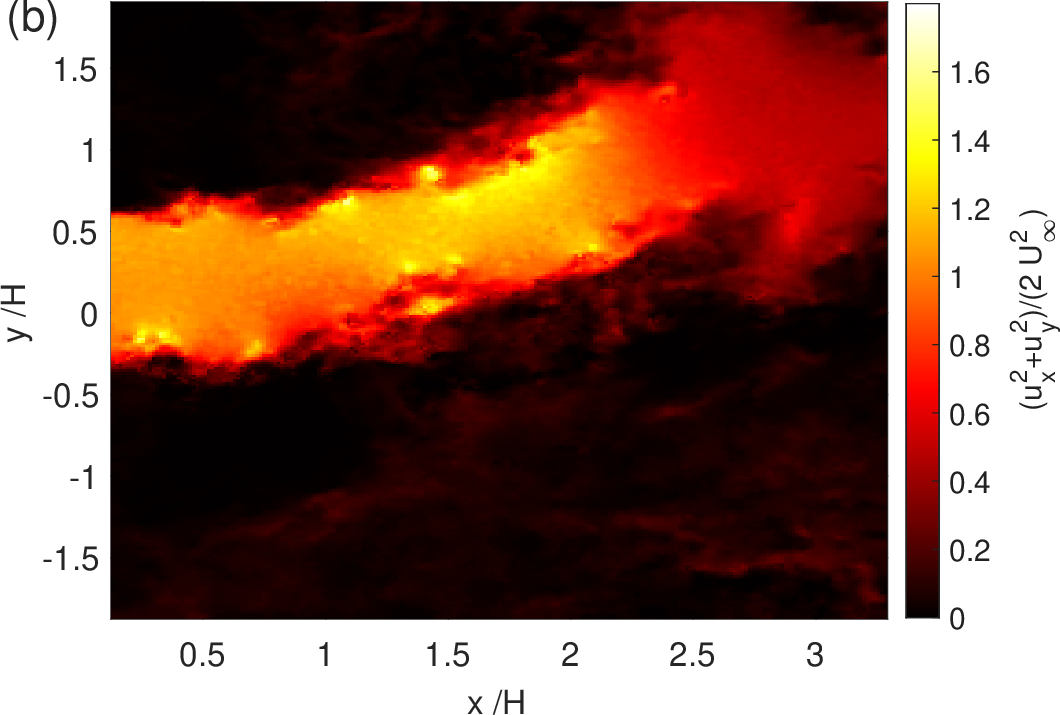}
\includegraphics[width=5.5cm]{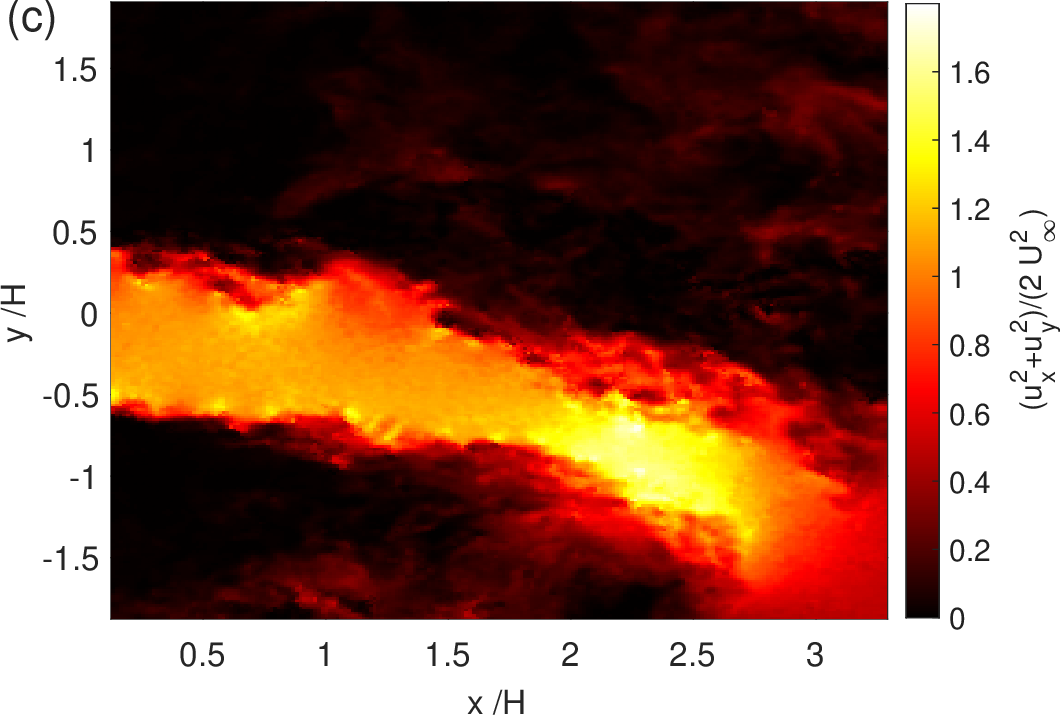}
}
\caption{From the velocity fields measured by means of PIV with a gap ratio of $G/H=2.2$ during the third experimental campaign:
(a) time series of the weighted position of the jet (the red dots indicate the corresponding instants in time corresponding to figures (b,c)).
Colour levels of the in-plane kinetic energy (b) at $t=100$s, (c) at $t=1000$s.
}
\label{vis2-2}
\end{figure}

The flow is again
bistable. This can be observed in the time series of $Y_m$ (Fig.~\ref{vis2-2} (a)):
the weighted $y$ position of the jet spends durations of $100$s to $1000$s fluctuating around $Y_m\simeq - H/2$ and $Y_m\simeq H/2$.
This corresponds to instants in time where the jet is pointing to the left (at $t=100$s Fig.~\ref{vis2-2} (b), red dot in Fig.~\ref{vis2-2} (a))
or to the right (at $t=1000$s Fig.~\ref{vis2-2} (c), second red dot in Fig.~\ref{vis2-2} (a)). 
We also note that transitions from left to right are now much rarer and that $Y_m$ is found in a wider range, comparable to the one spanned at $G/H=1.25$.
As a consequence, the histogram of $Y_m$ takes non negligible values in the same range and has two symmetric modes.

\subsubsection{Bifurcation from multistability to monostability at $G/H=2.65$, $G/H=2.7$ and $G/H=3.5$}\label{strns}


\paragraph{$2.3 \le G/H \le 2.6$}

No qualitative change compared to lower
gap ratios is indicated by the visualisations of in-plane kinetic energy, time series and histograms (Fig.~\ref{RES_disp}).
In that range of $G/H$, $Y_m$ fluctuates
in the same range. Quantitatively, 
transitions between left and right positions are becoming more common as
the gap ratio is increased.

\paragraph{$G/H=2.65$}

\begin{figure}[!ht]
\centerline{
\includegraphics[width=16cm,clip]{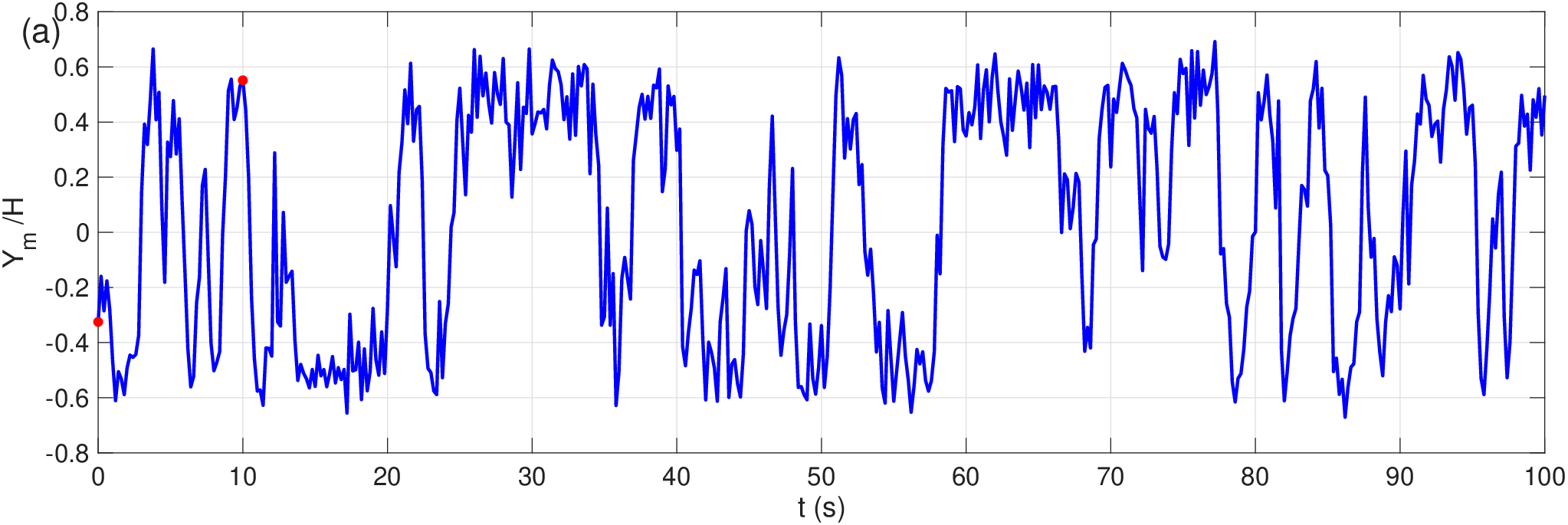}}
\vspace{1cm}
\centerline{
\includegraphics[width=5.5cm]{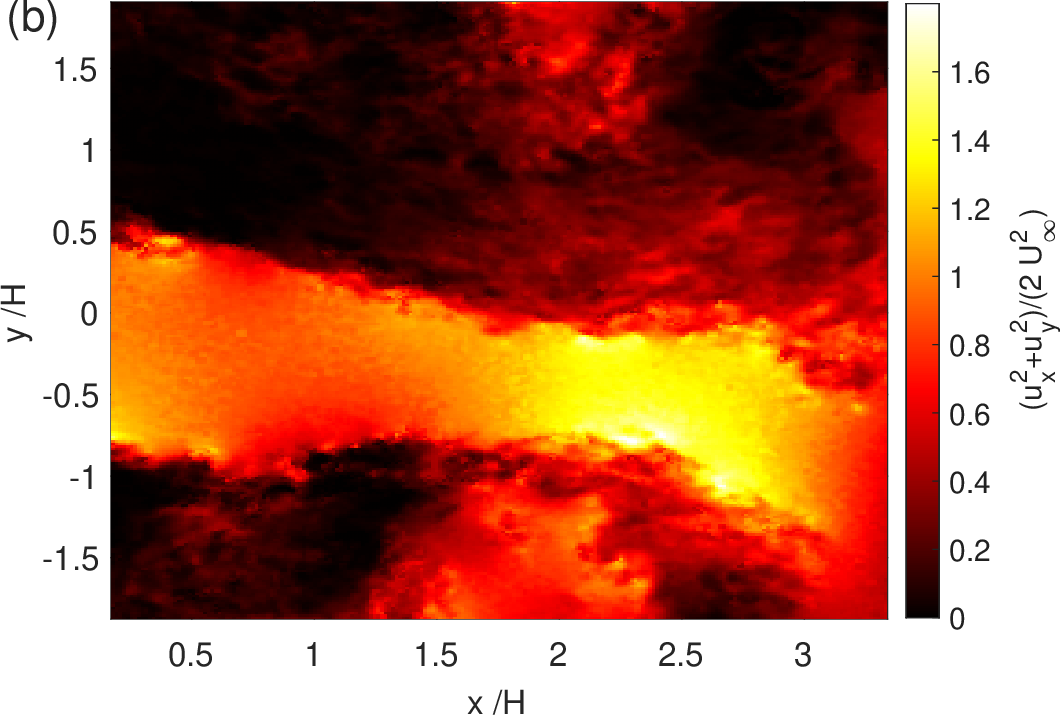}
\includegraphics[width=5.5cm]{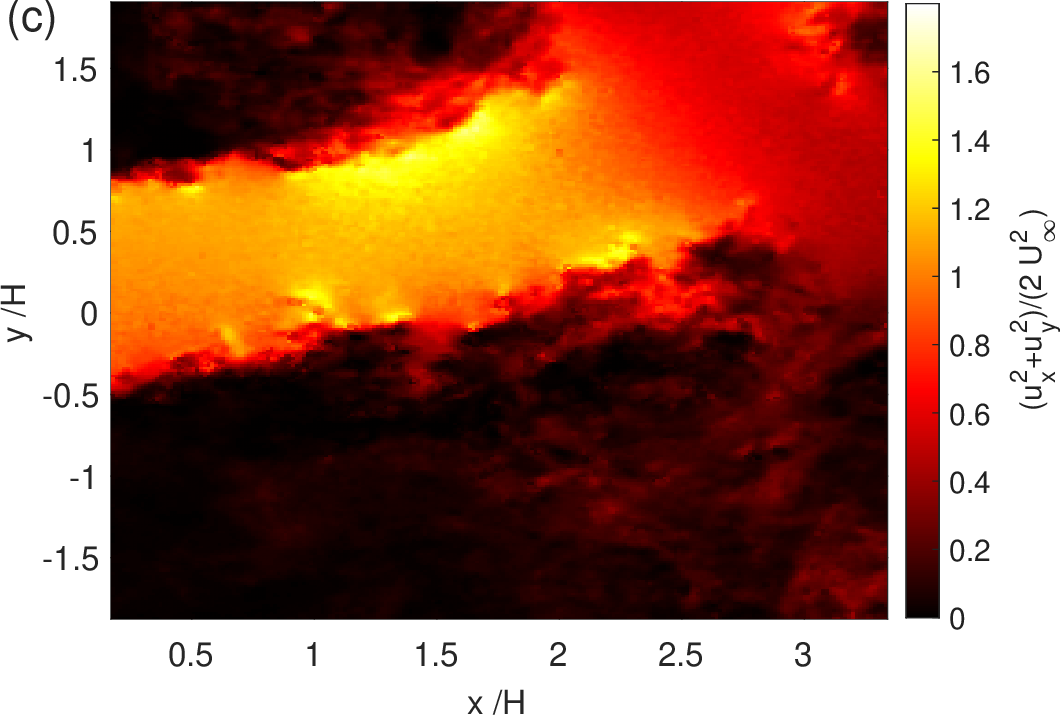}
}
\caption{From the velocity fields measured by means of PIV with a gap ratio of $G/H=2.65$ during the third experimental campaign:
(a) Time series of the weighted position $Y_m$ of the jet, zoom on the time interval $[0,100]$s
(the two red dots indicate the instants in time corresponding to figures (b,c)),
colour levels of the kinetic energy in the horizontal measurement plane obtained from the same measured velocity fields at times
(b) $t=0.2$s,
(c) $t=10$s.
}
\label{vis2-65}
\end{figure}

We describe the times series of $Y_m$ 
(Fig.~\ref{vis2-65} (a)), complemented
by two visualisations of the in-plane kinetic energy at $t=0.2$s (Fig.~\ref{vis2-65} (b)) and $t=10$s (Fig.~\ref{vis2-65} (c)), which are
indicated by red dots in the time series.
The jet position $Y_m$ fluctuates in a slightly narrower
range than at $G/H=2.2$ (Fig.~\ref{vis2-2} (a)).
The time series and the corresponding visualisations still indicate bistability,
with the jet residing on the right (Fig.~\ref{vis2-65} (b)) or on the left (Fig.~\ref{vis2-65} (c)).
As a consequence, the corresponding histogram of $Y_m$ still has two modes.
Note that we chose to zoom the time series of $Y_m$ in the time interval $[0,100]$s.
Indeed,  the characteristic residency time for $Y_m$ on either side is now much shorter than what was observed for smaller $G/H$ where multistability was observed.
This characteristic time is of the order of $10$s.

\paragraph{$G/H=2.7$}

\begin{figure}[!ht]
\centerline{
\includegraphics[width=14cm,clip]{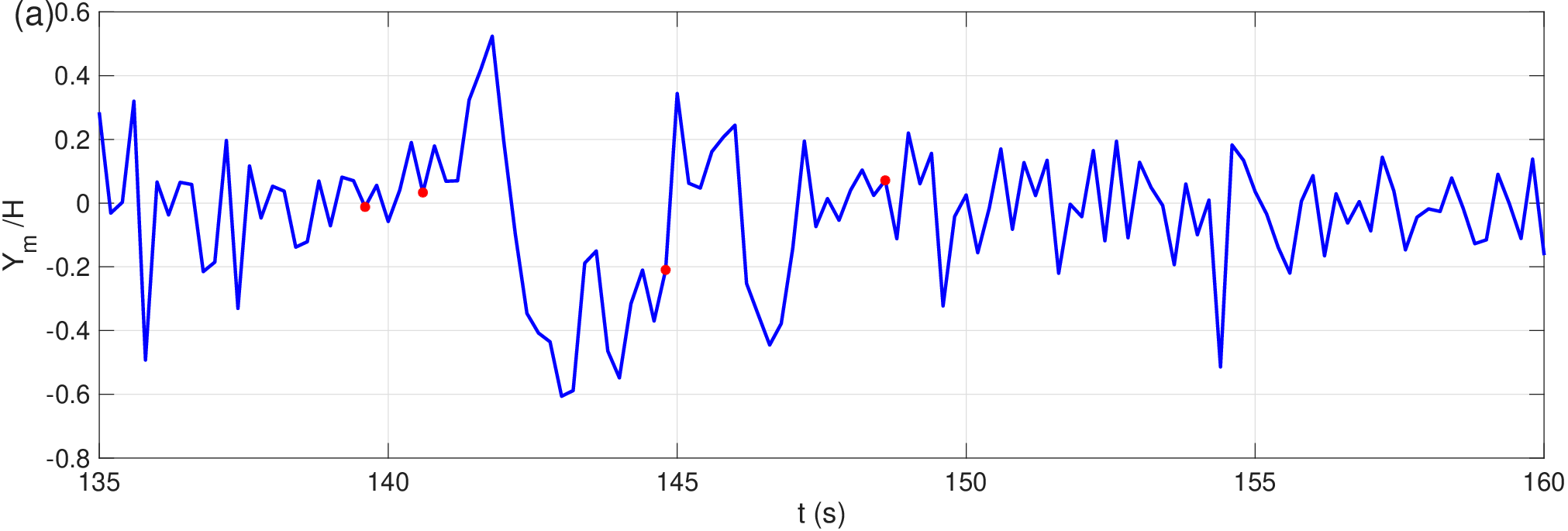}
\includegraphics[width=5.5cm,clip]{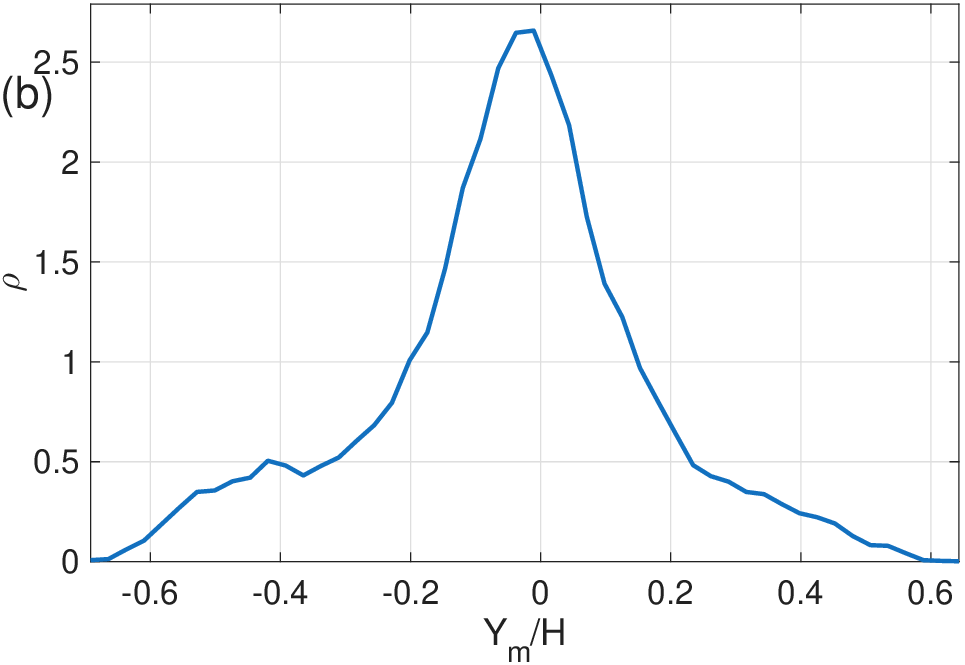}
}
\centerline{
\includegraphics[width=14cm,clip]{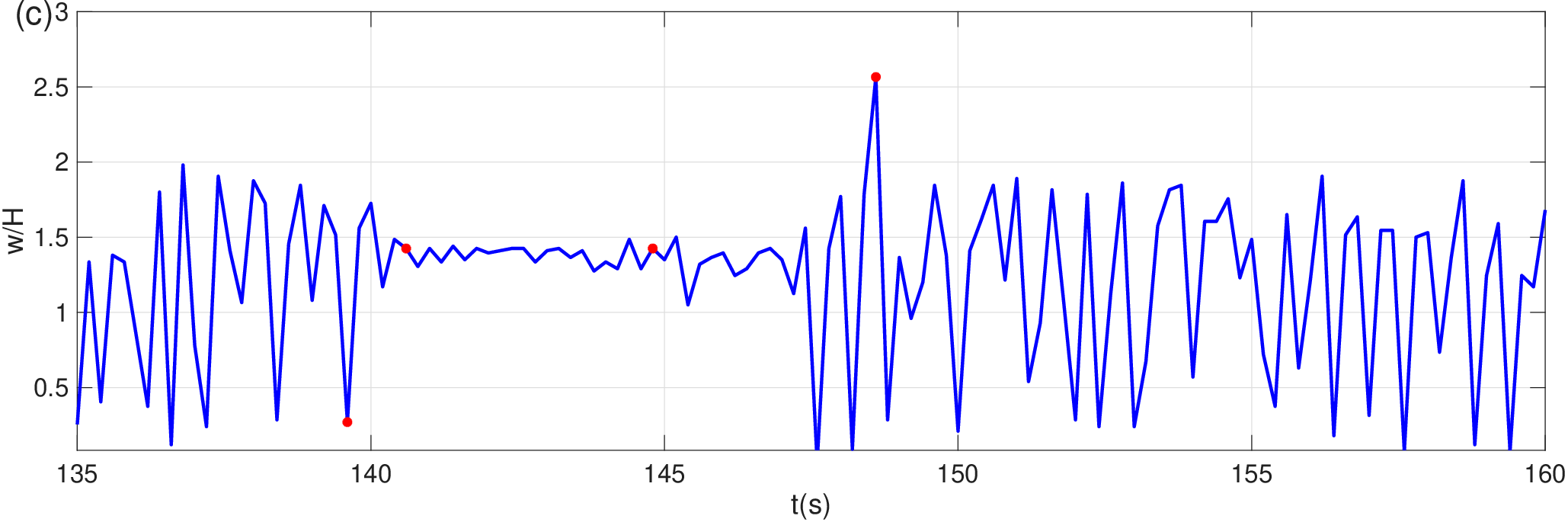}}
\vspace{1cm}
\centerline{
\includegraphics[width=4cm]{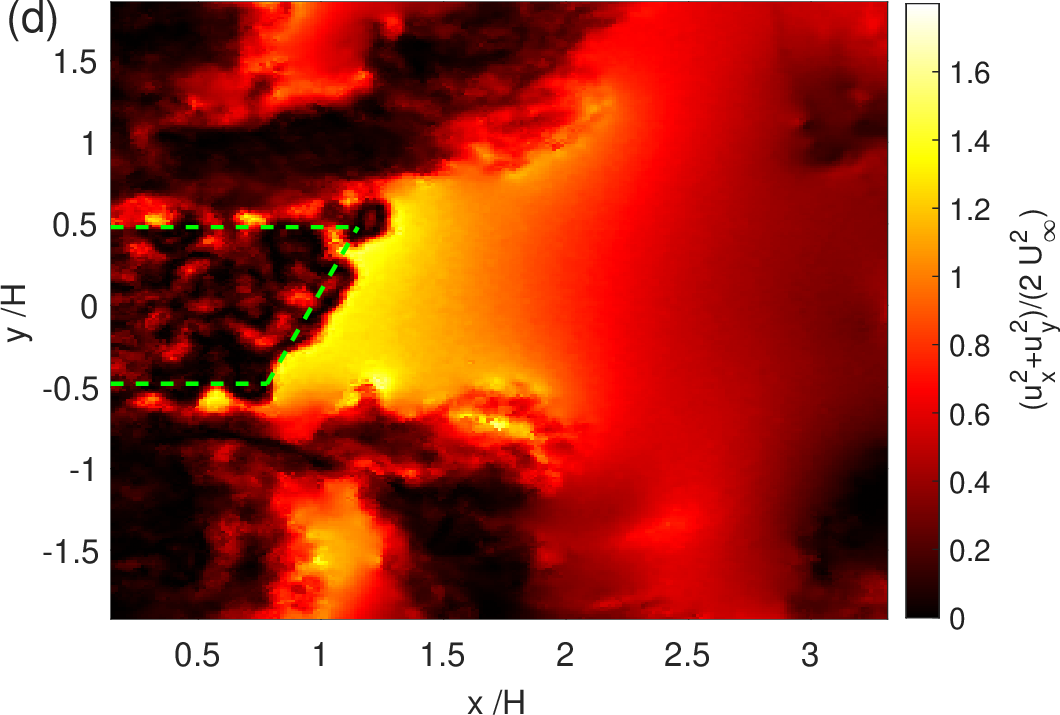}
\includegraphics[width=4cm]{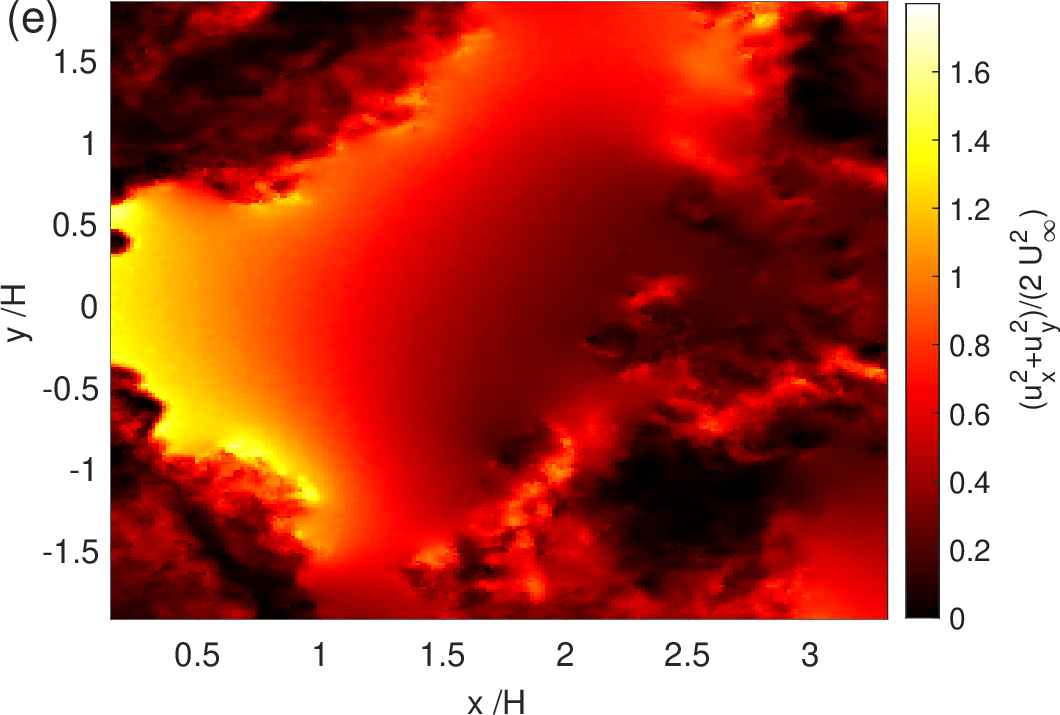}
\includegraphics[width=4cm]{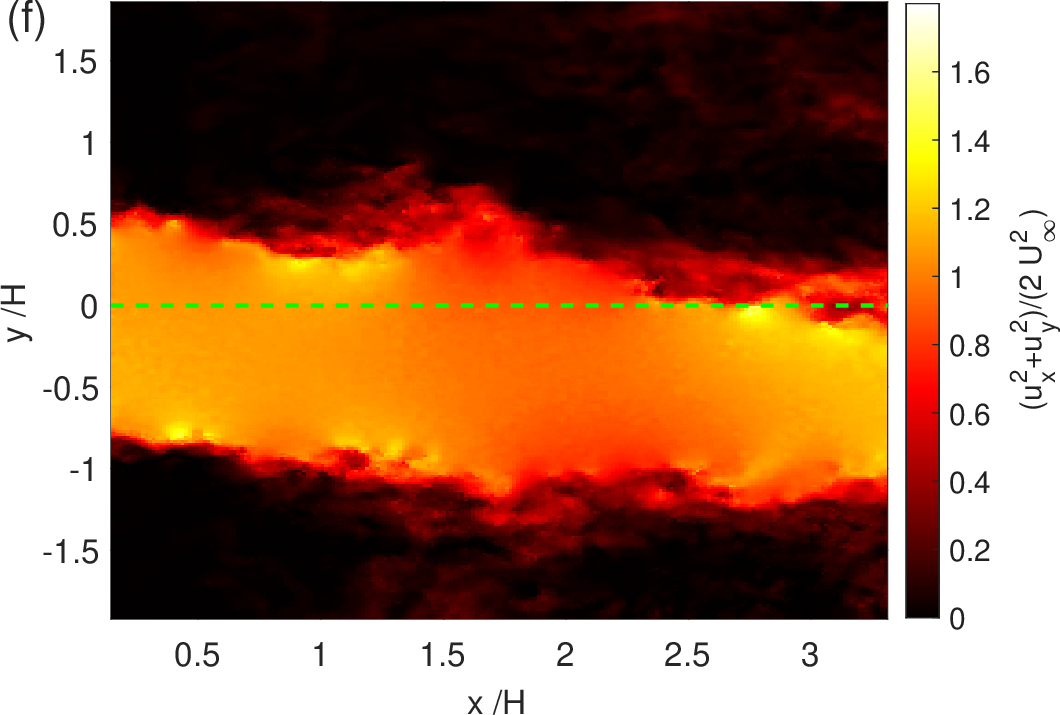}
\includegraphics[width=4cm]{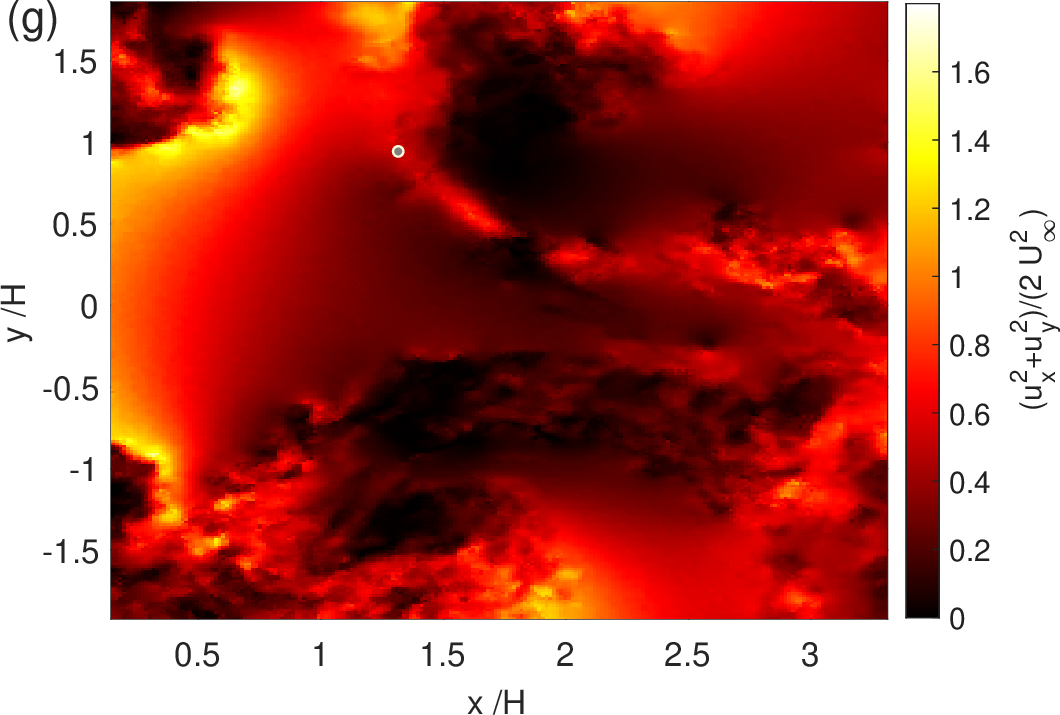}}
\caption{
From the velocity fields measured by means of PIV with a gap ratio of $G/H=2.7$ during the third experimental campaign:
(a) time series of the weighted position $Y_m$ of the jet zoomed on the time interval $[135, 160]$s,
(b) histogram of $Y_m/H$,
(c) time series of the jet width obtained from the same data set, in the same time interval
(on both time series, the four red dots indicate the instants in time corresponding to figures (d,e,f,g)),
colour levels of the kinetic energy in the horizontal measurement plane obtained from the same measured velocity fields at times
(d) $t=139.6$s, where a dashed green frame indicates the area where the jet is broken down,
(e) $t=140.6$s,
(f) $t=144.8$s (a dashed green line indicates $y=0$),
(g) $t=148.6$s.
}
\label{vis2-7}
\end{figure}

The situation becomes much more complex.
We describe the flow at this gap ratio using the time series of the weighted position of the jet $Y_m$ (Fig.~\ref{vis2-7} (a))
and the time series of the jet width $w$ (Fig.~\ref{vis2-7} (c)), both for the same zoomed time interval $t\in [135,160]$s.
We complement this description with the histogram of $Y_m$  (Fig.~\ref{vis2-7} (b)),  
and visualisations of the in-plane kinetic energy (Fig.~\ref{vis2-7} (d-g))
at instants marked by red dots in the time series (Fig.~\ref{vis2-7} (a,c)).
The qualitative change when comparing to observations at the slightly smaller gap ratios is striking.
We still observe short intervals of time where the jet is pointing to the left or to the right (for instance at $t=144.8$s, Fig.~\ref{vis2-7} (e)).
In these intervals, the width of the jet is steady and $|Y_m|$ is close to the value observed at $G/H=2.65$ (Fig.~\ref{vis2-65} (b,c)).
Note that up to $G\le 2.65$, the width of the jet increases affinely with $G/H$:
The width observed in figure~\ref{vis2-7} (e) is in line with this tendency.
However, we also observe instants in time where the jet is wider and is not pointing to any side
 at $t=140.6$s and $t=148.6$s (Fig.~\ref{vis2-7} (e,g)),
with corresponding values of $Y_m$ close to zero (Fig.~\ref{vis2-7} (a)).
The corresponding time intervals are also characterised by very rapid fluctuations of the jet width (Fig.~\ref{vis2-7} (c)).
We even observe instants (for instance at $t=139.6$ Fig.~\ref{vis2-7} (d))
where the jet is broken down.
This means that in-plane kinetic energy is close to zero in an area, indicated by the green dashed frame, at the core and the base of the jet.
Observation of the histogram (Fig.~\ref{vis2-7} (b)) shows that the recorded signal now leads to a single mode at $Y_m=0$.
While the histogram does not have heavy tails, it does have ``feet'' in the form of plateaus
in the ranges $[-2H/3,-H/3]$ and  $[H/3,2H/3]$.
The ``feet'' of the histogram are the consequence of the alternating of intervals of time where the jet
is either pointing left or right (Fig.~\ref{vis2-7} (f))
while the mode is the consequence of the intervals of time where the jet is agitated though more or less symmetric with respect to the centreline.
This behaviour of the jet is similar to what \cite{alam2011wake} termed the transition regime at higher Reynolds numbers.

\paragraph{$G/H=3.5$}
 
\begin{figure}[!ht]
\centerline{
\includegraphics[width=16cm,clip]{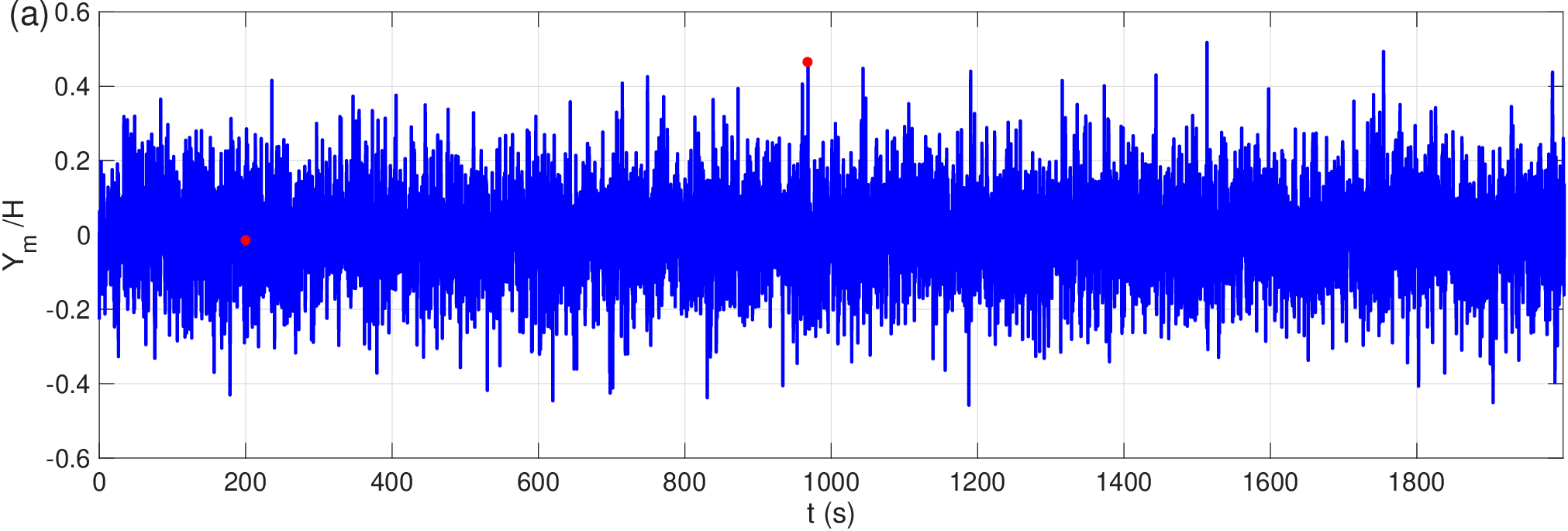}}
\vspace{1cm}
\centerline{
\includegraphics[width=5.5cm]{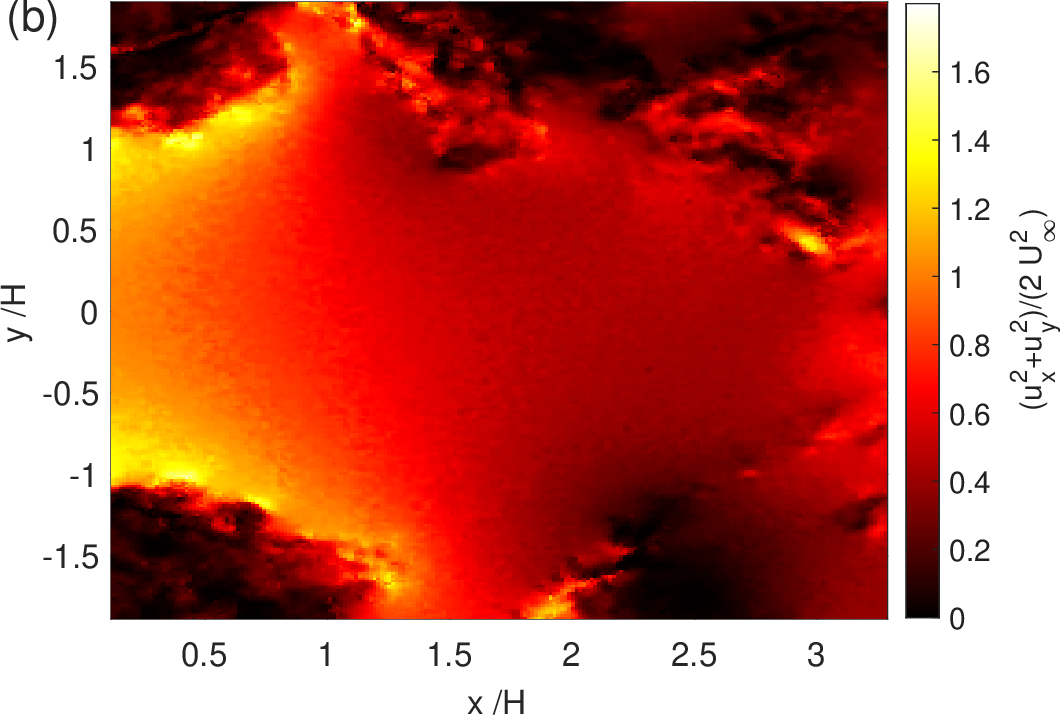}
\includegraphics[width=5.5cm]{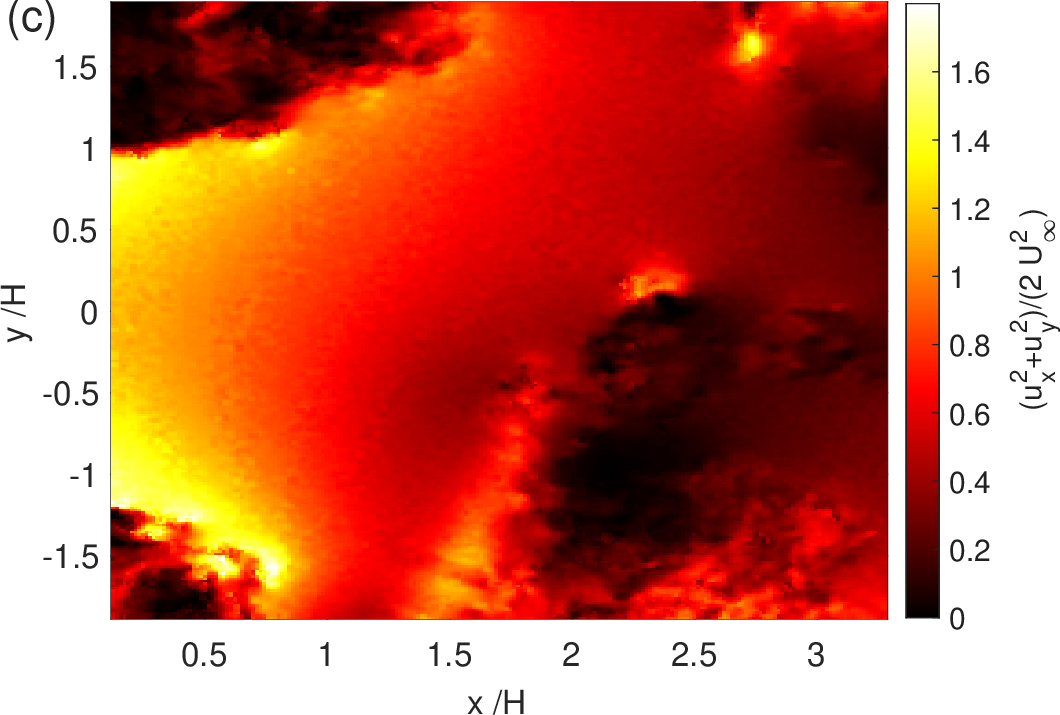}}
\caption{
From the velocity fields measured by means of PIV with a gap ratio of $G/H=3.5$ during the third experimental campaign.
(a) Time series of the weighted position $Y_m$ of the jet (the two red dots indicate the instants in time corresponding to figures (b,c)),
colour levels of the kinetic energy in the horizontal measurement plane obtained from the same measured velocity fields at times
(b) $t=200$s,
(c) $t=967.6$s.}
\label{vis3-5}
\end{figure}

The flow has become monostable at $G/H=3.0$.
We illustrate this using the time series of $Y_m$ (Fig.~\ref{vis3-5} (a)) as well as visualisations
of the in-plane kinetic energy at two successive times (Fig.~\ref{vis3-5} (b,c)).
The weighted position of the jet now fluctuates around zero, without any residency near strictly positive nor
strictly negative values.
Note that these fluctuations are not Gaussian and that the resulting histogram is spiked near $0$ (Fig.~\ref{RES_disp} (a)).  
This centred jet organisation had been observed at some instants for $G/H=2.7$ (Fig.~\ref{vis2-7} (d,f)), albeit with a thinner jet.
This corresponds to a wider centred jet which is neither left nor right leaning.
Larger values of $Y_m$ correspond to instants where left-right symmetry of the jet is imperfect (Fig.~\ref{vis3-5} (c)).
This jet type also has an almost monotonically increasing width (see \cite{chen2021turbulence}, Fig.~5 (e) for a global view).
We will examine this change of shape in more detail in a a future article.

\subsection{Model analysis}\label{srmod}

We now discuss the information that we
can extract from the finite time Kramers--Moyal coefficients and the fitted models
at specific values of $G/H$, in line with our goal of studying the bifurcation.
We divide this presentation into three parts: we first present the models
fitted in typical cases of multistability (\S~\ref{modtyp}),
then in situations where the left-right symmetry is broken but in an unusual way (\S~\ref{modpec}),
and finally throughout the bifurcation from bistable to monostable (\S~\ref{modtrans}).

\subsubsection{Tristable flow at $G/H=1.25$, bistable flow at $G/H=2.4$ and monostable flow at $G/H=3.5$}\label{modtyp}

\begin{figure}
\centerline{
\includegraphics[width=5.5cm]{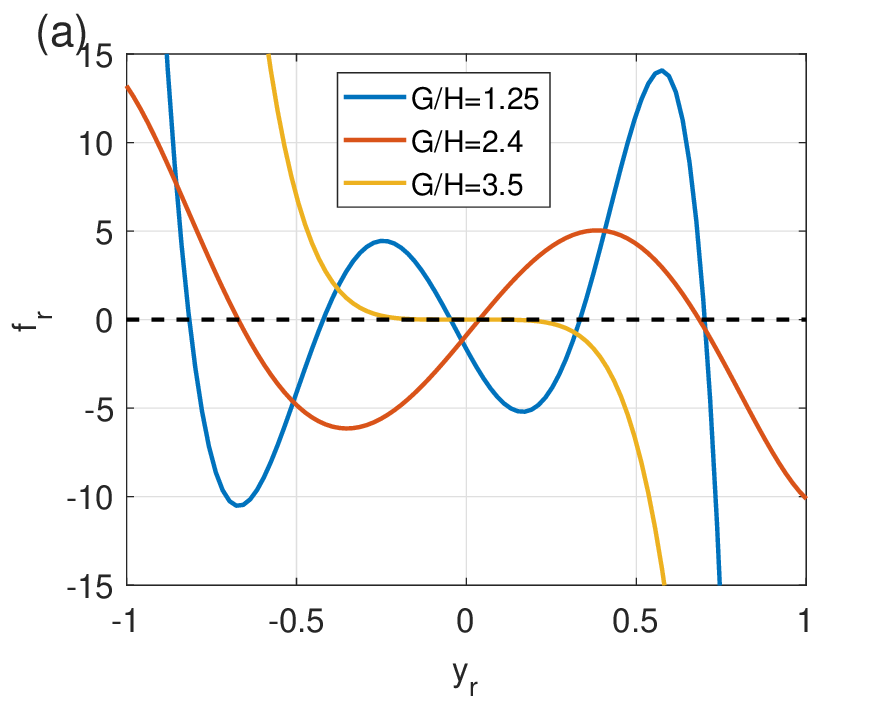}
\includegraphics[width=5.5cm]{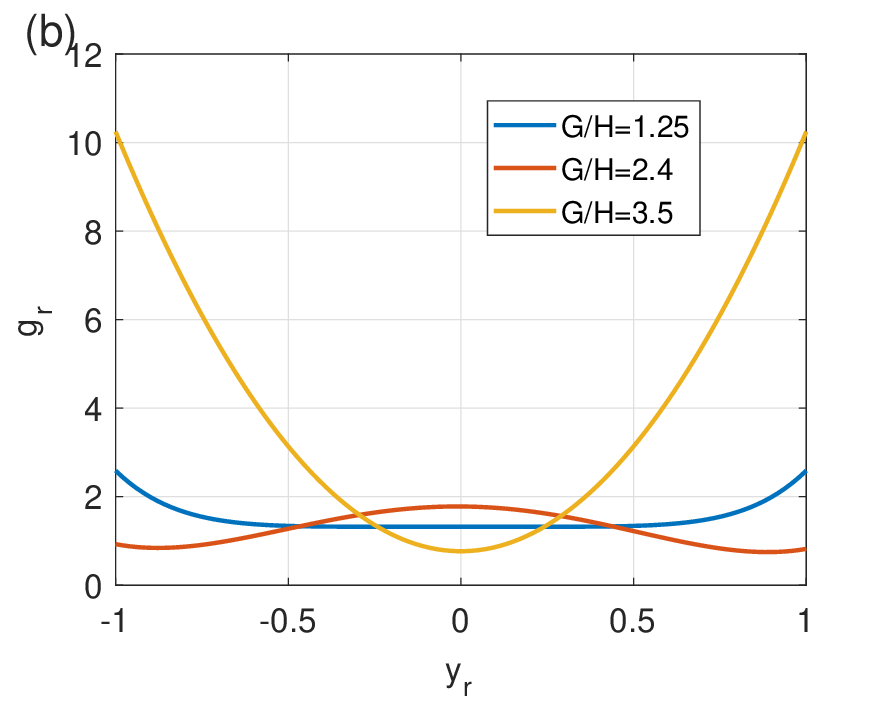}
}
\caption{
Models drift and diffusion determined from the Kramers--Moyal coefficients shown in figure~\ref{compaf},
computed from rescaled values of $Y_m$ sampled during the third experimental campaign for different values of $G/H$.
For typical multistability for $G/H=1.25$, $2.4$ and $3.5$:
(a) drift as a function of rescaled $Y_r$,
(b) diffusion as a function of rescaled $Y_r$.
}
\label{modGH}
\end{figure}

We start our description of the models constructed from the Kramers--Moyal coefficients by three values of gap ratio displaying
tristability (at $G/H=1.25$), bistability (at $G/H=2.4$)
and monostability (at $G/H=3.5$) (Fig.~\ref{modGH} (a,b)) from the third dataset.
The Kramers--Moyal coefficient and fitted drifts computed using the data from the first dataset, at these gap ratio,
convey the same information as the ones of the third dataset (albeit at a slightly larger $\tau$).

We had displayed a comparison of the model, finite time Kramers--Moyal coefficients computed from the model
and empirical Kramers--Moyal coefficients in (\S~\ref{ssemp},~\ref{fitfunc}) (Fig.~\ref{compaf} (a,b))
for $G/H=2.4$. Said Kramers--Moyal coefficients are attenuated
with respect to the model drift and diffusion, while retaining the same shape and the same
central features regarding multistability.
In particular the zeros of the drift and the first Kramers--Moyal coefficient
are at almost the same value of $Y_m$ with the same stability.
We do not display these coefficients again for $G/H=1.25$ and $3.5$
as they convey the same information as the model drifts and diffusion.
Indeed, the location and relative stability
of multistable points is mostly contained in
the drift $f_r$ (Fig.~\ref{modGH} (a)) and the respective first empirical Kramers--Moyal coefficient $f_{r,\tau}^e$. 
We recover the property often observed in stochastic multistable systems
that the position of the modes of the PDF corresponds to the zeros
of both $f_r$ and $f_{r,\tau}^e$, where the derivative $df_r/dy<0$
is negative.
Stated differently, as we had introduced in section~\ref{ssratio}, they correspond to the stable fixed points of the deterministic part of
the Langevin equation.
This can be seen for $G/H=1.25$,
where  $f_r$ has five zeros.
Three of those show $df_r/dy<0$ (at $y_r\simeq -0.82$, $-0.05$ and $0.7$),
where the three modes of the PDF were observed (Fig.~\ref{RES_disp}). 
Meanwhile, the two other zeros, where $df_r/dy>0$ (at $y_r\simeq -0.44$ and $0.33$),
are the unstable fixed points of the deterministic part of the Langevin equation and
correspond to the local minima of the pdf, in between the modes.
We then find the same property for $G/H=2.4$, which
is typical of the bistable jet:
we now have two zeros of $f_r$ and $f_{r,\tau}^e$ (at $y_r\simeq-0.66$ and $y_r\simeq 0.7$), Fig.\ref{compaf} (a)) where $df/dy<0$,
corresponding to the two modes, and one near $y_r=0$,
corresponding to the local minimum of the PDF (Fig.~\ref{compaf} (c)).
Then at $G/H=3.5$, the number of zeros is reduced to one at $y_r\simeq 0$,
with $df/dy<0$,  in line with the monostability of the jet along the centreline.
Note that the model is more curved than the finite time Kramers--Moyal coefficient (which is in turn almost linear),
in order to account for the non Gaussian PDF of $Y_m$.

For all three gap ratios $G/H=1.25$, $2.4$ and $3.5$ considered in this section,
the second Kramers--Moyal coefficients (see Fig.~\ref{compaf} (b) for $G/H=2.4$)
and the resulting diffusions (Fig.~\ref{modGH} (b))
bear a trace of multistability.
The  local minima of the diffusion are near the modes of the PDF and the stable zeros of the drift.
Meanwhile, the local maxima of $g$ are near the local minima of the PDF and the unstable zeros of the drift. However,
these variations should not be overinterpreted.
They do not necessarily command multistability.
Such variations can even appear in the second finite time Kramers--Moyal coefficient at the location
of the stable and unstable fixed points of the drift while the underlying diffusion is constant.
We will see later on that the shape, amplitude and steepness of $a_{r,\tau}^e$ and $g_r$
vary little with $G/H$ and that variations of relative stability
of the position will be controlled by the drift.

\subsubsection{Limiting cases $G/H=1.15$ and $1.8$}\label{modpec}

\begin{figure}
\centerline{
\includegraphics[width=5.5cm]{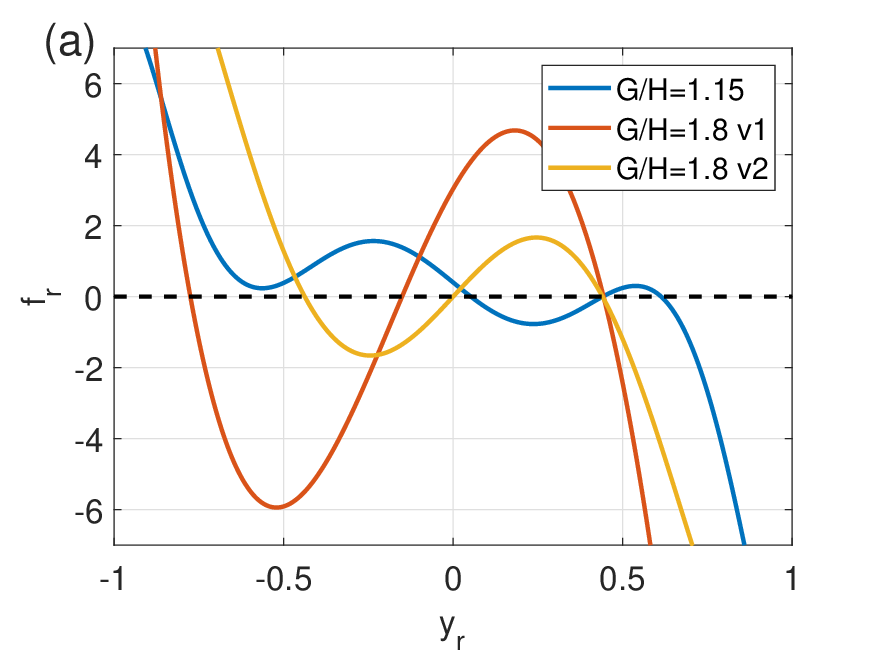}
\includegraphics[width=5.5cm]{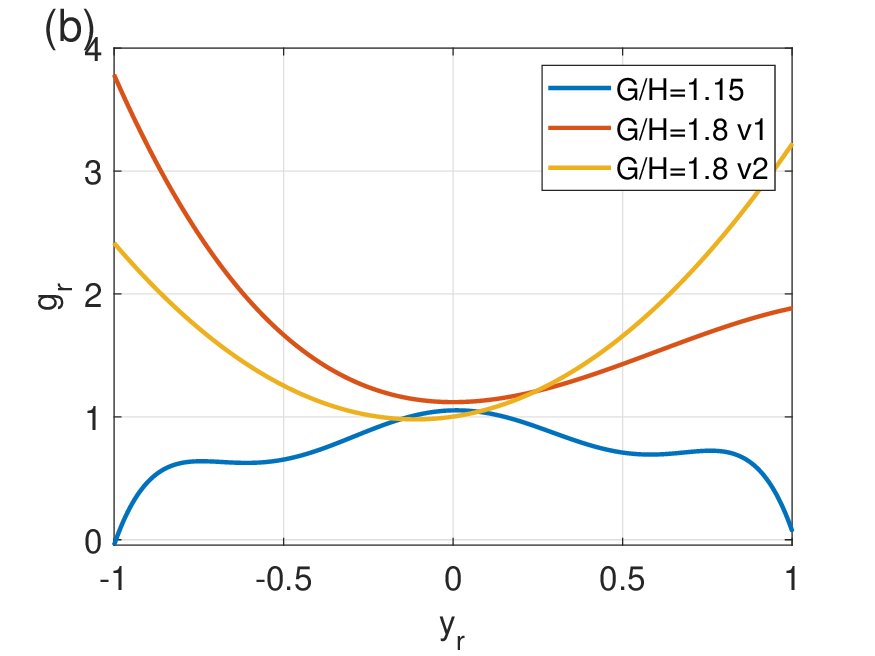}
}
\caption{
Models drift and diffusion determined from the Kramers--Moyal coefficients shown in figure~\ref{KMGH2},
computed from rescaled values of $Y_m$ sampled during the third experimental campaign for different values of $G/H$.
For undecided cases $G/H=1.15$ (for which we used $\lambda=100$), and two groups of runs at $G/H=1.8$:
(a) drift as a function of rescaled $Y_m$,
(b) diffusion as a function of rescaled $Y_m$.}
\label{modGH3}
\end{figure}

\begin{figure}
\centerline{
\includegraphics[width=5.5cm]{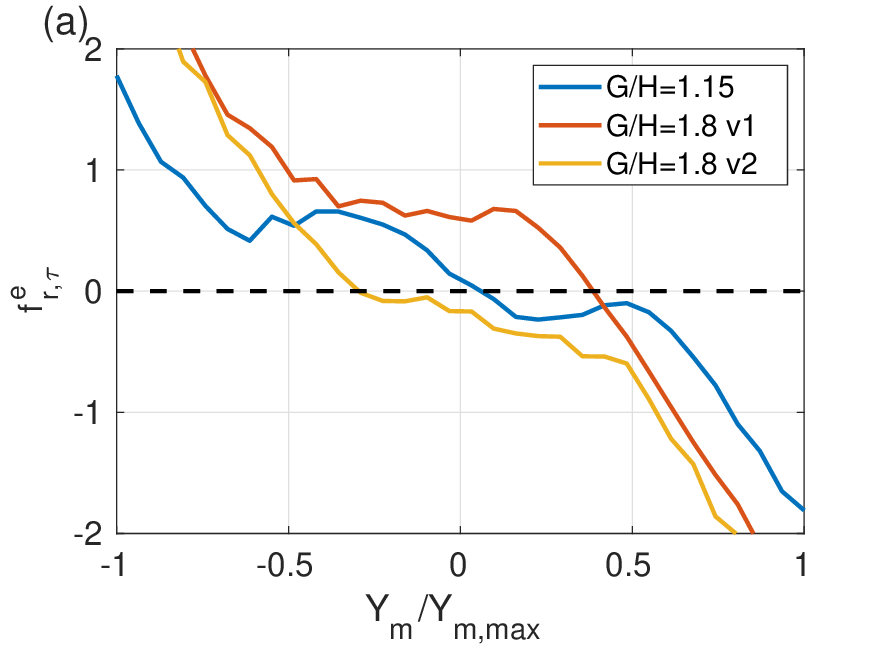}
\includegraphics[width=5.5cm]{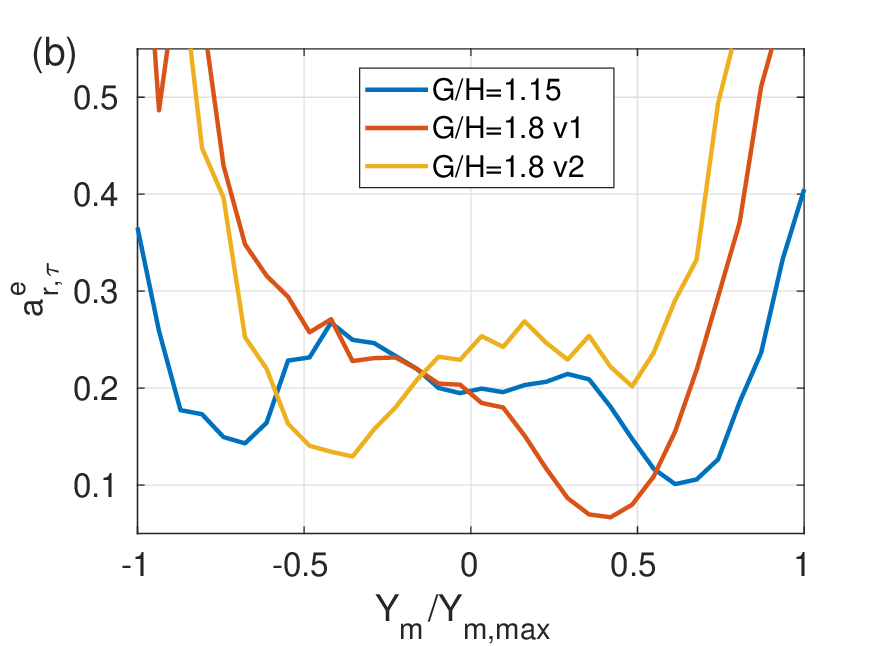}
}
\caption{
First and second Kramers--Moyal empirical coefficients computed from rescaled values of
$Y_m$ sampled during the third experimental campaign for different values of $G/H$.
In that case $\tau=0.2$s.
For undecided cases $G/H=1.15$, and two groups of runs at $G/H=1.8$:
(a) first Kramers--Moyal coefficient $f_{r,\tau}^e$ as a function of rescaled $Y_m$,
(b) second Kramers--Moyal coefficient $a_{r,\tau}^e$  as a function of rescaled $Y_m$.}
\label{KMGH2}
\end{figure}

We secondly describe the finite time empirical Kramers--Moyal coefficients (Fig.~\ref{KMGH2} (a,b)),
and the corresponding fitted model (Fig.~\ref{modGH3} (a,b)) computed from $Y_m$ sampled at $G/H=1.15$
and for the two groups of runs at $G/H=1.8$v1, v2.
We focus on these two gap ratios because they display a jet for which the left-right symmetry is broken, which is not monostable at the centre,
while not generating a bimodal histogram and signal.
The symmetry breaking is not simply readable from the histograms and its analysis benefits from
the examination of both the kramers--Moyal coefficients and models.
In this subsection and in the next, we examine them together to ascertain our conclusions, because the model fit
requires that the Lagrange multiplier is increased to $\lambda=100$ to have a precise enough
computation.

We first describe the Kramers--Moyal coefficients and fitted model at $G/H=1.15$ in view of those at $G/H=1.25$.
The second Kramers--Moyal coefficient (Fig.~\ref{KMGH2} (b)) is very similar
to what is sampled when the system is bistable
(see Fig.~\ref{compaf} (b) at $G/H=2.4$ and Fig.~\ref{modGH2} (b) at $G/H=2.6$ and $2.65$).
The first Kramers--Moyal coefficient (Fig.~\ref{KMGH2} (a)) is peculiar:
it displays a single zero for $|Y_m|/Y_{\max}\simeq 0$
and two local extrema, closer to zero at approximately
symmetric positions $Y_m/Y_{\max}\simeq -0.6$ and $Y_m/Y_{\max}\simeq 0.5$.
The premises of a symmetry breaking and a tristable flow are indicated by the local extrema and the stable $0$ of this function.
This is reflected in the fitted drift (Fig.~\ref{modGH3}).
It displays a single stable zero at $0$ as well as two local extrema at $y_r\simeq 0.66$ where $f_r$ is close to zero.
This model is close to displaying tristability. This is contained in the sign of the three odd model parameters,
and will be discussed again when we examine bifurcations diagrams (\S~\ref{sbifdiag}).
Using the model, We can discuss this closeness to tristability more precisely than using the time series and the histogram.

We then discuss the finite time Kramers--Moyal coefficients sampled for both groups
of runs at $G/H=1.8$ along with the fitted models.
In both cases, the first Kramers--Moyal coefficient (Fig.~\ref{KMGH2} (a))
has a single zero at $|Y_m|/Y_{\max}\simeq 0.4$, close to the single mode of the corresponding PDF. 
For $|Y_m/Y_{\max}|\ge 0.4$, the first Kramers--Moyal coefficients sampled in both groups of datasets are very close to one another. The same goes
for the second Kramers--Moyal coefficient (Fig.~\ref{KMGH2} (b)).
Inside the range $Y_m/Y_{\max}\in[-0.4,0.4]$, while the first Kramers--Moyal coefficients have a different sign,
they have approximately the same slope.
We note that these empirical finite time Kramers--Moyal coefficient give rise to
 fitted drifts (Fig.~\ref{modGH3} (a)) which now indisputably present a broken symmetry and two stable fixed points.
This could present a middle ground between the state of the flow and what is permitted by the polynomial models.
Meanwhile, the fitted diffusions (Fig.~\ref{modGH3} (b)) are convex, with a single minimum near $y_r=0$ and only
display a small asymmetry.
Again, we can use the empirical Kramers--Moyal coefficients and the fitted models to pinpoint the 
fact that the flow is close to bistability, with a symmetry breaking position having more weight than the other.



\subsubsection{Bifurcation from bistability to monostability in the range $G/H=2.6$, $2.65$ and $2.7$}\label{modtrans}

\begin{figure}
\centerline{
\includegraphics[width=5.5cm]{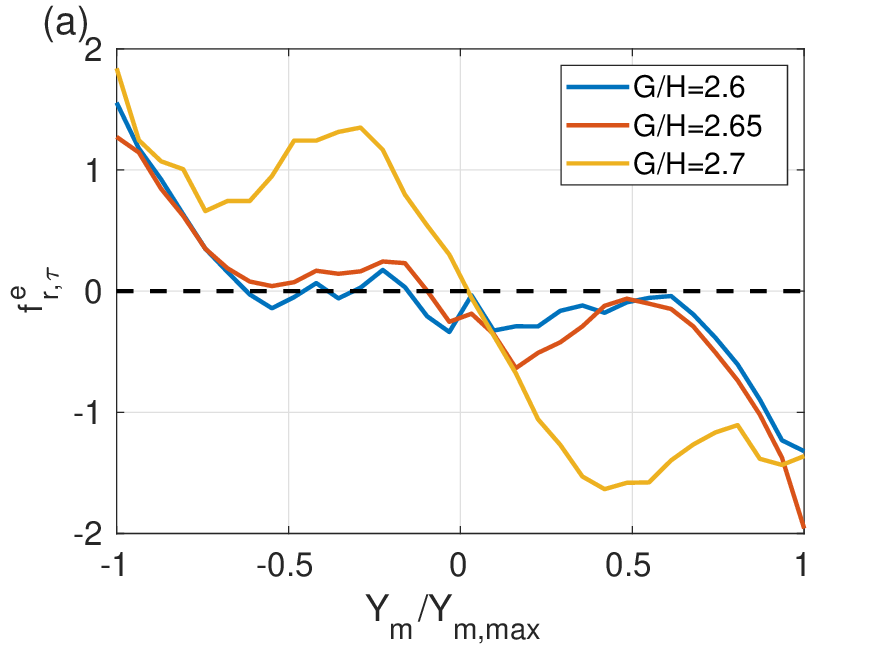}
\includegraphics[width=5.5cm]{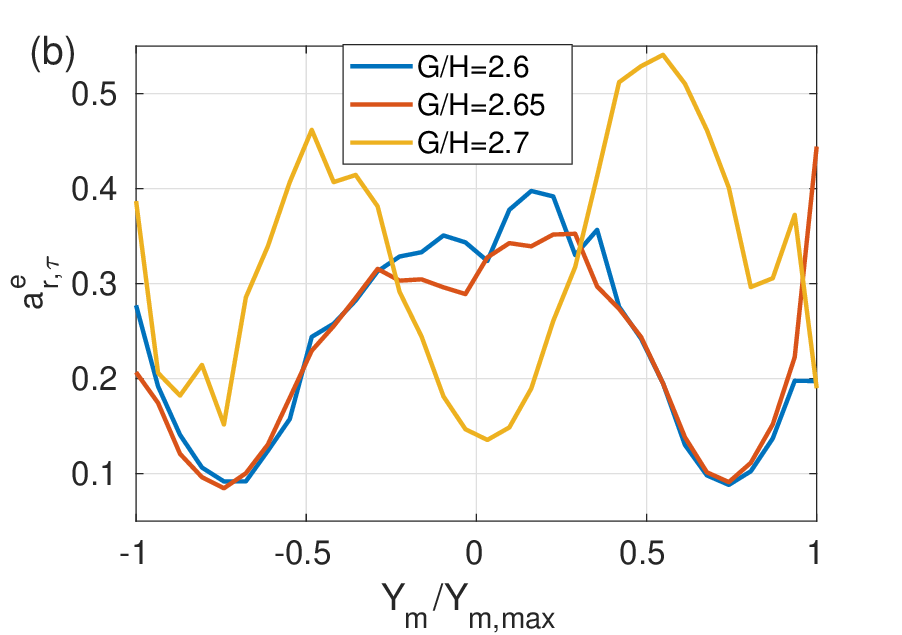}}
\caption{
First and second Kramers--Moyal empirical coefficients computed from rescaled values of
$Y_m$ sampled during the third experimental campaign for different values of $G/H$.
In that case $\tau=0.2$s.
Near the bifurcation for $G/H=2.6$, $2.65$ and $2.7$:
(a) first Kramers--Moyal coefficient $f_{r,\tau}^e$ as a function of rescaled $Y_m$,
(b) second Kramers--Moyal coefficient $a_{r,\tau}^e$  as a function of rescaled $Y_m$.
}
\label{KMGH}
\end{figure}

\begin{figure}
\centerline{
\includegraphics[width=5.5cm]{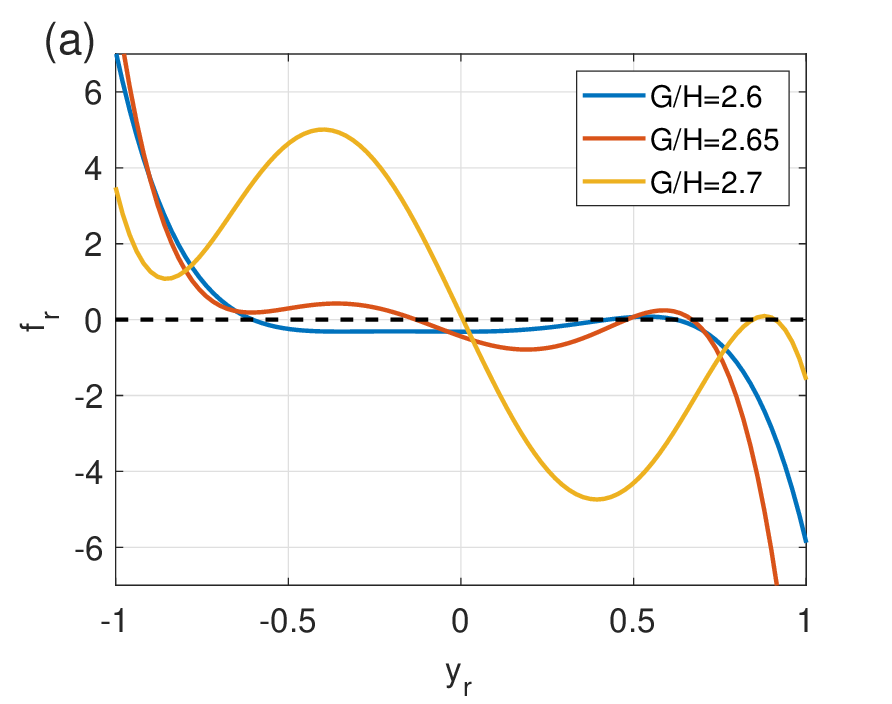}
\includegraphics[width=5.5cm]{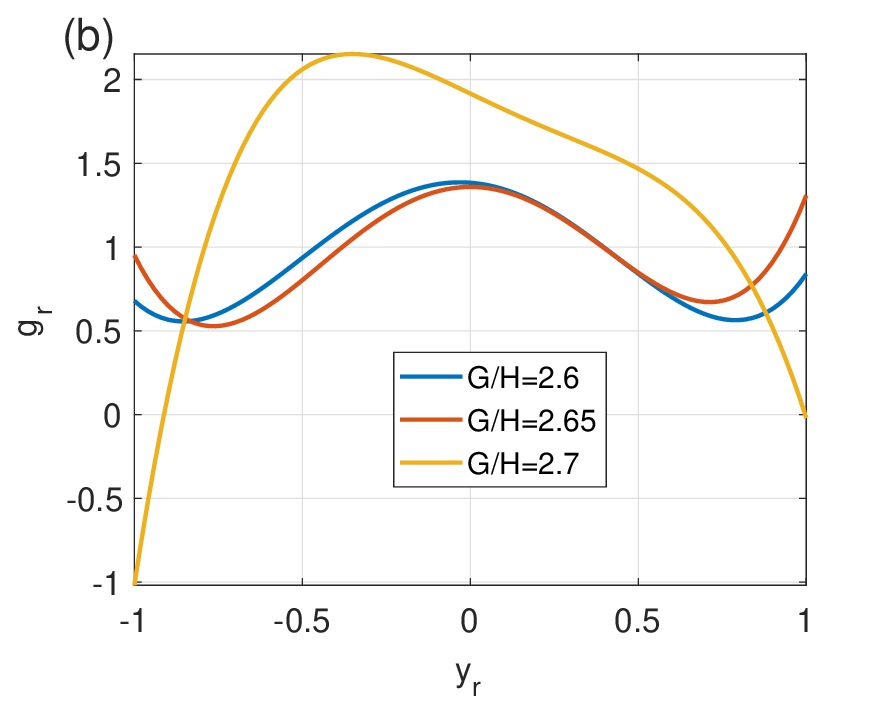}}
\caption{
Models drift and diffusion determined from the Kramers--Moyal coefficients shown in  figure~\ref{KMGH},
computed from rescaled values of $Y_m$ sampled during the third experimental campaign for different values of $G/H$.
Around the bifurcation for $G/H=2.6$, $2.65$ and $2.7$ (for which we used $\lambda=100$):
(a) drift as a function of rescaled $Y_m$,
(b) diffusion as a function of rescaled $Y_m$.
}
\label{modGH2}
\end{figure}

We finally describe how the Kramers--Moyal coefficients (Fig.~\ref{KMGH} (a,b)) and the models constructed from them (Fig.~\ref{modGH2} (a,b))
are modified throughout the peculiar bifurcation from bistable to monostable experienced by the system around $G/H=2.7$.
Compared to their counterparts at $G/H=1.25$ and $G/H=2.4$,
the first and second Kramers--Moyal coefficients at $G/H=2.6$ and $2.65$
have the same amplitude and shape for $\left|Y_m/Y_{\max}\right|>0.66$,
which corresponds to jet positions on the sides of multistable positions.
The first Kramers--Moyal coefficient is quantitatively changed
and is close to zero in the interval $Y_m/Y_{\max}\in[-0.66,0.66]$, while still negative for $Y_m>0$ (for both $G/H=2.6$ and 2.65)
and positive for $Y_m<0$ (for $G/H=2.65$).
Similarly, the fitted drifts (Fig.~\ref{modGH2} (a)) are also flat in the range $\left||Y_m|/Y_{\max}\right|<0.66$,
they tend to be smaller than what is computed inside that range for $G/H=2.4$.
At $G/H=2.6$, the model drift still cancels out at three points, and thus
still represent the two bistable positions. At $G/H=2.65$, the even flatter first finite time Kramers--Moyal coefficient
leads to a very sensitive model drift which has stable symmetry broken positions but where the central position becomes close to stable as well.  
For $\left|Y_m/Y_{\max}\right|<0.66$,
the second Kramers--Moyal coefficient is still close to what is estimated at $G/H=2.4$.
Similarly, for these two gap ratios, the fitted diffusions (Fig.~\ref{modGH2} (b)) are unchanged in shape and amplitude compared
to what is computed at $G/H=2.4$.

  At gap ratio $G/H=2.7$, the closest to the bifurcation from bistability to monostability,
the sampled first Kramers--Moyal coefficient (Fig.~\ref{KMGH} (a)) and the
corresponding fitted drift (Fig.~\ref{modGH2} (a)) have changed brutally compared to those fitted at $G/H=2.65$.
They now both have a stable $0$ at $Y_m\simeq 0$ and $y_r\simeq 0$ with $df_r/dy<0$.
The first empirical Kramers--Moyal coefficient has local minima at $\left|Y_m/Y_{\max}\right|\simeq 0.66$,
where the multistable points were. Meanwhile the drift touches the line $f=0$, for $|y_r\simeq0.75|$, in a similar range.
For larger values of $\left|Y_m/Y_{\max}\right|>0.66$,
this coefficient is similar to what was observed at other gap ratios.
This is indicated by the collapse of the three curves of $f^e_{r,\tau}$ for the three gap ratios in those intervals of $Y_m/Y_{\max}$.
Conversely, we have a bifurcation of the drift as $G/H$ is increased from $2.65$ to $2.7$:
it goes from two bistable positions at $y_r\simeq 0.66$ to one monostable position at $y_r\simeq 0$.
At $y_r\simeq 0.66$, we now have two slow zones which can be viewed at the ghosts of the former fixed points:
that is to say zones in $y$ where the drift is relatively small but non zeros and which contained actual fixed points at a slightly different
value of the control parameters.
The change in the second Kramers--Moyal coefficient is less drastic (Fig.~\ref{KMGH} (b)):
it retains the same order of magnitude as for other gap ratios and has
a modulation in line with the position of fixed points.
Similarly, the fitted diffusion retains the same order of magnitude as for smaller gap ratios.
Taken together, this drift and diffusion account for the peculiar dynamics and histogram observed at $G/H=2.7$.
The stable fixed point at $Y_m=0$ represents the PDF mode (Fig.~\ref{vis2-7} (b)) and the fact that the
jet is mostly found to be along the centreline (Fig.~\ref{vis2-7} (a,e,g)).
Meanwhile the ghosts of the fixed points represent the possibility for the jet to visit the multistable positions (Fig.~\ref{vis2-7} (a,f))
and generate the ``feet'' of the histogram (Fig.~\ref{vis2-7} (b)).
Finally, the shape of the diffusion accounts for the fact that the felt noise is more intense near $y_r\simeq 0$.


\subsection{Bifurcation diagrams}\label{sdiag}

Throughout the former sections, we have had case by case  views of the regimes and regime changes
occurring in the flow as the gap ratio is increased.
In order to have a global view of the flow states in the range $G/H\in[1.15,3.5]$,
we construct bifurcations diagrams from selected quantities from the models and averages of our flow diagnostics.
In these diagrams, we use the values computed from all groups of runs in the first and third datasets
as well as the continuous groups of runs in the second dataset.   
We separate this presentation in two parts: we firstly present diagrams constructed from the cumulants of
scalars we have sampled in our study, complemented with refined information from the models (\S~\ref{sbdiag}).
We then secondly complement this view with bifurcation diagrams for model parameters (\S~\ref{sbifdiag}),
so as to illustrate how the regime changes can be read from the examination of a few values.

\subsubsection{Bifurcation diagram from flow diagnostics}\label{sbdiag}

\begin{figure}[!ht]
\centerline{
\textbf{(a)}
\includegraphics[width=12cm]{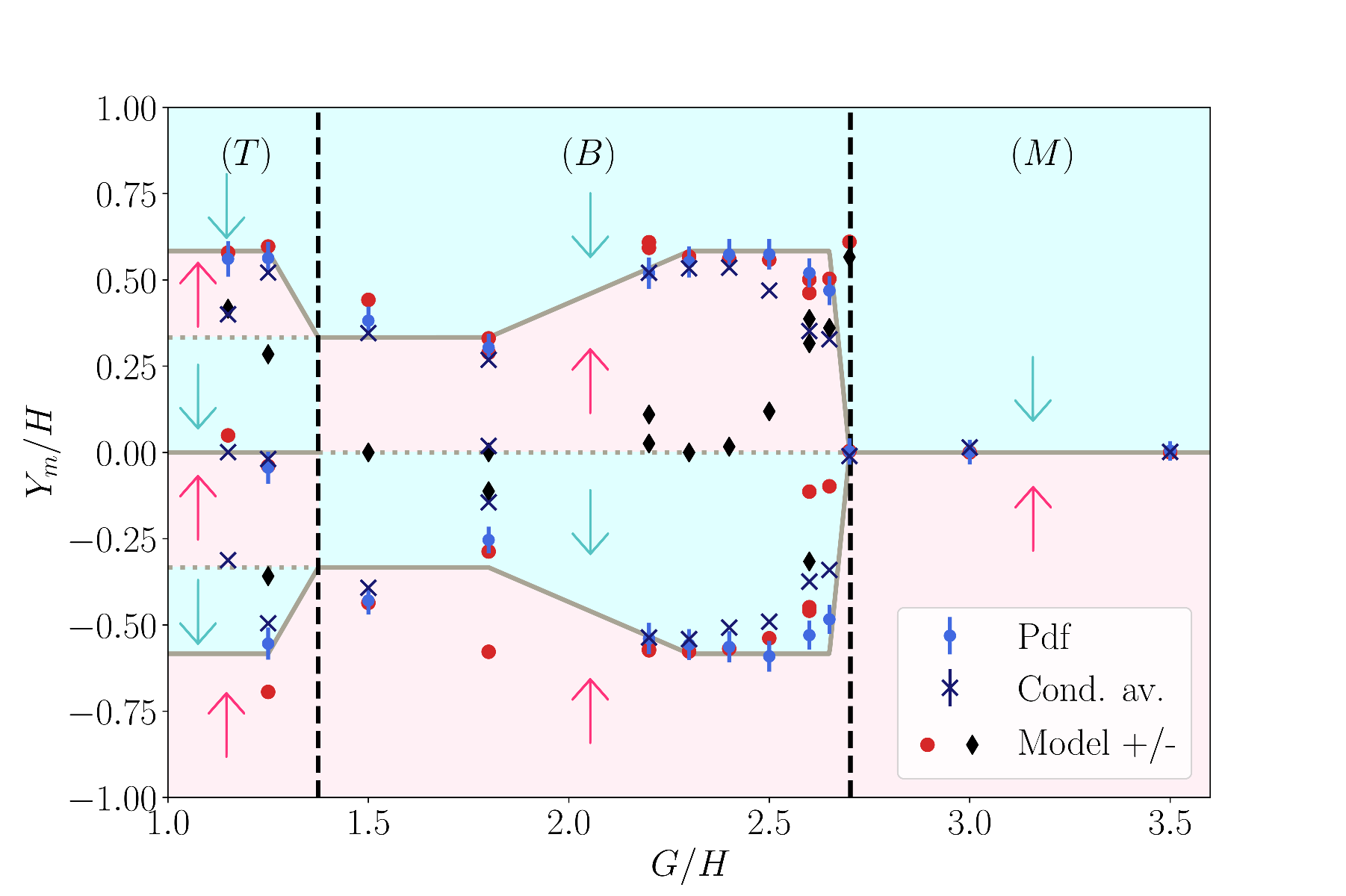}} 
\centerline{
\includegraphics[width=6cm]{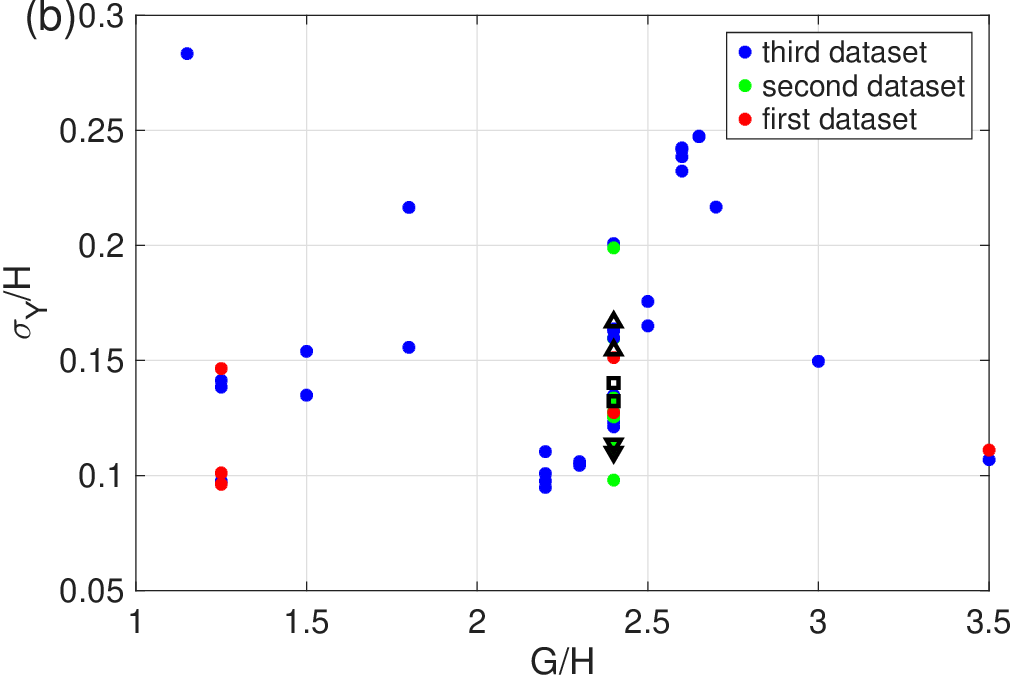}
\hspace{1cm}
\includegraphics[width=6cm]{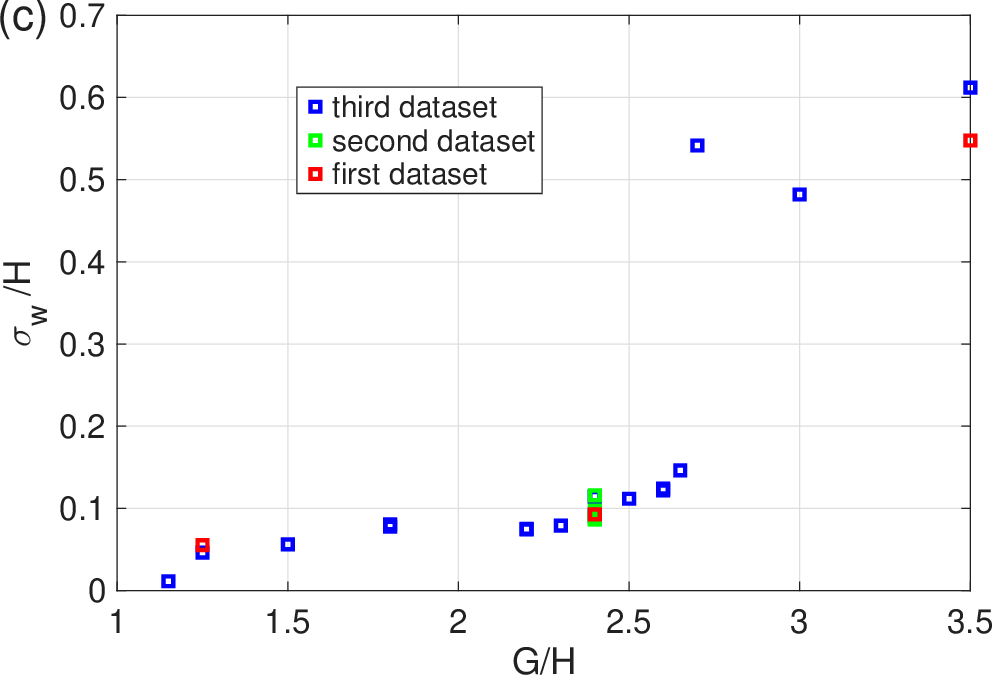}
}
\caption{
Using time series of $Y_m$, determined using the time series computed from the
velocity fields obtained during the three experimental campaign:
(a) generic bifurcation diagram as a function of $G/H$ presenting the modes of the PDF of $Y_m$ (red dots),
the average of $Y_m/H$ conditioned to being in $Y_m>0$ or $Y_m<0$ states (blue dots) along with the stable (black crosses) 
and unstable zeros (black diamonds) 
of the fitted models.
(all points are obtained with $\lambda=10$, except for $G/H=2.2$, $3.0$
where $\lambda=1$ and $G/H=1.15$ and $G/H=2.7$ where $\lambda=100$).
Each regime type: tristability (T), bistability (B) and monostability (M) is separated by black dashed lines.
Furthermore, the sign of the model drift and the corresponding attracting direction is indicated by the filling colour 
(\textcolor{cyan}{cyan} : negative, \textcolor{pink}{pink} : positive) and the arrows. 
(b) Standard deviation of $Y_m/H$ conditioned to being in $Y_m>0$ or $Y_m<0$ states as  function of $G/H$.
(c) Standard deviation of the jet width computed from the velocity fields obtained in the first and the third experimental campaigns
as function of $G/H$. The same colour code as in (a,b,c) is used to designate the dataset.}
\label{bifdiag}
\end{figure}

We present a bifurcation diagram for the jet multistability as a function of $G/H$ in figure~\ref{bifdiag} (a)
using three diagnostics of the multistable states: the average of $Y_m$ conditioned on being on the left,
right (and centre when necessary) positions, (see \S~\ref{stated}),
the modes of the histogram (see \S~\ref{ssemp}), again distinguishing left, right and centre positions
and the stable zeros of the drift of the fitted models (see \S~\ref{ssratio}, The unstable zeros are added to the plot for further analysis).
The ordinate of our diagram is in units of $Y_m/H$.
We complement this diagram with vertical black dashed lines separating the tristable (T), bistable (B) and monostable (M) regimes.
We indicate the sign of the drift using the colour filling (\textcolor{cyan}{cyan}: negative, \textcolor{pink}{pink}: positive) and the arrows. 
The gray lines, which approximately correspond to the position of the metastable states separate these areas vertically.
For each value of $G/H$, the values of conditional average of $Y_m$ presented here are obtained as an average over the groups
of runs of the average within a group of runs. The error bars correspond to the standard deviation $\Sigma$ of this estimate: 
if we denote $\sigma_{Y,i}$ the standard average of  $Y_m$ within each group of runs $i$, $N_i$ the number of time frames in that group of runs
and $\mathcal{N}$ the number of groups of runs at that gap ratio, 
we have $\Sigma^2 =\frac{1}{\mathcal{N}}\sum_{i=1}^\mathcal{N}\frac{\sigma_{Y,i}^2}{N_i}$.

We first note that each of these diagnostics give a similar estimate of the positions of the multistable states, for almost all values of $G/H$.
The exception is one of the peculiar cases: at $G/H=1.15$ the jet dynamics are not yet settled in tristability (\S~\ref{modpec}).
At that gap ratio, the zero of the model is at the central position
while the histogram mode and the average of $Y_m$ are located at the left symmetry broken position.
The correspondence between the position of the modes and that of the conditional average
can be expected from the weight of the histogram around the mode.
The correspondence of these measures with the stable zeros of the drift confirms that they control the
multistable positions.

Using this bifurcation diagram, we can precisely describe the multistable states as $G/H$ is increased.
In the range of gap ratio $[1.15,1.8]$,
several distinct regimes of multistability were observed (\S~\ref{sover}):
we go from a single mode at $G/H=1.15$ (where the histogram is wide),
to tristability, with three modes (whose position depends little on the dataset) at $G/H=1.25$,
to bistability (with two modes) at $G/H=1.5$.
We can only compute a single mode and a single most probable value for
each group of runs obtained at $G/H=1.8$ 
(see the histograms in figure~\ref{vis1-8} (b)).  
Both these values are nevertheless placed on the bifurcation diagram because
they quantify the degree of symmetry breaking at that gap ratio.
We are encouraged to do so by the fact that for both groups of runs, the models have two symmetric zeroes, as visible in the diagram. 
In terms of the amplitude of the modes,
we observe a monotonic decrease of the absolute value of the mode amplitude with $G/H$ in the range $[1.15, 1.8]$
(with the presence of the maximum at $0$ at $G/H=1.25$).

In the range of gap ratios $[2.2,2.65]$, we observe bistability for all gap ratios, with two symmetric modes.
In term of amplitude, the modes display a moderate increase then decrease in absolute value,
while the conditional average is only slightly decreasing
and the stable zeros reproduce this decrease in amplitude.
Bistability then disappears: the amplitude of the mode goes from approximately
$H/2 $ to zero as $G/H$ is increased by $0.05$ to $G/H= 2.7$.
This is the largest change of mode amplitude observed, especially for such a small change in gap ratio.
Note that the ghost of the fixed points at this gap ratio can lead the drift to cross the line $f=0$ (Fig.~\ref{modGH2} (a))
and can lead to the detection of a pair of stable and unstable zeros at $Y_m\lesssim 2H/3$m.
For larger gap ratios $G/H=3.0$ and $3.5$, the only mode and significant stable fixed points are near $0$.
This could mean that the bifurcation from bistability to monostability is discontinuous.

Note that at $G/H=1.25$, $2.2$, $2.4$, $2.6$, and $3.5$ the position of the modes and conditional average depends little on the group of run.
This is shown by the moderate size of the corresponding error bars.
This is not only the case within a dataset,
but also from a dataset to the other at $G/H=1.25$, $3.5$ (first and third datasets) and $2.4$ (all three datasets).
For those diagnostics of multistability, there is no sensitivity on the dataset used. 

Given the consistency between the position of PDF maxima and the conditional average of $Y_m$,
we further use the conditional averaging detailed in section~\ref{stated} and examine the standard deviation of $Y_m$
conditioned on being on the left or right state (Fig.~\ref{bifdiag} (b)).  
As $G/H$ is increased in the range $[1.15,1.8]$, there is no monotonous trend of $\sigma_Y$,
as we observe several distinct regimes of multistability for each value of $G/H$.
We start with a large standard deviation at $G/H=1.15$,
when the jet has not selected a precise multistable position, and the histogram is very flat.
The standard deviation is then smaller  at $G/H=1.25$,
for which the flow is tristable. We note that the standard deviation is more sensitive to the left or right position and the group of runs
than the modes, averages or zeros were.
The standard deviation remains moderate at $G/H=1.5$, where the flow has become bistable and increases
at $G/H=1.8$, where the jet settles either on the left or on the right.
This increase is caused by a smaller relative stability of the symmetry breaking position and
regular excursions of the jet toward the centreline.
As $G/H$ is increased to $2.2$, the standard deviation is decreased,
because the changes of direction of the jet are now much scarcer than at lower gap ratios.
In the range of gap ratio $[2.2,2.65]$, we now observe an increase of $\sigma_Y$ when the jet is bistable,
from $G/H=2.2$ to $2.65$, for which it is maximum.
This increase of the standard deviation accounts for the fact that transitions from left to right
are becoming increasingly common as $G/H$ is increased.
The standard deviation of the position then decreases from $G/H=2.7$ to $3.5$,
where the jet has become monostable. The large value at $G/H=2.7$
accounts for the frequent changes of state from bistable positions to a symmetric jet. 
This highlights the maximum of standard deviation right at the bifurcation.
The standard deviation is more sensitive than the modes and averages to experimental imperfections which are different from
one realisation of the group of runs measurement to another.
At $G/H=2.2$, $2.6$ and $3.5$, the conditional standard deviation depends
less on the group of runs than it was the case at $G/H=1.25$.
At $G/H=2.4$, seven of the twelve computed $\sigma_Y$ are within 5\% of $0.0038$. Four of the five outliers
originate from the groups of runs v1 and v3, where the bars are less precisely set. 

We have computed the time average of the jet width as a function of $G/H$ (not shown here).
It is an affine function of $G/H$. We find that $\langle w\rangle \simeq 0.86 (G-H)$,
the jet is slightly thinner than the gap.
The disappearance of multistability does not leave a strong mark in
the dependence on $G/H$ of the width of the jet at the base.
The values computed in distinct experimental campaigns are close to one another.
The standard deviations of the jet width is presented in figure~\ref{bifdiag} (c) as a function of $G/H$.
This standard deviation increases slowly when the flow is multistable from almost zero at $G/H=1.15$ to a value
of approximately $0.17H$ for $G/H= 2.65$.
Right at the bifurcation, at $G/H=2.7$, the standard deviation increases to more than $H/2$.
This brutal change is the effect of the bifurcation to monostability through a very intermittent scenario.
The standard deviation of the jet width then has no notable increase for $G/H\ge 3.0$.
Time series of the width indicate that this corresponds to moderate fluctuations which scale with the increasing jet width.
Finally, we remark that for this observable, there is little dependence on the group of runs and on
the dataset,
the largest difference being found at $G/H=3.5$ between the first and the third dataset.

\subsubsection{Detection of regime changes and experimental imperfections using model coefficients
}\label{sbifdiag}

In this section, we complement the bifurcation diagrams using information coming from the leading odd model parameters of
the fitted drift and even model parameters of the fitted diffusion, in line with our rationale (\S~\ref{ssratio}).

\begin{figure}
\centerline{
\includegraphics[width=5.5cm]{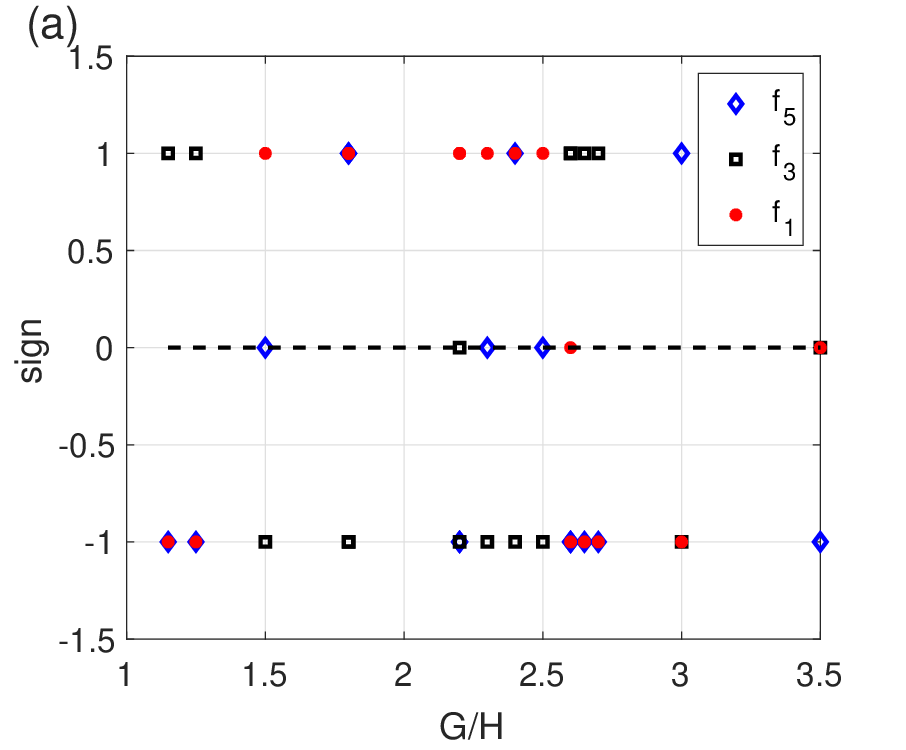}
\includegraphics[width=5.5cm]{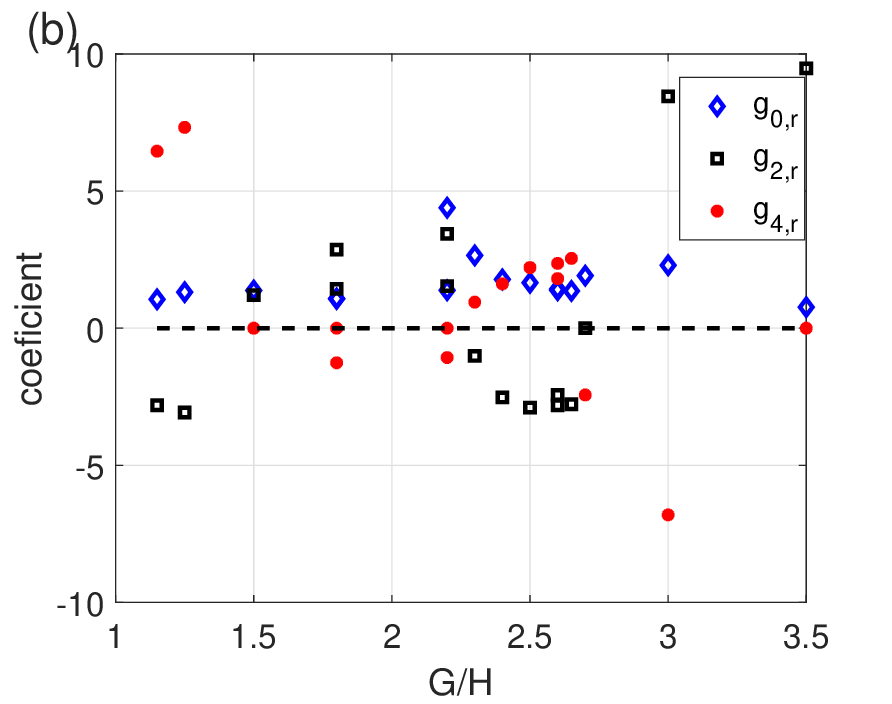}
\includegraphics[width=5.5cm]{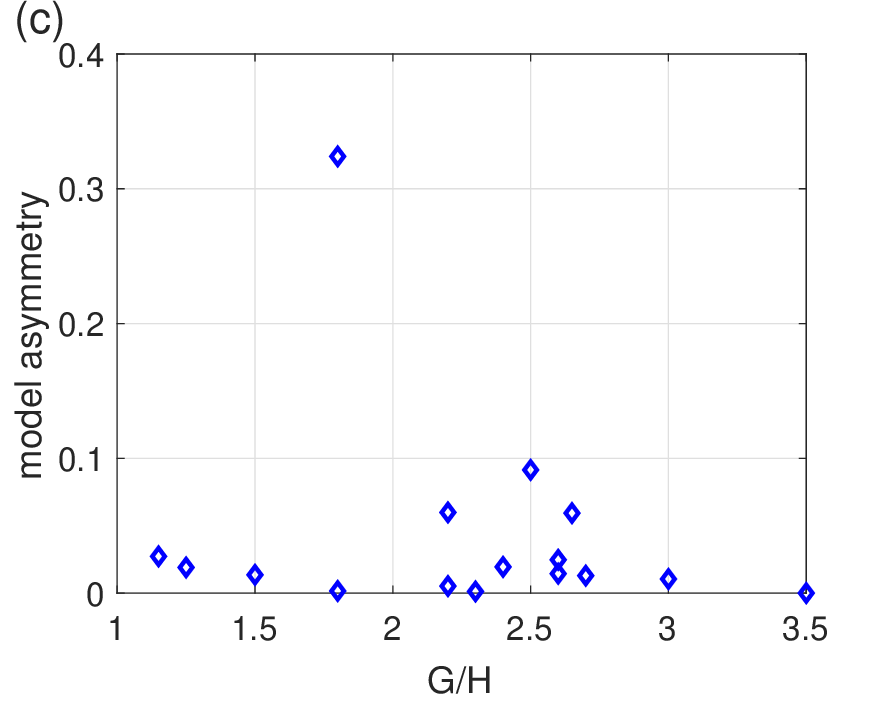}
}
\caption{
Properties of model parameters computed using the third dataset as a function of the gap ratio $G/H$.
(a) Sign of even model parameters $f_1$, $f_3$, $f_5$ of the trained drift (see \S~\ref{ssratio}, Eq.~(\ref{oddODE})).
(b) amplitude of the odd model parameters of the rescaled diffusion (Eq.~(\ref{leq}),~(\ref{ediffg}),~(\ref{resfg})).
(c)  Measured relative asymmetry of the model $I_{\rm asym}$ (Eq.~(\ref{dia})) as a function of the gap ratio.
}
\label{bifdiagmod}
\end{figure}

We display the sign of the three leading odd model parameters $f_1$, $f_3$ and $f_5$ of the fitted drift (Fig.~\ref{bifdiagmod} (a)).
Using this, we can follow the sequence of regime changes as the gap ratio is increased.
As we had argued in section~\ref{ssratio},
the odd terms of the drift are dominant given the (broken) left-right symmetry.
Given the observation that the multistable positions correspond to the stable zeros of the drift (Fig.~\ref{bifdiag} (a)),
we will further check that we have a correspondence between the sign of the three odd model parameters
and the type of multistability (\S~\ref{ssratio}).
We first discuss the structure of selected models in the range $G/H\in[1.15,1.8]$,
where we have a succession of several, sometimes complex cases.
At $G/H=1.15$ and $G/H=1.25$, the model both have $f_1<0$, $f_3>0$ and $f_5<0$,
the difference in amplitude in model parameters distinguishes the tristable case of $G/H=1.25$ and the flat case of
$G/H=1.15$. For $G/H\ge 1.5$, the fitted model drifts then have $f_1>0$ up to $G/H=2.6$.
All the selected drifts have a bistable structure, where we find $f_3<0$ for the cubic model parameter,
except for one model at $G/H=2.2$, where this role is assumed by the quintic term with $f_5<0$ and $f_3=0$ (leading to a stiffer shape for the model drift).
In the range $G/H\in [2.3, 2.6]$, the quintic model parameter $f_5$ is mostly positive and tunes the shape of the selected model.
The transition from monostability to bistability is naturally detected in the model parameters, where
we go from $f_1=0$ at $G/H=2.6$ to $f_1\le 0$ for $G/H> 2.6$.
It is then the combination of $f_3$ and $f_5$ (as well as even terms for $G/H=2.65$)
that distinguishes the limiting bifurcating cases for $G/H=2.65$ and $2.7$ from the monostable cases
of $G/H\ge 3.0$.
At $G/H=3.0$ with $f_3<0$ and $f_5>0$, the model drift at this gap ratio accounts for a single stable fixed point at $y_r=0$ and slower decrease
of the first empirical Kramers--Moyal coefficient for $y_r>0.6$.
Then, at $G/H=3.5$, some fitted drifts are led by $f_5$ (see the effect in Fig.~\ref{modGH} (a)):
this can express the spiked PDF at $y=0$.
This is another indication that the drift contains almost all information on the relative multistability.

We then examine the magnitude of the three leading even model parameters of the diffusion (Fig.~\ref{bifdiagmod} (b)).
We firstly note that the three model parameters do not vary much in absolute value, indicating that the
relative noise felt by the jet depends little on the gap ratio. In particular, the constant model parameter $g_{0,r}$ remains between $1$ and $2$.
As observed in the selected test cases, the signs of the model parameters $g_{2,r}$ and $g_{4,r}$ are governed by the state of multistability:
the $y_r$ variations of the diffusion are secondary and generally slaved to those of the drift.
We typically have $g_{2,r}<0$ and $g_{4,r}>0$ when the jet is bistable ($G/H=1.5$ and $G/H\in [2.3, 2.65]$), reflecting smaller values of
the diffusion near the bistable points.
Note that at $G/H=1.5$ and in groups of runs at $G/H=2.2$, where the symmetry is broken, we sometimes find a diffusion with a single minimum at $y_r\simeq 0$, where
the bistable positions are not visible, with not effect on bistability.
We find the opposite for most other gap ratios: the relative amplitude of $g_{2,r}$ and $g_{4,r}$
then distinguishes the monostable cases $G/H\ge 3.0$ from the others.

In these bifurcation diagrams, we noted that there was little difference in the models
fitted from different groups of runs at $G/H=2.2$, while the
models fitted at $G/H=2.6$ are more sensitive in terms of model parameter sign because of the closeness to the bifurcation.

We will present in an article currently in preparation how the near constant amplitude of the model parameters of the diffusion and the
varying amplitude of the model parameters of the drift results in the variations of the transition probability
in bistable states with $G/H$.
This further indicates that the shape and variations of the diffusion has less effect on the behaviour of the model than that
of the drift.

We finally observe the relative weight of the asymmetry of the model.
We measure this asymmetry with $I_{\rm asym}$ (Eq.~(\ref{dia})), the relative weight of the even coefficients in the square norm of the drift,
as introduced in appendix~\ref{app_sdei}.\ref{mod_det}.\ref{mesasym}.
We present this measure in figure~\ref{bifdiagmod} (c).
We note that the trained models can be divided in three category: those which are almost totally symmetric ($I_{\rm asym}\lesssim 0.03$),
those with moderate asymmetry $0.05\lesssim I_{\rm asym}\lesssim 0.1$ and one case with strong asymmetry $I_{\rm asym}\simeq 0.33$.
The strongest asymmetry is found at $G/H=1.8$ and is entirely in line with our observations:
we only find a single maximum in the histogram at that gap ratio and the drift trained for the v1 dataset has all its zeros shifted towards negative $y$.
This is the only trained drift which is not close to odd.
This further leads us to think that at that gap ratio, 
the flow is particularly sensitive to the experimental imperfections and that the monomodal histograms are not a consequence of finite time sampling.
When the measured asymmetry is moderate, the even coefficients of the drift do 
not account for a global shift of this function, but rather for differences in slope near the zeros of the drift 
as well as of the absolute value of the local extrema.
A notable case is found at $G/H=2.2$. The amplitude of the modes of the histogram are different by a factor of ten, owing to a shorter time
spent on one side rather than the other.
However, this difference is not starkly visible in the Kramers--Moyal coefficients and in the drift and finally, in our measure of asymmetry.
This indicates that this asymmetry of the histogram is a more likely caused by a sampling bias than an asymmetry of the bar configuration.

\section{ Conclusion: Summary, discussion and perspectives}\label{sconc}

\subsection{Summary}

In this article, we have presented an experimental study and a modelling of
the multistable jet in the wake between two parallel square bars.
Velocity fields in the horizontal plane normal to the bars
were measured by means of PIV right behind the bars at Reynolds number $Re=(HU_\infty\rho)/(\mu)=9.8\cdot 10^3\pm 4\cdot 10^2$
for a wide range of gap ratios $G/H\in[1.15,3.5]$ in three distinct experimental campaigns.
From these velocity fields, we have
estimated two scalars: the spanwise position of the jet $Y_m$, which is closely related to its angle with respect to
the centreline, and the width of the jet.
From the time series of $Y_m$, we have determined the time intervals in which the jet was in one multistable
position or another. Using this mask, we have computed conditional cumulants of $Y_m$.
The diagnostics were used to characterise the different regimes of multistability.
Furthermore they were used to follow the
changes of regimes occurring when $G/H$ is varied.
In particular, we have used these multiple diagnostics to shed light on the specific way multistability
disappears as $G/H$ is increased above $2.7$ at $Re=10^4$.
A second key point of this work is the automated construction of analytic stochastic models
from the time series of $Y_m$.
The generated Langevin equations, and in particular the drift functions, help us further pinpoint the multistable positions,
their relative stability, the level of noise felt
and finally, the peculiar change of regime from bistability to monostability as $G/H$ is increased.

\subsection{Discussion: Studying the changes of regime at $G/H=2.7$ using our diagnostic and modelling}

The key point of this article was the study of the changes of multistability regime, in particular the complex one at $G/H=2.7$.
This study was based on our diagnostics ($Y_m$ and $w$) and strongly aided by the investigation of the trained models.
We now compare our findings with other studies of the wake of the double bar at different control parameters, with other
studies of bistable flows, and with other studies of model properties at complex regime changes.

Throughout this article, we monitored quantities computed from $Y_m$, the
weighted position of the jet.
This included  its conditional average, the position of the modes of its PDF and more importantly, the features of the models 
(such as its stable fixed points or areas where the drift was close to zero).
These quantities provided a precise order parameter for the bifurcations of the jet as $G/H$ was varied, by being non zero
as soon as the left-right symmetry of the flow was broken and by having an amplitude
varying monotonously with the amplitude of the symmetry breaking.
Using these quantities, we were able to construct phase diagrams indicating the two to three multistable positions
and pinpointing the successive changes of regime (Fig.~\ref{bifdiag}).
In this diagram, the change of multistability regimes or the complex cases were easier to discuss because we could access the information brought by the trained models.
In particular, we could  envision the brutal disappearance of bistability,
with a mechanism similar to what had been observed at higher Reynolds numbers
by \cite{alam2011wake}. 
They reported the regime change at $Re=1.7\cdot 10^4$  and $G/H=2.4$ as well as at $Re=4.7\cdot 10^4$ in the range $2.4\le G/H\le 2.9$.
In that range of gap ratio and at those Reynolds numbers the presence of both the bistable characteristics of low gap ratio and
the monostable characteristic of high gap ratio was measured in Strouhal numbers and force coefficients.
However, \cite{alam2011wake} present no PDFs that could help distinguish between intermittency and tristability in that range of
gap ratios.
In our measurements, at $Re=10^4$, intermittency between a thicker jet at the centre and a thinner bistable jet
is measured as the gap ratio is increased to $G/H=2.7$ (while the jet is only bistable for $G/H\le 2.65$).
This does not correspond to tristability between a jet symmetric with respect to the centreline (like the one seen at $G/H=3.0$ and $3.5$) 
and a jet breaking the left right symmetry (like the one seen in the range $G/H\in[2.3,2.65]$). 
Such a configuration would not only leave its clear trace in PDFs in the form of three modes, but would also be directly readable from the trained drift.
At $G/H=3.0$ only monostability is observed in our measurements, indicating that
the range of gap ratios where intermittency is found ends before this gap ratio in our measurements.
We finally note that there is a 10\% difference in the gap ratio where the bifurcation is observed
at these different Reynolds numbers ($Re=10^4$ on the one hand and $Re\ge 1.7\cdot 10^4$ on the other hand), 
in different experimental conditions.
The regime change gap ratio can be independent of the Reynolds number for $500\lesssim Re\lesssim 20000$
for the case of circular cylinders \citep{xu2003reynolds}, in fixed experimental conditions.
The Reynolds number dependence has been less thoroughly investigated
for square bars, so that the difference reported here is not necessarily a discrepancy.
Changes of regime are sensitive: in our case, 
it could be impacted by the interaction with the boundary layers at the top and bottom of the wind tunnel.

The standard deviation of $Y_m$ provided a companion measure to the order parameters to follow the transition at $G/H=2.7$.
Indeed, turbulent flows often display a maximum of fluctuations, sometimes called a fluctuation crisis \citep{cortet2010experimental}, at the values of control
parameters where the symmetry is broken and where multistability (dis)appears \citep{cortet2010experimental,rolland2018finite,norman2011unsteady}.
Our case of the wake of the parallel bars is peculiar because the fluctuation crisis happens
for a discontinuous bifurcation.
A similar fluctuation crisis also naturally occurs in the fitted Langevin equation at the bifurcation.
Moreover, we noted that the inclusion of the standard deviation of the jet width
greatly helped giving a synthetic view of the bifurcation: it appeared not
only through a disappearance of the left and right multistable position but
also through a spatial reorganisation the jet.
Since this standard deviation does not display a maximum, but a sudden jump instead,
this indicates that this reorganisation occurs through an increase of disorder. 

The description of the bifurcation at $G/H=2.7$ using the Kramers--Moyal coefficients and the fitted models
was a fundamental consequence of the modelling procedure.
Constructing this specific, enlightening model at that gap ratio was possible because of the freedom of functional form given by the procedure.
The constructed model displayed a stable position at $y=0$, for the jet along the centreline.
This model also retained zones of very low drift where the former bistable states were.
This represents the dynamics of the jet at this value of the control parameter,
which can spend some time fluctuating at the centre, but can also be thrown from
time to time towards the more steady bistable positions that existed at smaller values of $G/H$. 
This brings a connection to the transition to chaos \emph{via} type 1 temporal intermittency described by \cite{pomeau1980intermittent}.
That type of transition to chaos concerns few degrees of freedom dynamical systems.
As a control parameter is varied several things happen in phase space.
A limit cycle disappears through a global bifurcation.
In its place, we find a ghost of the former corresponding fixed point of the Poincar\'e map: 
a region of phase space where the dynamics are very slow but never steady.
Finally a chaotic repellor is generated in another region of phase space. 
One can then observe intermittency in time between chaotic evolution of the system
and the former periodic behaviour.
This intermittency is possible because of two features of the phase space.
Firstly, chaos is repelling and connections to the ghost of the limit cycle still exist.
Secondly, the zone where the limit cycle was is very slow
and the system can spend a long time there before being driven back to the chaotic zone.
In that type of transition to chaos, the ghost of the limit cycle gradually disappears, leading to a fully chaotic dynamics.
In our case the equivalent of the laminar, ordered, phase is the bistable regime, while that of the chaotic phase
is the early monostable regime
(which is indeed characterised by a higher variance in shape of flow patterns and width of the jet).
The system is not thrown out of a chaotic regime by deterministic fluctuations,
but out of a mode by turbulent fluctuations that we model by the noise.
Meanwhile, we still have slow zones in the model, where the multistable points were at lower gap ratio.
Again, the noise of the model is the effect that can kick the system back to the central position.

We finally note that we have constructed a distinct model for each values of $G/H$. 
As a consequence,
we have not proposed an explicit parametric dependence of the selected terms and their values.
For some systems, a model with a built in parametric dependence was proposed \citep{rolland2011pattern}.
The fitted parameters are then relevant over a range of control parameters which includes the bistability regime change.
To some extent, proposing a control parameter dependence in our model could be feasible.
In the range $[2.2,2.65]$, the model parameters grow in absolute value as $G/H$ is decreased from the threshold $G/H=2.7$.
This growth occurs  in a concomitant manner that keeps the metastable positions, the zeros
of the drift, at the same position (another difference with a continuous bifurcation scenario).
Moreover, both in our case and in earlier studies, a proposed
parametric dependence would not be able to account for the non trivial bifurcation from bistability to monostability.
In our case, the rapid change of sign of model parameters at $G/H=2.7$ is
harder to relate to the smoother variations that are observed in the range $[2.2,2.65]$
with a simple functional form.
Thus a parametric dependence that cannot represent the multistability regime change is less interesting.
Again, while we may have lost in not having a parametric dependence in the modelling procedure, we gained in ability to describe
the complex regime change.

\subsection{Perspectives}

Some questions important for multistability have not been directly addressed in this article.
There is firstly the question of the physical mechanisms that can maintain the  multistable positions.
This matter has already been discussed in the literature,
and the possibility of a Coanda effect has been proposed early on \citep{ishigai1972experimental}.
This is a tempting interpretation given that a difference in base pressure is systematically observed on the cylinders \citep{kim1988investigation}.
Another important question is that of the series of events
that occur when the jet changes direction.
It has been noted in simulations that the vortex shedding is modified when the jet changes direction \citep{afgan2011large}.
This is possible because of the time scale separation between the two phenomena.
The change of direction occurs over  a characteristic time scale of $\mathcal{O}(0.5)$s in our experiments.
Meanwhile, in our experiments, the vortex shedding period is of order $\mathcal{O}(0.025)$s (in line with the value of the Strouhal number \S~\ref{app_sdei}.\ref{mod_det}.\ref{stmark}).
Each successive cycle can take place in a different configuration of the wake.
However, the question of whether we have a one way coupling from the change of direction to the vortex shedding,
or whether the turbulent vortex shedding can trigger the change of direction we observe in experiments is open.
The turbulent wake could be a trigger for direction change, since they are observed in LES where the wake is the only source of noise.
In experiments there are other sources of noise such as the incoming rate of turbulence 
(which can play a major role in the rare fluctuations of blunt body wakes \citep{lestang2020numerical}) or boundary layers at the top and bottom of the wind tunnel.
As a consequence, numerical simulation is an interesting complementary tool
to address the two questions mentioned above.
It has been shown that LES can represent the bistable flow, and it is possible
to model more complex boundary conditions to test their effect.
Moreover, they can help us investigate effects of the vertical extend of the flow: 
three dimension, three components views are difficult to access in large enough domains in experiments.

Another question we will address in a future article is the sensitivity of the configuration
to experimental imperfections.
Indeed, compared to the flow in the wake of two parallel circular cylinders,
our case of the wake of two square based bars can lead to departure to the ideal case,
even if one bar is turned by a small angle.
This can easily happen from one realisation of the experiment to the other as the bars
are reassembled.
Using scalar diagnostics, we have seen that at most gap ratios (and particularly at $G/H=2.4$),
the properties related to the average (PDF modes of the position, conditionally averaged position, average and standard
deviation of the jet width) were only slightly perturbed by slightly different
experimental conditions. Quantities measuring the fluctuations, such as the conditional variance are quantitatively more sensitive,
even if they do not display a qualitatively different behaviour.  
At some gap ratios (like $G/H=1.8$) where the amplitude of the symmetry breaking
is smaller, experimental imperfections can lead to one position or the other
losing its metastability, while the left right symmetry is still broken.
This can raise the question of whether the symmetry breaking is actually partial.
We will delve in more details in that question using the flow patterns in an article currently in preparation.

This work opens several wider perspectives for the study of multistability in fluid flows,
such as wakes and atmospheric flows.
Indeed, the simple geometry of the parallel bars flow means that it can be assembled in an experimental facility
or simulated numerically with simple means and moderate sensitivity to finer tuning parameters.
It can be used to test novel experimental,
numerical or modelling approaches of study of multistability in turbulence.
In turn the methodology already tested in this article could be applied to geophysical records and relax hypotheses on
the form of the model used to study tippings, as was done by \cite{ditlevsen2023warning}, who prescribed the functional form of their model.
Indeed, in our study, the margin given to the modelling procedure was fundamental to elucidate the behaviour of the flow 
at the change of regime: such a refinement could also be done when studying tipping scenarios in climate science.
It could also be used to propose stochastic models for the drag crisis.
Indeed, from the point of view of bifurcations and multistability, this
change of regime is subcritical, however the PDFs of the drag coefficient
and not simply modelled \citep{norman2011unsteady}, while the drag coefficient signals
display a bistability between a quiet state and a strongly fluctuating one \citep{deshpande2017intermittency}.
This is reminiscent of the intermittent behaviour presented in this article.

\section*{Acknowledgments}

The equipment used in this work is funded under the project \emph{Recherche et Innovation en Transport et Mobilit\'e Eco-responsable et Autonomes (RITMEA)},
CPER 2021--2027 (previously CISIT and Elsat2020),
co-funded by European Regional Development Fund (FEDER), the French state and Hauts-de-France Regional Council.
The authors thank Jiangang Chen for providing his experimental dataset and several figures.
Indra Kanshana participated in the data gathering for the second and third experimental dataset under the supervision of P. Bragan\c ca and C. Cuvier.
The authors thank for Giorgio Krstulovic for stimulating remarks on multistability,
Jean-Christophe Loiseau for discussions concerning the Langevin regression,
Jacques Bor\'ee, \'Edouard Boujo and Luc Pastur for discussions on wake flows,
and Yves Pommeau for pointing the case of the drag crisis.

For the purpose of Open Access, a CC-BY public copyright licence has been applied by the authors to the present
document and will be applied to all subsequent versions up to the Author Accepted Manuscript arising from this submission.


\appendix

\section{Details on Experimental campaign}\label{app_exp}

In this appendix, we give the technical details on the PIV set up (\S~\ref{app_exp}.\ref{compiv}), practical choices on sampling and experimental conditions 
for the three datasets (\S~\ref{app_exp}.\ref{app1std},~\ref{app_exp}.\ref{100hz},~\ref{app_exp}.\ref{sup_param}).

\subsection{Common features for PIV measurements}\label{compiv}

The value of velocity $U_\infty=5\text{m}\cdot{s}^{-1}$ of the flow arriving on the bars is set 
at the beginning of each experimental campaign by setting the value of the velocity 
at the exit of the convergent at $U_{\rm exit}=4.8\text{m}\cdot\text{s}^{-1}$.
The velocity $U_{\rm exit}$ is maintained within $\pm 0.5\%$ of the prescribed value throughout all experiments of a campaign,
using a control loop that commands the rotation velocity of the wind tunnel fan.
It has been checked with a Pitot tube measurement at the downstream position of the bars that setting this smaller
value of $U_{\rm exit}$ indeed led to the desired value of $U_{\infty}$.
We have $U_{\infty}>U_{\rm exit}$ because of the development of the boundary layers on the walls of the tunnel, at constant flow rate.

The bars have a length of $0.909$m and are set on two rails of thickness $4.5\cdot 10^{-2}$m clamped at the top and bottom of the
tunnel approximately $5.6$m downstream of the entrance of the wind tunnel test section.  
The horizontal distance between the bar corners is measured with a precision up to $10^{-5}$m
using a digital caliper.

For the measurements by means of PIV, the flow is seeded with drops of a mix of water and poly-ethylen glycol 
of average diameter of $10^{-6}$m (typical diameters are found
between $10^{-7}$m and $2\cdot 10^{-6}$m).
The concentration ensures that the target of 0.04 particle per pixel are found in the images,
which ensure the precision of displacement measurements \citep{foucaut2004characterization}.
All images are acquired on sCMOS censors digital camera, diaphragmed at a $f\sharp$ of $8$,
which leads to particle images of $1.7$ - $1.8$ pixels.
The velocity vectors are computed on a grid with step sizes ${\rm d}x$ and ${\rm d}y$.
In all measurement campaigns, multi-pass is performed to minimise the smallest resolved scale \cite{soria1996investigation},
given by the smallest interrogation window of size $\delta x \times \delta x$ ($24\times 24$ pixels).
The corresponding smallest resolved length $\delta x$ is below $2$mm for each PIV set up. This corresponds to approximately $2.4 {\rm d}x$ and
an overlap of $58\%$.

\subsection{Technical details of the first dataset}\label{app1std}

The field of view measured by means of PIV by \citep{chen2021turbulence} has a length of $\Delta x=7.41\cdot 10^{-1}$m and a width of $\Delta y=1.58\cdot 10^{-1}$m
and it starts at a distance to the bars of $x_c-(\Delta x/2)=1.6\cdot 10^{-2}$m
downstream of the bars.
These velocity fields were measured using a setup sketched in figure~\ref{skint} (d),
using a system of four sCMOS cameras to obtain the large field of view, lighted by a BMI laser.
The laser was placed below the wind tunnel, upstream of the bars and the FoV is lighted thanks to the transparent bottom wall of the tunnel.
The velocity is recorded on a grid of spatial step ${\rm d}x={\rm d}y=7.61\cdot 10^{-4}$m.
These fields have a spatial resolution controlled by the smallest interrogation window of $\delta x\simeq1.6\cdot10^{-3}$m.
Velocity fields are measured at a frequency of $4$Hz.
In that campaign, groups of runs are $5000$s long and comprise of 10 runs of $500$s each sampled $10$s after the other.
This thus represents measurement duration of approximately  $5090$s.
Data was sampled for the three gap ratios $G/H=1.25$, $G/H=2.4$ and $G/H=3.5$.
In all three datasets, the measured atmospheric pressured was in the range $[1005,1009]$hPa, at a temperature of $288$K
(leading to a dynamic viscosity of $\mu=1.82\cdot 10^{-5}\text{kg}\cdot\text{m}^{-1}\cdot\text{s}^{-1}$).
The Reynolds number was noted to be equal to $1.0\cdot10^4$ for all measurements at this velocity \citep{chen2021turbulence}.

\subsection{Technical details of the second dataset}\label{100hz}

In this campaign, the second dataset was measured in a field of view of length of $\Delta x=9.69\cdot 10^{-2}$m and width $\Delta y=1.02\cdot10^{-1}$m.
The FoV starts  $x_c-(\Delta x/2)=2.8\cdot 10^{-3}$m  downstream from the bars (so that $x_c=5.2\cdot 10^{-2}$m).
The velocity fields are given on a grid of spatial step ${\rm d}x={\rm d}y=6.29\cdot 10^{-4}$m, and have a spatial resolution of $\delta x\simeq 1.5\cdot 10^{-3}$m.
The data was sampled at $100$Hz, meaning that a change of direction of the jet spans 25 to 50 frames.
Each run has a duration of $15.5$s and was sampled at the single gap ratio $G/H=2.4$.
The measured atmospheric Pressure was in the range $[1020,1026]$hPa at a temperature of $293$K.
This leads to a dynamic viscosity of  $\mu=1.85\cdot 10^{-5}\text{kg}\cdot\text{m}^{-1}\cdot\text{s}^{-1}$ and a Reynolds number in the range $[9.8\cdot 10^3,9.9\cdot 10^3]$
(similar to the one probed by \cite{chen2021turbulence}).
These datasets were measured using a setup sketched in figure~\ref{skint} (e) with a single camera.
Note that for this campaign and the next, the field of view is now lighted by
the laser from the side.
A Quantronix Darwin duo laser was used with a Miro M340 camera.
As to the optical system mounted on the camera, the magnification $M$ was 0.16 and the aperture used was $f\sharp= 8$, 
leading to a particle size of 1.2 pixels on images.
The same PIV processing method was used here as in \citep{chen2021turbulence} (multi-pass, multi-grid with image
deformation and final interrogation window of 24 by 24 pixels, corresponding of $1.51 \text{mm}\times 1.51 \text{mm}$).
The time separation between two consecutive images for velocity vector computation was tuned to obtain a typical displacement of particles of 10 pixels
inside the jet.

For this high frequency dataset, runs were selected in three different manners:
\begin{itemize}
\item 101 runs were recorded and processed indiscriminately (with a straight bars configuration, see Fig.~\ref{skint} (b) of the article).
Using these control parameters, the mean first passage time before the jet changes direction
is several times longer than the run duration.
As a consequence only a small fraction of these runs displays a change of direction.
\item In order to minimise the number of processed images and processing time per recorded
change of direction of the jet,
24 runs were recorded and processed under the condition that they displayed a change of direction of the jet
(with a straight bars configuration, see Fig.~\ref{skint} (b) of the article).
In practice, images for $15.5$s long runs were recorded. The velocity fields were computed from two pairs of images
at the beginning and the end of the run. The images were systematically processed for the whole run if the direction
of the jet was different in the two velocity fields. Otherwise the images were discarded. This led to a reduction of
the computational expense for PIV analysis while retaining the same number of examples showing changes of direction of the jet.
\item Finally, in order to test the sensitivity of the flow to the bar assembly, five groups of 10 runs, each run $15.5$s long, were measured,
one with straight bars, and four with one bar turned around its axis (as exemplified in the sketch of Fig.~\ref{skint} (c) of the article).
The rotation of the bars can be quantitatively characterised by a single parameter: the angle $\alpha$ of rotation.
This angle was
deduced from the measured distance of the bar corners to the opposite bar side $D_1$ and $D_2$ as $|\sin(\alpha)|=\frac{|D_2-D_1|}{H}$.
The sign depends on which bar is turned.
The four cases consist of two cases where only the left bar is turned and two cases where only the right bar is turned.
The left bar is turned by a positive angle of $8.7\cdot10^{-3}$rad 
(of $0.50^{\rm o}$, with $D_1=4.240\cdot 10^{-2}\text{m}<D_2=4.266\cdot 10^{-2}\text{m}$),
and then by a negative angle of $-1.1\cdot10^{-2}$rad (of $0.65^{\rm o}$, with  $D_1=4.286\cdot 10^{-2}\text{m}>D_2=4.252\cdot 10^{-2}\text{m}$).
The right bar is turned by a negative angle of $-1.9\cdot10^{-2}$rad (of $-1.1^{\rm o}$ 
with $D_1=4.218\cdot 10^{-2}\text{m}<D_2=4.276\cdot 10^{-2}\text{m}$), and then by a positive angle of $+7\cdot 10^{-3}$rad
(of $+0.42^{\rm o}$, with $D_1=4.274\cdot 10^{-2}\text{m}>D_2=4.252\cdot 10^{-2}\text{m}$).
When neither bar are turned, we have $D_1=D_2=G-2\frac{H}{2}=H\left(\frac{G}{H} -1\right)=4.2\cdot 10^{-2}\text{m}$.
\end{itemize}

\subsection{Technical details of the third dataset}\label{sup_param}

In this campaign, the third dataset was measured in 
a field of view of length of $\Delta x=9.51\cdot 10^{-2}$m and of width $\Delta y=1.13\cdot10^{-1}$m. The FoV starts $x_c-(\Delta x/2)=3.5\cdot 10^{-3}$m downstream from the bars.
The velocity fields are given on a grid of step ${\rm d}x={\rm d}y=4.5\cdot 10^{-4}$m and have
a spatial resolution of $\delta x\simeq1.2\cdot10^{-3}$m, controlled by the size of the smallest interrogation window. The data was sampled at $5$Hz,
with the same arrangement of camera and laser as in the second dataset (Fig.~\ref{skint} (e)).
A sCMOS 
camera from Lavision (using a magnification of $M=0.22$ and an aperture of $f\sharp=8$) 
and an Innolas spitlight PIV compact 400 laser were used.
The images were processed with the same PIV algorithm as in the first two datasets.
Velocity fields were measured in groups of $N$ runs, each run 2000s long, separated by 10s.
In this dataset, the measured atmospheric pressure was in the interval $[990.4,1024.8]$hPa, and temperature was set to $293$K
(leading to a dynamic viscosity of $\mu=1.85\cdot 10^{-5}\text{kg}\cdot\text{m}^{-1}\cdot\text{s}^{-1}$).
For all measurements, the Reynolds number is a few percent below $10^4$, in the range $[9.5\cdot 10^3,9.9\cdot 10^3]$,
as it was the case for the other two datasets.
The values of $\frac{G}{H}$ for which data was sampled and the corresponding properties of the datasets (Reynolds numbers and number of runs per group)
are given in table~\ref{tabparam}.

\begin{table}
\begin{center}
\begin{tabular}{|c|c|c|c|c|c|c|c|c|c|c|c|c|c|}
\hline $\frac{G}{H}$& 1.15 &1.25 &1.5 &1.8 &2.2 &2.3 &2.4&2.5&2.6&2.65&2.7&3.0&3.5 \\ \hline
$Re~(\times 10^3)$ &9.9&9.6&9.5&9.6&9.7&9.9&9.7&9.7&9.7&9.7&9.7&9.7&9.6\\ &&&&9.6&9.7&&9.9&&&&&& \\ &&&&&&&9.7&&&&&& \\ &&&&&&&9.7&&&&&& \\ \hline
$N$&10&10&3&10&2&3&2&2&1&2&2&3&1 \\ &&&&2&3&&3&&2&&&& \\ &&&&&&&1&&&&&& \\ &&&&&&&2&&&&&& \\ \hline
\end{tabular}
\end{center}
\caption{Table giving the gap ratios $\frac{G}{H}$ for which velocity measurements were performed for the third dataset,
along with the Reynolds number $Re$ of the flow and the number of consecutive runs $N$ per group.
Several groups of measurements were performed for $\frac{G}{H}=1.8$, $2.2$, $2.4$ and $2.6$,
for a given gap ratio, they are labelled as v1, v2 \emph{etc}.
The properties of said groups are placed on top of another in each line.}
\label{tabparam}
\end{table}

Some experimental imperfections that occurred during this experimental campaign should be reported.
Firstly it has been noted during this campaign that one of the ``left'' bar had a small bend leading to a shift of $3\cdot 10^{-4}$m
between one
end of the bar and the other (corresponding to a bending angle of $3\cdot 10^{-4}$rad or $0.02^{\rm o}$).
The left and right bars were swapped when recording the group of runs at $\frac{G}{H}=2.4$ v4, to check whether this had an effect on the flow.
Since the same equipment has been used during the three campaign and left unused outside of the campaign,
it is not clear whether this bend was present during the first two campaigns.
Note that other modifications of the set up were performed for velocity measurements at $G/H=2.4$.
The bars were shifted forward by 1mm for the group of runs $G/H=2.4$v3, while the mounting
of the bars was perturbed by small random shifts for the group of runs $G/=2.4$v1.
Secondly the time step between the pair of two consecutive images used for computation of the velocity vectors at $\frac{G}{H}=2.7$
we incorrectly set to $1.25\cdot 10^{-4}$s instead of $5\cdot 10^{-5}$s. This lead to typical estimated displacements of 25 pixels, 
instead of 10 pixels in all the other computations. This means that the velocity fields computed by
the PIV procedure had to be rescaled by a factor $\frac25$.
However, this does not lead to excessive errors on the computation of velocity fields thanks to the multi-pass PIV algorithm,
because the typical particle displacement from one image to another
was about 25 pixels inside the jet.

\section{Measurement of passage time and their distribution}\label{mmfpt}

\begin{figure}
\centerline{\includegraphics[width=5.5cm]{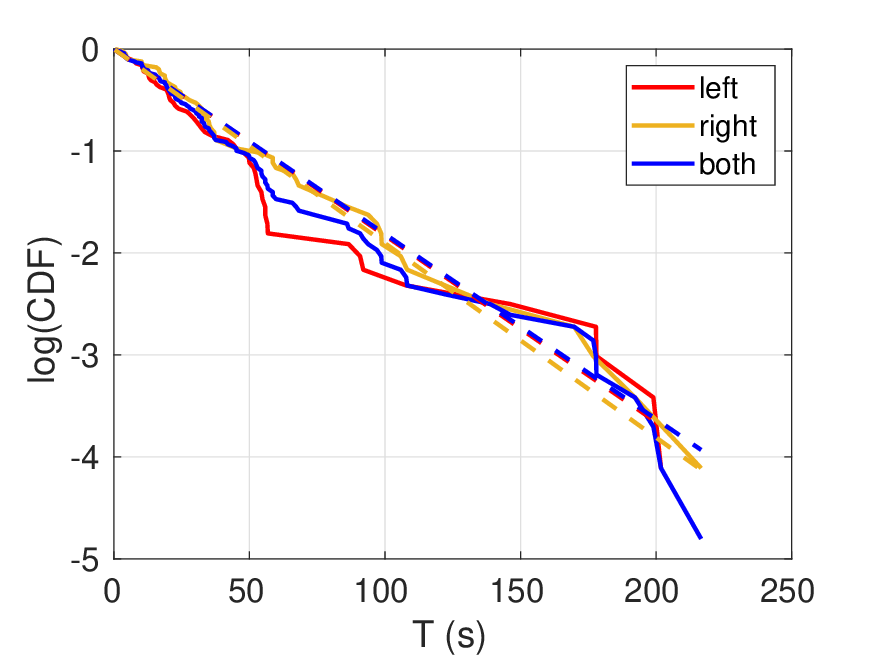}
}
\caption{
Cumulated density function of exit times out of states $Y_m>0$, $Y_m<0$ and both in logarithmic scale obtained from
time series of $Y_m$ computed from velocity fields sampled at $\frac{G}{H}=2.4$ during the third experimental campaign.
The linear regressions of these cumulated density functions are superimposed with dashed lines and the same colour code.}
\label{maskcdf_b}
\end{figure}

Using the state detection based on the time series of $Y_m$, we perform a measure of the times spent in either left or right state.
From the mask determined from the time series (Fig.~\ref{maskcdf_a}, \S~\ref{stated}), we
can determine a set of durations spent in both left and right states.
These durations are called \emph{residency times} or passage times.
The counter of these durations starts when the jet enters the left state and stops when it enters the right state (and vice versa).
For each value of $\frac{G}{H}$, we compute the arithmetic average of these durations.
It is also very enlightening to compute the cumulated distribution of these duration.
In the literature, two types of scaling laws are found for the PDF of residency times
and the corresponding  cumulated density function (CDF):
\begin{itemize}
\item an exponentials law \citep{kim1988investigation,rolland2011pattern,grandemange2013turbulent} where the PDF is
\begin{equation}
p(t)=\frac{1}{\tau}\exp\left(-\frac{t}{\tau}\right)\,,
\end{equation}
and the
 cumulated density function, the probability that the residency time is larger than $T$, is
 \begin{equation}
 \mathbb{P}(t>T)=\int_{s=T}^{\infty}p(s)\,{\rm d}s=\exp\left(-\frac{T}{\tau} \right)\,.
 \end{equation}
The PDF and the CDF are parameterised by $\tau$  the mean first passage time.
\item power laws \citep{pereira20191}: the PDF is then proportional to $t^{-\beta}$, with $\beta>0$.
These distributions often have a cut--off at large $t$. Without it, this PDF can only be normalised provided $\beta >1$ and cumulants only start to be defined for $\beta>2$.
\end{itemize}
In practice, this CDF is estimated in the following manner:
the CDF at time $T$  is the proportion of measured passage times
above $T$ out of a group of runs.
Displaying the CDF in logarithmic scale
indicates that it is an exponential (Fig.~\ref{maskcdf_b}).
The least squares linear fit of the logarithm of the CDF is a good match.
In the exponential case the mean first passage time is also the opposite of the inverse of the decay rate $-\frac{1}{\tau}$ (the slope
in logarithmic scale).
Processing the time series of $Y_m$ sampled at $\frac{G}{H}=2.4$v2 in the third dataset,
we find a mean first passage time of $44$s on the left side from the arithmetic average and $54$s from the
linear fit, while we find a mean first passage time of $52$s on the right side from the arithmetic average and $55$s from the linear fit.
We find a mean first passage time out of either state of $48$s from the arithmetic average and $52$s from the linear fit.
The arithmetic average is almost always within $10\%$ of the opposite of the inverse of the slope of the fit of the CDF.
The quality of the fit and the closeness of the two types of estimates of $T$
is also verified for all gap ratios for which we have sampled enough durations (not shown here).
Note that this distribution type is directly related to the shape of the spectra of $Y_m(t)$ (see \cite{pereira20191}
and the discussion in section~\ref{app_sdei}.\ref{mod_det}.\ref{stmark}).
In our case, the exponential distribution is directly related to the white range at low frequencies observed in the spectrum (Fig.~\ref{correlsp} (d)).
We will reuse this measure to discuss residency times and probabilities of transition in a future article.

\section{Procedure for the Stochastic Differential Equation identification}\label{app_sdei}

In this appendix, we complement the core of the article on stochastic processes (\S~\ref{app_sdei}.\ref{mod_det}) and
on the procedure used to construct the analytic Langevin equations (\S~\ref{app_sdei}.\ref{appmodcomp}).

\subsection{Bases of modelling by a Stochastic Differential equation}\label{mod_det}

In the procedure we follow, we need to compute PDF and finite time Kramers--Moyal coefficient from our models.
These computations are performed following partial differential equations for probabilities: the Fokker--Planck equation and its adjoint/backward version 
(\S~\ref{app_sdei}.\ref{mod_det}.\ref{app_FP}).
In that procedure, we make a Markov hypothesis. It is verified on data in subsection~\ref{app_sdei}.\ref{mod_det}.\ref{stmark}.
In the analysis of the computed models, we will extensively use the properties of normal forms and bifurcations,
which are reminded in subsection~\ref{app_sdei}.\ref{mod_det}.\ref{app_normform}.

\subsubsection{The Forward and Backward Fokker--Planck equation solved for PDF and Kramers--Moyal coefficients computation}\label{app_FP}

A time series of $Y(t)$ starting from an initial condition $Y_0$, following the Langevin equation~(\ref{leq}), 
represents a realisation of the corresponding stochastic process.
Each realisation is random and independent of the others.
The range of all time evolutions possible is represented by the probability density function $p(y,t|Y_0,0)$, the probability
of having $Y(t)=y$ at $t$ given the initial condition $Y_0$ at $t=0$.
The random variable described by the PDF is denoted by a lower case $y$. This $y$ is time independent.
This probability density function is the solution of the Fokker--Planck equation \citep{gardiner2009stochastic} \S~5.2
\begin{equation}
\frac{\partial p}{\partial t}=-\frac{\partial ~}{\partial y}(fp)+\frac{\partial^2~}{\partial y^2}(ap)\,,\label{fpeq}
\end{equation}
where we use $a(y)=g^2(y)/2$ the covariance of the noise, and where the initial condition of $p$ is written
with the Dirac delta function as $p(y,t=0|Y_0,0)=\delta (y-Y_0)$.
The steady PDF $\rho(y)$ in such a system is the solution of the steady version of Fokker--Planck equation (Eq.~(\ref{fpeq})),
as
\begin{equation}
0=-\frac{\partial ~}{\partial y}(f\rho)+\frac{\partial^2~}{\partial y^2}(a\rho)\,.\label{stfpeq}
\end{equation}
Alternatively, given the ergodicity of the dynamics, it is the limit at infinity of the transition probability
taken from any initial condition $\rho(y)=\lim_{t\rightarrow \infty} p(y,t|Y_0,0),\forall Y_0$.
If the system has one degree of freedom, then $\rho$ can be written analytically using $f$ and $a$ (or alternatively $g$).
Using an arbitrary value $y_0$, we have,
\begin{equation}
\rho(y)=\frac{2}{Zg^2(y)}\exp\left(\int_{y'=y_0}^{y}\frac{2f(y')}{g^2(y')}\,{\rm d}y' \right)\,,
\end{equation}
where $Z$ is a normalisation constant
\begin{equation}
Z=\int_{\tilde{y}=-\infty}^{\infty} \left(\frac{2}{g^2(\tilde{y})}
\exp\left(\int_{y'=y_0}^{\tilde{y}}\frac{2f(y')}{g^2(y')}\,{\rm d}y' \right)\right){\rm d}\tilde{y}\,.
\label{respdf}
\end{equation}
Note that the resulting PDF is independent of $y_0$.

In the core of the article we had stated that the finite time Kramers--Moyal coefficients could be obtained from well chosen solutions of the backward Fokker--Planck equation.
Similarly to \citep{callaham2021nonlinear}, we follow \citep{lade2009finite,honisch2011estimation} for this computation.
This is based on the fact that the finite time Kramers--Moyal coefficients (Eqs.~(\ref{atau}))   
can be obtained from the ensemble average $\left\langle (Y(\tau)-y_0)^n \right\rangle$ (up to a proportionality factor),
conditioned on the fact that $Y(0)=y_0$.
In that expression, we could shift $Y(t)$ in time with respect to equations~(\ref{atau}).   
For sparsity of the computation, we rewrite this cumulant as a function of the moments $\langle Y(\tau)\rangle_{Y(0)=y_0}$
and $\langle Y^2(\tau)\rangle_{Y(0)=y_0}$ conditioned on $Y(0)=y_0$.
Indeed, we have
\begin{align}
\langle (Y(\tau)-y_0)^2\rangle_{Y(0)=y_0}=\langle Y^2(\tau)\rangle_{Y(0)=y_0}-2y_0\langle Y(\tau)\rangle_{Y(0)=y_0}+y_0^2\,,\label{cml2}\\
\langle Y(\tau)-y_0\rangle_{Y(0)=y_0}=\langle Y(\tau)\rangle_{Y(0)=y_0}-y_0\,,\label{cml1}
\end{align}
In turn, these two ensemble averages can be written using $p(y,\tau|y_0,0)$ the probability to have $y$ at $\tau$, given
that we have $y_0$ at $0$. For any $n\in \mathbb{N}$, and in particular for the cases of $n=1$ and $n=2$ that interest us, 
the ensemble average of the $n^{\rm th}$ moments of $Y(t)$ is then
\begin{equation}
\left\langle Y(\tau)^n \right\rangle=\int_{y} p(y,\tau|y_0,0)y^n\,{\rm d}y\,.\label{condm}
\end{equation}
This conditional probability is the solution of the Fokker--Planck equation~(\ref{fpeq})
with the initial condition $\delta (y-y_0)$.
This solution can only be computed analytically for simple expressions of the drift and the diffusion.
When their expression is more complex, we have to turn to a numerical integration.
However, we cannot use a Dirac as an initial condition in such a resolution:
this will call for a rewriting of the problem.
For this matter, the linear Fokker--Planck equation~(\ref{fpeq}) is viewed using the infinitesimal generator of transitions $L^\dag$.
The Fokker--Planck equation~(\ref{fpeq}) then reduces to $\partial p/\partial t=Lp$, where  $L^\dag$
is the adjoint operator (with respect to the standard inner product) of
$L[\cdot]=-\partial (f\cdot)/\partial y+\partial^2 (a\cdot)/\partial y^2$.
Given that the Fokker--Planck equation is linear, we can write that $p(y,\tau|y_0,0)=e^{\tau L}[\delta(y-y_0)]$
(where the square brackets indicate that the operator is applied to the Dirac).
Inserted in equation~(\ref{condm}), this tells us that, using the properties
of an adjoint, we have
\begin{equation}
\left\langle Y(\tau)^n \right\rangle_{Y(0)=y_0}=\int_{y} y^ne^{\tau L}[\delta(y-y_0)]\,{\rm d}y
=\int_{y}\delta(y-y_0)\left(e^{\tau L^\dag}[ y^n]\right)\,{\rm d}y\,.\label{condm2}
\end{equation}
The moment $\left\langle Y(\tau)^n \right\rangle_{Y(0)=y_0}$ can easily be computed numerically thanks to this rewriting.
The first integral contained the application of differential operators to a Dirac, which was very inconvenient.
The second integral now contains the integration of a regular function $\left(e^{\tau L^\dag}[ y^n]\right)$ against a Dirac, which
just results in the selection of its value at $y_0$.
Stated differently, equation~(\ref{condm2}) tells us that the $n^{\rm th}$ conditioned moment $\left\langle Y(\tau)^n \right\rangle_{Y(0)=y_0}$
is $q(y_0,\tau)$, the solution of the Backward Fokker--Planck equation $\partial q/\partial t=L^\dag q$,
with initial condition $q(y,0)=y^n$, taken at $y=y_0$ at $t=\tau$, for any $n\in \mathbb{N}$. In this article, we will use $n=1$ and $2$.
The advantage of using the moments is that a single resolution of the Backward Fokker--Planck equation
\begin{equation}
\frac{\partial q}{\partial t}=f\frac{\partial q}{\partial y}+a\frac{\partial^2q}{\partial y^2}\,,\label{bfp}
\end{equation}
can then be used to recover the cumulants (Eqs.~(\ref{cml1}),~(\ref{cml2})).

\subsubsection{Time scales and tests of the Markovianity hypothesis}\label{stmark}

When we train a stochastic model using empirical Kramers--Moyal coefficients sampled at a time step $\tau$,
we need to verify the Markov hypothesis of independence of increments of $Y_m$ at that time scale.
One way to ensure this is to check that the signal has lost correlations at that time scale.
We test that assumption by estimating the time correlation functions of the signal of $Y_m(t)$.
Indeed, we had stated in section~\ref{slfpeq} of the article that increments of
a Markov process discretised with a time step $\tau$
are independent of the past history of the process.
If time correlations are found on a scale larger than $\tau$, this hypothesis would be contradicted.

We compute these time correlation functions using time frames where the jet is pointing in a single direction.
Indeed, correlation functions computed from longer time series of $Y_m(t)$, which contain one or more change of direction
(see Fig.~\ref{skint} (c) of the article), would highlight a much longer
correlation time, which is of the order of the mean first passage time.
The reason for this is that, on a long time scale, the time average of $Y_m(t)$ is zero,
and the position of the jet retains some time correlation as long as the jet points in the same direction.
We will present the passage time distribution in section~\ref{mmfpt} and we will discuss the mean first passage times in more details in
another article.

\begin{figure}
\centerline{
\includegraphics[width=5.5cm]{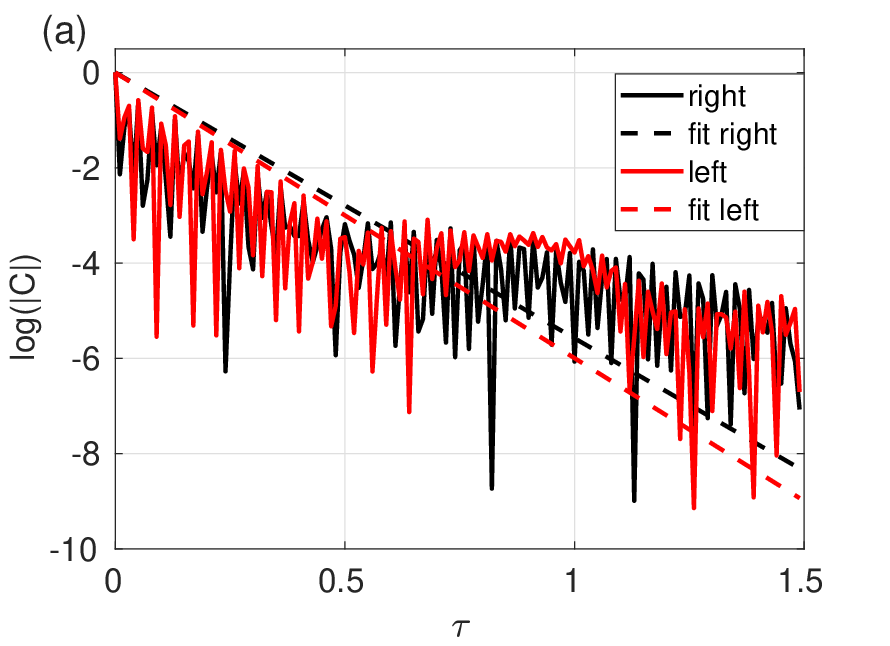}
\includegraphics[width=5.5cm]{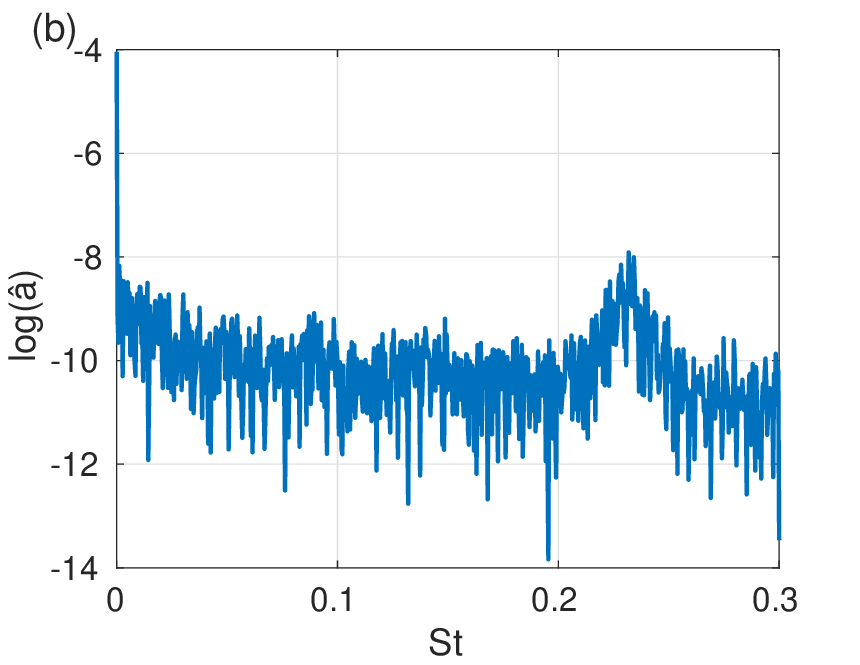}
}
\centerline{
\includegraphics[width=5.5cm]{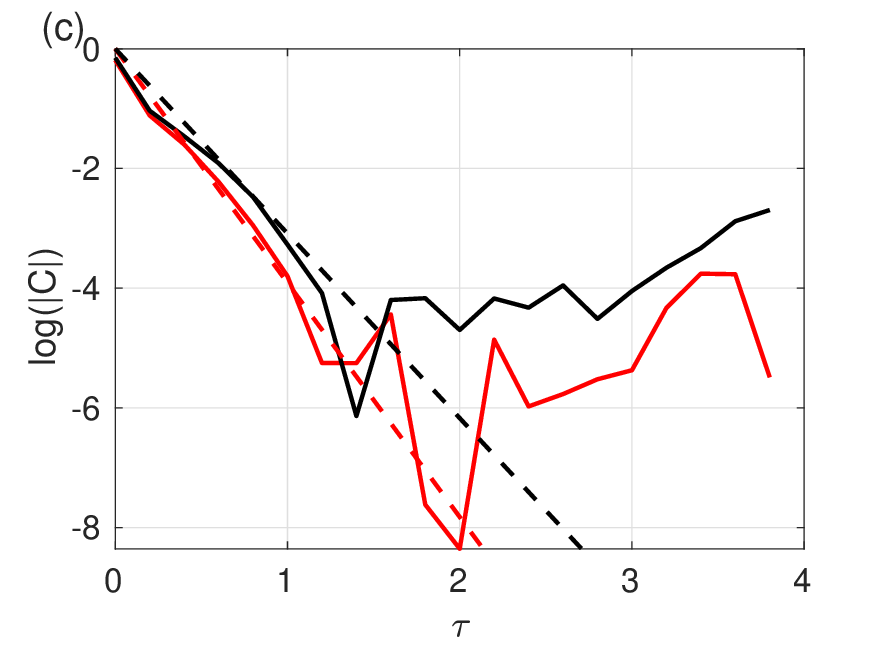}
\includegraphics[width=5.5cm]{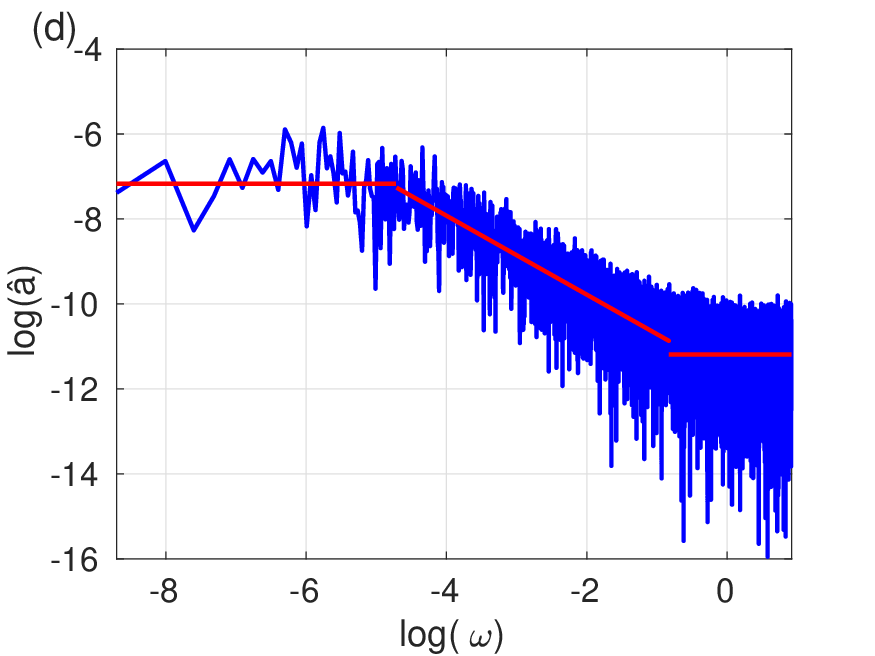}
}
\caption{(a) Logarithm of the correlation function of $Y_m(t)$ computed from PIV fields sampled at $100$Hz,
using runs when the jet is either only on the left ($C^l$, (Eq.~(\ref{eqCL}))) or only on the right ($C^r$, (Eq.~(\ref{eqCR}))).
Linear fits of the logarithm of the correlation function (up to $0.5$s) are added as dashed line for both cases.
(b) logarithm of the absolute value of the Fast Fourier Transform of $Y_m$ computed from a run of PIV field sampled at $100$Hz.
(c) Logarithm of the correlation function of $Y_m(t)$ computed
from PIV fields sampled at $5$Hz at $\frac{G}{H}=2.4$ (third campaign second group of runs),
using time intervals of length at least $4$s where the jet is either only on the left or only on the right, among those
detected using the procedure presented in section~\ref{stated} of the article, figure~\ref{maskcdf_a}.
(d) Logarithm of the absolute value of the Fast Fourier Transform of $Y_m$ as a function of the logarithm of the absolute value of
the frequency  computed using the whole time series of $Y_m$ sampled at $5$Hz at $\frac{G}{H}=2.4$ (third campaign second group of runs).
The lines indicate the average in the frequency interval $[1.67\cdot 10^{-4}\text{Hz},0.0089\text{Hz}]$,
the linear fit in log-log scale in the frequency interval $[0.0089\text{Hz},0.436\text{Hz}]$
and the average in the frequency interval $[0.436\text{Hz},2.5\text{Hz}]$.
}
\label{correlsp}
\end{figure}

We first compute the time correlation functions of $Y_m$ using each of the PIV runs measured at $100$Hz (section~\ref{dat_expc} 
of the article, see also \S~\ref{app_exp}.\ref{100hz} of the appendix),
where the jet points
solely to the left (denoted by $\{Y_{m,j}^l(t)\}_{1\le j \le M_{\rm left}}$) or solely to the right (denoted by $\{Y_{m,j}^r(t)\}_{1\le j \le M_{\rm right}}$).
For this matter, for each run containing $N_t=1550$ fields, we compute the time average
\begin{equation}
\langle Y_{m,j}^{l,r}\rangle_t=\frac{1}{N_t}\sum_{t=1}^{N_t}Y_{m,j}^{l,r}(t)\,,
\end{equation}
and the time variance
\begin{equation}
{\sigma_{m,j}^{l,r}}^2=\frac{1}{N_t}\sum_{t=1}^{N_t}{Y_{m,j}^{l,r}(t)}^2-\langle Y_{m,j}^{l,r}\rangle_t^2\,.
\end{equation}
For each run, we have a time correlation function
\begin{equation}
f_j^{l,r}(\tau=i \delta t)=\frac{1}{N_t+1-i}\sum_{t=1}^{N_t+1-i}
\frac{\left(Y_{m,j}^{l,r}(t)-\langle Y_{m,j}^{l,r}\rangle_t \right)\left(Y_{m,j}^{l,r}(t+1-i)-\langle Y_{m,j}^{l,r}\rangle_t \right)}{{\sigma_{m,j}^{l,r}}^2}\,,\,
\end{equation}
with $\delta t=0.01$s.
We then average over the $M_{\rm right}=65$ correlation functions computed from runs where $Y_m<0$ to obtain a time correlation function
for the jet pointing to the right
\begin{equation}
C^r(\tau)=\frac{1}{M_{\rm right}}\sum_{j=1}^{M_{\rm right}}f_j^{r}(\tau)\label{eqCR}
\,.
\end{equation}
 Similarly, we averaged over the $M_{\rm left}=30$ correlation functions computed from runs where $Y_m>0$
to obtain a time correlation function for the jet pointing to the left
\begin{equation}
C^l(\tau)=\frac{1}{M_{\rm left}}\sum_{j=1}^{M_{\rm left}}f_j^{l}(\tau)\label{eqCL}
\,.
\end{equation}
In figure~\ref{correlsp} (a), we display the logarithm of the time correlation functions of $Y_m$
when the jet is pointing to the left and the correlation function of $Y_m$ when the jet is pointing to the right.
In both cases, the time correlation function comprises rapid fluctuations on top
of an exponential decay (shown by the linear decay in logarithmic scale).
This exponential decay is highlighted by the linear fit (using data up to $0.5$s of time lag)
of the logarithm of both time correlation functions.
Both time correlation functions have approximately the same decay rate,
and the inverse of the slope gives the correlation time of the signal, of order $\mathcal{O}(0.18)$s, which is smaller than
the time step of $0.2$s that we used for the computation of finite time Kramers--Moyal coefficients.
We also observe that the time correlation functions display fluctuations on
a time scale of a few $\mathcal{O}(10^{-2})$s.
In order to highlight the origin of these fluctuations,
we display the logarithm of the spectrum of $Y_m$ for one run in figure~\ref{correlsp} (b),
as a function of the Strouhal number $St=\frac{Hf}{U}$ (the frequency rescaled by $\frac{U}{H}=167$Hz).
We note a maximum at Strouhal number $St\simeq 0.23$.
Firstly, these fast oscillations correspond to the oscillations observed in the time correlation function.
Secondly, they are at a Strouhal number which is commensurate with the (several) vortex shedding frequencies observed
in this type of wake \citep{alam2011wake}, which report a Strouhal number of approximately $0.2$ for the largest vortex shedding frequency,
toward the end of the regime of bistability.

We then perform a similar time correlation function computation using the data sampled at $5$Hz at $\frac{G}{H}=2.4$ during the third experimental campaign.
We use the procedure presented in section~\ref{stated} of the article to select the intervals of time where the jet is either only on the left
or only on the right for at least 4 seconds.
Using those, we compute the time correlation function of $Y_m$ conditioned to be either on the left or on the right (Fig.~\ref{correlsp} (c)),
with the same method as for the $100$Hz data.
Similarly to the former computation, these functions are computed in each time interval, then averaged over the intervals where $Y_m$ has a given direction.
These give us approximate correlation times of $0.26$s and $0.32$s.
These values are consistent with what is obtained from high frequency data.
In order to place these correlation functions and the time frames in which they are computed in a bigger picture,
we compute the spectrum of the whole time series and we display it in figure~(\ref{correlsp}) (d).
We note that the spectrum in linear by parts in three intervals of frequencies.
We therefore add three red lines in three successive frequency intervals to indicate the main tendencies of the spectrum.
For $\omega\in [1.67\cdot 10^{-4}\text{Hz},0.0089\text{Hz}]$
(bounded below by the smallest frequency allowed by the duration of the run),
the spectrum is more or less constant.
For $\omega\in [0.0089\text{Hz},0.436\text{Hz}]$, the spectrum decreases approximately like $\frac{1}{\omega}$.
For $\omega \in [0.436\text{Hz},2.5\text{Hz}]$ (bounded above by the largest frequency allowed by the sampling rate) the spectrum is again approximately constant,
taking values of the same order of magnitude as what can be observed for the highest frequencies seen in figure~\ref{correlsp} (b).
The first interval of frequency is directly related to the exponential distribution of residency time on the left and on the right,
the cut-off being directly related to the transition rate.
Meanwhile the last interval represents the rapid fluctuations of the wake.
Note that due to the sampling rate of $5Hz$, we cut off the oscillations at frequencies close to those of vortex shedding.
This is also reflected in the near monotonous decrease of the correlation function at short lags.

From the examination of time fluctuations of our data, we can draw some conclusions as to the relevance of our modelling hypotheses.
We firstly note that the correlation time of $Y_m$, conditioned on the jet being on the left or on the right is always approximately
smaller that the lag of $\tau=0.2$ used to compute the Kramers--Moyal coefficients, regardless of the sampling rate of velocity fields.
This rapid enough decorrelation indicates us that representing $Y_m$ by a Markov process
at scales larger than $\tau=0.2$s is a good approximation.
The white range at low frequency of the spectrum, that we will confront to the passage time distributions in appendix~\ref{mmfpt},
gives weight to another modelling hypothesis, the Gaussianity (or whiteness) of the noise.
In relation to that, Kramers--Moyal coefficients of order 3 and 4 have been computed (not shown here):
they are approximately 20 times smaller than the first and second coefficients.
This further confirms that $Y_m$ is close to a process undergoing a Gaussian white noise, where the coefficients of order strictly larger than 2 are zero.

\subsubsection{Properties of normal forms}\label{app_normform}

The fixed points of equation~\ref{oddODE} are
obtained by cancelling the derivative $dy/dt=0$.
The equation $f(y)=0$, written as $0=y(f_1 +f_3 y^2+f_5 y^4)$ can have
one, three or five real roots:
\begin{equation}
y=0\,\text{and, if real}\,
y=\pm\sqrt{\frac{-f_3\pm\sqrt{f_3^2-4f_1 f_5}}{2f_5}}\,.\label{yzerodd}
\end{equation}
We then have three typical types of solution given by the signs of $f_1$, $f_3$ and $f_5$:
\begin{enumerate}[i)]
\item\label{forms} If we primarily have $f_1 \le 0$ and secondarily have, $f_3\le 2\sqrt{f_1 f_5}$, $f_5\le 0$,
the system has a single monostable fixed point at $y=0$.
The prototype of such a drift is linear $f(y)=-|f_1| y$.
\item If we primarily have $f_1 >0$ and $f_3\le 0$ and secondarily have $f_5\le 0$,
we find bistable positions at $y=\pm \sqrt{|f_1/f_3|}$ with possible corrections in $f_5$,
while the transition state, the unstable fixed point, is at $y=0$.
The prototype of such a drift is cubic $f(y)=|f_1|y-|f_3|y^3$. This drift is similar to the
one visible in figure~\ref{compaf} (a).
Note that, if we go from $f_1 <0$ and $f_3< 0$ (mainly described in point~\ref{forms}) to $f_1 >0$ and $f_3< 0$, 
the ordinary differential equation experiences a supercritical pitchfork bifurcation.
\item Finally, if we have the combination of signs $f_1 <0$, $f_5<0$ and  $f_3\ge 2\sqrt{f_1 f_5}\ge 0$
the system is tristable, with a stable fixed point at $y=0$
and a pair of symmetric stable fixed points and unstable fixed points respectively at
\begin{equation}
y=\pm\sqrt{\frac{-f_3-\sqrt{f_3^2-4f_1 f_5}}{2f_5}}
\,\text{and}\,
 y=\pm\sqrt{\frac{-f_3+\sqrt{f_3^2-4f_1 f_5}}{2f_5}}\,.
 \end{equation}
They play the role of transition states in between.
The prototype of such a drift is necessarily quintic $f(y)=-|f_1|y+|f_3|y^3-|f_5|y^5$.
This drift is similar to the one visible in figure~\ref{modGH} (a) for $G/H=1.25$.
Note that this ODE can be viewed as the result of a subcritical pitchfork bifurcation.
\end{enumerate}
The bifurcations can then be detected by changes of sign of the three model parameters $f_1$, $f_3$ and $f_5$.

\subsubsection{Measure of the asymmetry of the drift}\label{mesasym}

We can measure the relative weight of the relative asymmetry in the model 
with the help of the even coefficients of the drift.
If we have $f_r=\sum_{k=0}^{N_f}f_{k,r}y_r^k$, then the  square norm of $f_r$ in $[-1,1]$ is
\begin{align}
\notag
I= \int_{y_r=-1}^1f_r(y_r)^2\,{\rm d}y=\int_{y_r=-1}^1\left(\sum_{k=0}^{N_f}f_{k,r}y_r^k \right)^2\,{\rm d}y
= \sum_{k=0}^{N_f}\sum_{l=0}^{N_f}f_{k,r}f_{l,r}\left[\frac{y_r^{k+l+1}}{k+l+1}\right]_{y_r=-1}^1\,.
\end{align}
We have $\left[\frac{y_r^{k+l+1}}{k+l+1}\right]_{y_r=-1}^1=\frac{1-(-1)^{k+l+1}}{k+l+1}$. 
It is equal to $2$ if $k+l+1$ is odd (if $k$ and $l$ have the same parity) and $0$ if it is even ($k$ and $l$
have different parity).
We  introduce the two sums
\begin{align}
I_{\rm even}=2\sum_{k=0}^{\frac{N_f-1}{2}}\sum_{l=0}^{\frac{N_f-1}{2}}\frac{f_{2k,r}f_{2l,r}}{2(k+l)+1} \,, \,
I_{\rm odd}=2\sum_{k=0}^{\frac{N_f-1}{2}}\sum_{l=0}^{\frac{N_f-1}{2}}\frac{f_{2k+1,r}f_{2l+1,r}}{2(k+l)+3} \,.
\end{align}
We reindexed the sums because we always take $N_f$ odd (we used $5$ and $7$ in practice).
We then have
\begin{equation}
I=I_{\rm even}+I_{\rm odd}\,.
\end{equation}
In the square norm of the drift, we can separate the weight of the odd coefficient (which respect the left-right symmetry) and
the even one (which break it).
We then measure the relative asymmetry of the model using 
\begin{equation}
I_{\rm asym}=\frac{I_{\rm even}}{I_{\rm even}+I_{\rm odd}}\,.\label{dia}
\end{equation}

\subsection{Optimisation of the model}\label{appmodcomp}  

In the core of the article, we have briefly described the procedure that tests polynomials for
the drift and diffusion and the cost function minimisation within it.
In this subsection of the appendix, we firstly detail how quantities are rescaled (\S~\ref{app_sdei}.\ref{appmodcomp}.\ref{app_resc}).
We then give numerical details on the cost function minimisation, given a set of monomials in the drift and diffusion (\S~\ref{app_sdei}.\ref{appmodcomp}.\ref{costmin}).
We explain how the optimisation of the polynomial type is performed in subsection~\ref{app_sdei}.\ref{appmodcomp}.\ref{optimshape} 
and exemplify this optimisation in subsection~\ref{app_sdei}.\ref{appmodcomp}.\ref{sselmod}.
We finally present a parallelisation procedure for this polynomial  type selection (\S~\ref{app_sdei}.\ref{appmodcomp}.\ref{spar}).

\subsubsection{Rescaling of time series, estimates and model properties}\label{app_resc}

In order to bound all model parameters that will appear in our procedures and keep their value close to one,
all our procedures are applied to rescaled time series of $Y_m$.
For each group of runs, we compute the maximum over the time series of the absolute value of $Y_m$.
the rescaled time series are then given by this maximum
\begin{equation}
Y_r(t)=\frac{Y_m(t)}{Y_{\max}}\,,\, Y_{\max}=\max_t(|Y_m(t)|)\,. \label{resy}
\end{equation}
Additionally, the rescaled empirical Kramers--Moyal coefficients $f_{r,\tau}^e$, $a_{r,\tau}^e$ are related to the original empirical
Kramers--Moyal coefficients \emph{via} a multiplicating factor
\begin{equation}
f_{r,\tau}^e(Y_r)=\frac{f_\tau^e(Y_rY_{\max})}{Y_{\max}}\,,\,a_{r,\tau}^e(Y_r)=\frac{a_\tau^e(Y_rY_{\max})}{Y_{\max}^2}\,.
\end{equation}

We term $\mu_r$ the histogram of the rescaled time series $Y_r$.
Similarly, we will use the PDF $\rho_r$ for the rescaled random variable $y_r=y/Y_{\max}$.
Note that owing to the constraint of keeping a constant normalisation of the histogram and the PDF,
the histogram of the rescaled time series is given by  $\mu_r(Y_r)=Y_{\max}\mu\left(Y_rY_{\max} \right)$,
while the PDF is given by $\rho_r=Y_{\max} \rho$.

We will only perform the fit for Kramers--Moyal coefficients and histograms computed from
 $Y_r$ (Eq.~(\ref{resy})), the time series of $Y_m$ rescaled by $Y_{\max}$ the maximum of $Y_m$ in absolute value.
The rescaled drift and diffusion of the models are respectively written as
\begin{equation}
f_r(y_r)=\sum_{k=0}^{N_f}f_{r,k}y_r^k\,,\, g_r(y_r)=\sum_{k=0}^{N_g}g_{r,k}y_r^k\,.\label{resfg}
\end{equation}
Similarly to the empirical Kramers--Moyal coefficients, there is a rescaling factor between the model for $y$
and the for model for $y_r$.
We have that
\begin{equation}
f_r(y_r)=\frac{f\left(\frac{y}{Y_{\max}}\right)}{Y_{\max}}\,,\,g_r(y_r)=\frac{g\left(\frac{y}{Y_{\max}} \right)}{Y_{\max}}\,,
\end{equation}
which results in a conversion factor of model parameters 
before and after rescaling $f_k=\frac{f_{r,k}}{Y_{\max}^{k-1}}$, $g_k=\frac{g_{r,k}}{Y_{\max}^{k-1}}$.
The rescaled finite time Kramers--Moyal coefficients $f_{r,\tau}$ and $a_{r,\tau}$ are obtained through the resolution
of the backward Fokker--Planck equation with rescaled $y_r$, drift $f_r$  and diffusion $g_r$.
The advantage of this rescaling is that the rescaled model parameters $f_{r,k}$ and $g_{r,k}$
are much closer to one another and all much closer to $1$. This will regularise and accelerate the optimisations.

\subsubsection{Optimising  the objective function for a given model}\label{costmin}

In order to measure the mismatch between the analytical model and the sampled data measured by the cost function $V$ (Eq.~(\ref{eqcost})),
we need to compute several functions corresponding to the model (Eq.~(\ref{resfg})):
\begin{enumerate}[a)]
\item we need to compute the finite time Kramers--Moyal coefficients at time $\tau$
corresponding to the model.
As stated in section~\ref{sKM}, this is done by solving the backward (or adjoint) Fokker--Planck equation (Eq.~(\ref{bfp})).
\item We need to compute the probability density function corresponding to the model (Eq.~(\ref{resfg})).
This is done by solving the corresponding forward Fokker--Planck equation on the same grid:
we compute the steady solution.
\end{enumerate}
In both cases, the computations are performed numerically.
The $Y_r$ are discretised on the same grid as the one where the empirical Kramers--Moyal coefficients are computed.
For both  PDE resolutions, the differential operators present in equations~(\ref{fpeq}),~(\ref{bfp})
are discretised using centred finite differences for variable $y$.
Some models lead to strong requirements of numerical stability of the resolution.
This is why we chose to discretise in time equation~(\ref{bfp}) using an eighth order Runge--Kutta method with an adaptive time step.
In practice,  for our cases, this led to numerical integrations more stable than the computation the exponential of the operator used by \cite{callaham2021nonlinear}.
These resolutions yield a numerical approximation of the finite time
Kramers--Moyal coefficients $\{ f_{\tau,i}^m\}_{1\le i \le N_y}$ and $\{ a_{\tau,i}^m\}_{1\le i \le N_y}$
resulting from the value of the model parameters $f_{r,k}$ and $g_{r,k}$.
These computations are performed at each step of reduction of $V$, since each modification of the model parameters $f_{r,k}$, $g_{r,k}$ leads to different PDFs
and finite time Kramers--Moyal coefficients.

When training models, we have adjusted the importance given to the match between the histogram and the PDF of the model,
measured by the Kullback--Leibler divergence in the cost function (Eq.~(\ref{eqcost})) and weighed by the Lagrange multiplier $\lambda$.
We have tested values of $\lambda=1$, $10$ and $100$ to vary the weight given to this constraint.
The value of $\lambda$ does not modify the shape of the selected drift and diffusion,
nor did it change the position of PDF modes in the large majority of cases.
Note that increasing $\lambda$ leads to a better rendition of the PDF tails for $|Y_m|$ large.
Indeed, with $\lambda=1$, the tails of the resulting PDF are often too heavy (Fig.~\ref{compaf} (c)).
Unless stated otherwise, we will present results obtained with $\lambda=10$.
The cost function (Eq.~(\ref{eqcost})) is minimised using a gradient free Nelder--Mead method,
from the {\sc optimize} library of {\sc scipy}.
In that approach each new step of the optimisation is computed using the value of the loss at a finite
number of former step.
That step leads to a new value of the model parameters, a new value of the cost, which is in turn
used for the next optimisation step.
We typically place a cap at $10000$ for the maximum number of iterations of the optimisation.
The initial guess for all optimisations is obtained from a least square fit of the empirical finite time
Kramers--Moyal coefficients by the full polynomials (Eq.~(\ref{resfg})).
This generally leads to a correct sign for the model parameters and relevant relative amplitude,
while the exact value will be adjusted from the cost minimisation.

\subsubsection{Optimising over function shape}\label{optimshape}

The former section presented how to adjust a fixed function form for the
drift and diffusion of the Langevin equation, so as to best match its dynamical properties
and its statistics.
However, we are not ensured that we have the best function form.
We seek a balance between sparsity (where we would have a few terms,
which is as expressive as possible) and precision of the fit,
which degrades when too few monomials are used in the models.
Following \citep{callaham2021nonlinear}, the model adjustment procedure of section~\ref{fitfunc} is included in an iterative process
that will propose a model close to our optimal, in a systematic way.
This approach is similar to \emph{structural risk minimisation} that can be performed in parametric machine learning \citep{shalev2014understanding}.

In practice, for a total number of monomials $N_f+N_g+2$, we perform the following operations.
\begin{enumerate}
\item We first adjust the full original model.
This gives us values for the model parameters and a cost $V_0$
\item We then proceed with $N_f+N_g$ steps of monomial removal.
At step $k\ge 1$
We receive a model with $K=N_f+N_g+3-k$ monomials and a cost $V_{k-1}$ from the former step.
At this step we will adjust at most $K$ models.
Each of these models consist in almost the model received from step $k-1$, at the exception of one monomial of
the drift or of the diffusion which is removed (unless the drift or the diffusion only has a single monomial left).
We will store the resulting minimal cost  $\tilde{V}_{k,i}$ and model parameter values for each of these reduced model
(with $i$ between $1$ and at most $K$ designating the reduced model form).
We then determine the model $i'$ that leads to the smallest cost increase  $\tilde{V}_{k,i'}-V_{k-1}$
compared to that of the former step: this indicates which of the monomials mattered the least.
We retain this model at this step, set $V_k=\tilde{V}_{k,i'}$ and pass it to the next step.
\end{enumerate}
At the end of this procedure, the last fitted model has one monomial for the drift and one monomial for the diffusion.
As $k$ is increased in this iterative procedure, the cost $V_k$ may first slightly decrease with $k$
(too many monomials results in a slightly cumbersome model type) but will then steadily increase.
This indicates that the models that decrease in complexity also decrease in precision.
In order to select the model realising a trade--off between precision and simplicity,
we select the model resulting from the step $k$ that results in a sudden increase in cost.
For instance, one can choose the step $k$ such that
\begin{equation}
V_{k}-V_{k-1}\ge 2\underset{2\le j\le k-1}{\max}(V_{j}-V_{j-1})\,.
\end{equation}
The procedure can be naturally parallelised over tested model, with reasonable performance (see appendix \S~\ref{app_sdei}.\ref{appmodcomp}.\ref{spar}). 

The examples of drift, and first Kramers--Moyal coefficients (Fig.~\ref{compaf} (a)), on the one hand,
and diffusion and second Kramers--Moyal coefficients (Fig.~\ref{compaf} (b)), on the other hand,
that we have presented throughout section~\ref{smodc} come out of the proposed procedure, where $\lambda=10$ is used.
We note that we have a match of the two first  Kramers--Moyal coefficients
at finite time $\tau=0.2$.
The model which is eventually computed also displays a good match between its PDF and the sampled histogram of $Y_m$ (Fig.~\ref{compaf} (c)).
We present the details of that model section in the next subsection of the appendix (\S~\ref{app_sdei}.\ref{appmodcomp}.\ref{sselmod}).

\subsubsection{Example of successively selected models for model construction at $\frac{G}{H}=2.4v2$}\label{sselmod}

\begin{figure}[!ht]
\centerline{\setlength{\tabcolsep}{0.5pt}
\begin{tabular}{ccccccccccccccc}
$\sharp 1$,&$f_r=$&$f_{r,0}$&$+f_{r,1} y_r$&$+f_{r,2}y_r^2$&$+f_{r,3} y_r^3$&$+f_{r,4}y_r^4$&$+f_{r,5} y_r^5$&
,&$g_r=$&$g_{r,0}$&$+g_{r,1} y_r$&$+g_{r,2}y_r^2$&$+\beta_3 y_r^3$&$+g_{r,4}y_r^4$\\
$\sharp 2$,&$f_r=$&$f_{r,0}$&$+f_{r,1} y_r$&$+f_{r,2}y_r^2$&$+f_{r,3} y_r^3$&$+f_{r,4}y_r^4$&$+f_{r,5} y_r^5$&
,&$g_r=$&$g_{r,0}$&$+g_{r,1} y_r$&$+g_{r,2}y_r^2$&&$+g_{r,4}y_r^4$\\
$\sharp 3$,&$f_r=$&$f_{r,0}$&$+f_{r,1} y_r$&$+f_{r,2}y_r^2$&$+f_{r,3} y_r^3$&&$+f_{r,5} y_r^5$&
,&$g_r=$&$g_{r,0}$&$+g_{r,1} y_r$&$+g_{r,2}y_r^2$&&$+g_{r,4}y_r^4$\\
$\sharp 4$,&$f_r=$&$f_{r,0}$&$+f_{r,1} y_r$&$+f_{r,2}y_r^2$&$+f_{r,3} y_r^3$&&&
,&$g_r=$&$g_{r,0}$&$+g_{r,1} y_r$&$+g_{r,2}y_r^2$&&$+g_{r,4}y_r^4$\\
$\sharp 5$,&$f_r=$&$f_{r,0}$&$+f_{r,1} y_r$&$+f_{r,2}y_r^2$&$+f_{r,3} y_r^3$&&&
,&$g_r=$&$g_{r,0}$&&$+g_{r,2}y_r^2$&&$+g_{r,4}y_r^4$\\
$\sharp 6$,&$f_r=$&&$f_{r,1} y_r$&$+f_{r,2}y_r^2$&$+f_{r,3} y_r^3$&&&
,&$g_r=$&$g_{r,0}$&&$+g_{r,2}y_r^2$&&$+g_{r,4}y_r^4$\\
$\sharp 7$,&$f_r=$&&$f_{r,1} y_r$&&$+f_{r,3} y_r^3$&&&
,&$g_r=$&$g_{r,0}$&&$+g_{r,2}y_r^2$&&$+g_{r,4}y_r^4$\\
$\sharp 8$,&$f_r=$&&$f_{r,1} y_r$&&$+f_{r,3} y_r^3$&&&
,&$g_r=$&$g_{r,0}$&&&&$+g_{r,4}y_r^4$\\
$\sharp 9$,&$f_r=$&&$f_{r,1} y_r$&&$+f_{r,3} y_r^3$&&&
,&$g_r=$&$g_{r,0}$&&&&\\
$\sharp 10$,&$f_r=$&&&&$f_{r,3} y_r^3$&&&
,&$g_r=$&$g_{r,0}$&&&&
\end{tabular}}
\vspace{0.5cm}
\centerline{\includegraphics[width=7cm,clip]{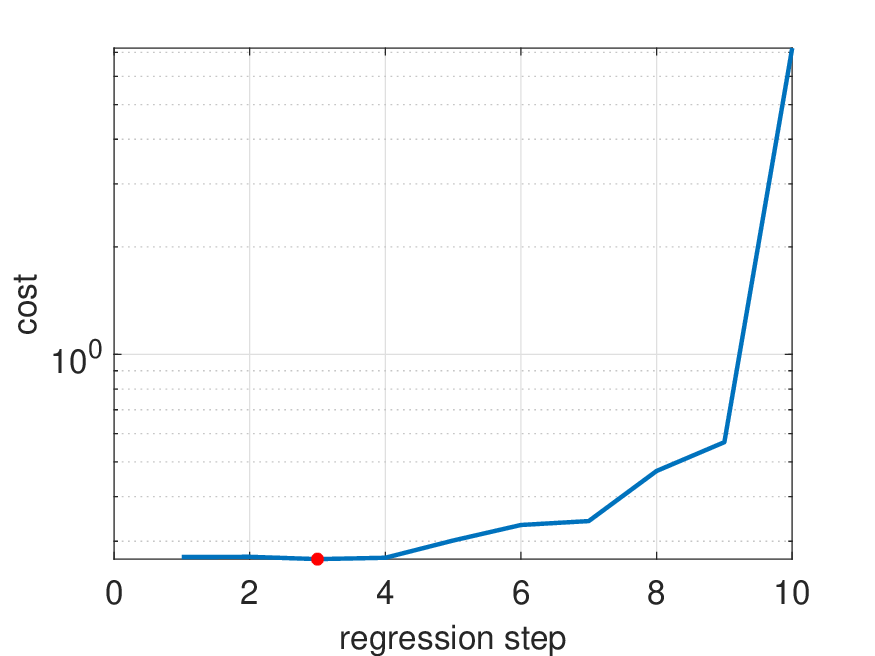}}
\caption{Top: Successive drifts $f_r$ and diffusions $g_r$ obtained at each step
of the iterative procedure when fitting against the data sampled at $\frac{G}{H}=2.4$v2.
Bottom: Cost of the model selected at each step as a function of the step number
when fitting against the data sampled at $\frac{G}{H}=2.4$v2 in log-lin scale.
The red dot indicates the model chosen using our systematic criterion.}
\label{selmod}
\end{figure}

In this section, we present the succession of models  (Fig.~\ref{selmod}, top)
selected at each step of the iterative procedure
when fitting against the data sampled at $\frac{G}{H}=2.4$v2, along
with the corresponding cost function (Fig.~\ref{selmod}, bottom).
A red dot in that plot indicates the cost for the model selected using our systematic criterion.
The model resulting from the application of the criterion proposed in section~\ref{app_sdei}.\ref{appmodcomp}.\ref{optimshape}
was presented in figure~\ref{compaf} and figure~\ref{modGH}.
In line with the symmetries of the finite time Kramers--Moyal coefficients
(approximately odd first coefficient and approximately even second coefficient),
the procedure rapidly removes the even powers of $y_r$
in the drift $f_r$ and the odd powers of $y_r$ in the diffusion.
We note that for steps 1 to 4 of the procedure, the growth of the cost $V$ is small and the
precise functional shape of the selected model will depend on the chosen systematic selection criterion as to the increase of $V$.
Different systematic quantitative criterion always retain the same leading monomials but can lead to different choices as
to the retained monomials modulated by a smaller coefficient.
This model retains even terms in the drift and odd terms in the diffusion to account for small asymmetries in the data, 
such as the $15\%$ difference in mode amplitude in the histogram of $Y_m$ at that gap ratio.
The effect of these terms is visible in the measure of the asymmetry presented in figure~\ref{bifdiagmod} (c).
As stated throughout the text, the model will retain the asymmetry contained in the data, whose source is both finite time sampling
and minute to non negligible imperfections in the experimental set up.
In the case we display, the large increases of $V$ happen when even terms are removed from the diffusion, from model $8$ on,
and eventually the odd term $y_r$ is removed from the drift.
Our systematic criterion on the cost function may seem very conservative for the case presented here.
However, it has proven efficient in training models when the dynamics is more complex.

\subsubsection{Performance of the parallelisation of the model construction}\label{spar}

\begin{figure}[!ht]
\centerline{
\includegraphics[width=7cm]{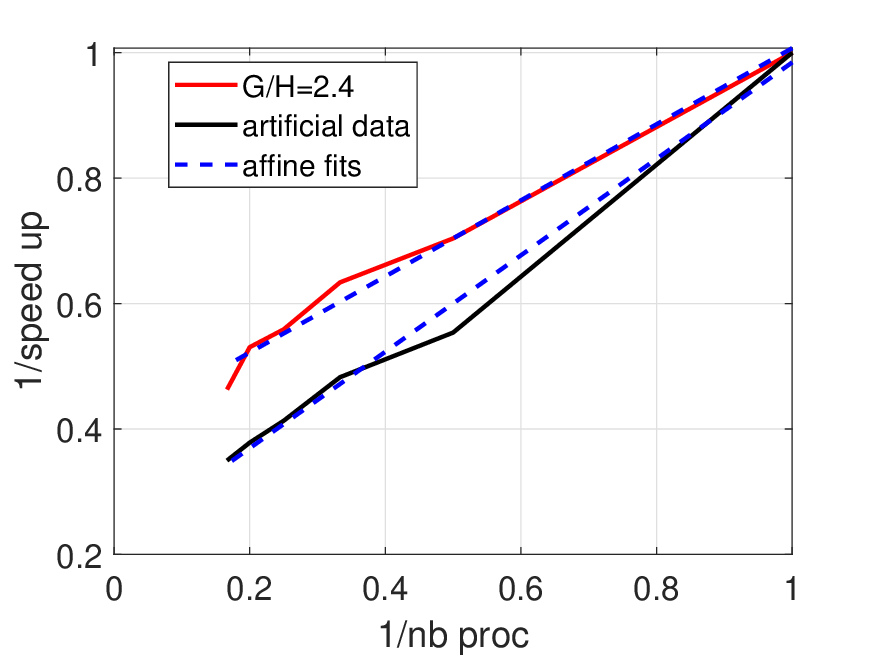}
}
\caption{Inverse of the speed up of calculations as a function of the inverse of the number of processors used,
when construction a model for data sampled at $\frac{G}{H}=2.4$v2 (third dataset) and for artificial data generated by a simple bistable Langevin equation.
Affine fits are added.}
\label{fspar}
\end{figure}

In this section, we present the test of the parallelisation of the model construction
presented in section~\ref{app_sdei}.\ref{appmodcomp}.\ref{optimshape} and detailed in section~\ref{app_sdei}.\ref{appmodcomp}.\ref{sselmod}.
During the successive steps,
the adjustment of each model is entirely independent of the others.
The task of adjusting $K$ models can then be distributed to $n_b<K$ processors.
We test the efficiency of such a parallelisation for two cases of model construction:
from the time series sampled at $\frac{G}{H}=2.4$ v2 in the third dataset (Fig.~\ref{posyw} (d), Fig.~\ref{compaf}),
and from an artificial time series sampled from the Langevin equation
\begin{equation}
dY=(Y-Y^3){\rm d}t+\frac12 {\rm d}W\,.
\end{equation}
We integrated this equation with a simple Euler--Maruyama method with a time step ${\rm d}t=2\cdot 10^{-3}$ during a duration of $10^3$ time units.
The resulting time series display approximately twenty transitions between the left and the right states.
In both cases, we use $N_f=5$, $N_g=4$ and $N_y=32$.
We define the speed up of our computation as the ratio of the computation time on one processor $T_{1}$
to the computation time on $n_b$ processor $T_{n_b}$.
These durations are measured using the {\sc time} library for the {\sc python} implementation of
the code.
This duration is measured between the very beginning of the computation
(which consist of serial tasks of data reading, Kramers--Moyal coefficient computation and solver setting, followed by the parallelised model construction)
and the very end of the code execution, after the computed models are written.
In the computations presented in this section, where we optimise at most $11$ models per procedure step,
we have observed a reasonable speed--up $\frac{T_1}{T_{n_b}}$ for up to $n_b=6$ processors.
We make a similar observation for all other model computation presented in this article.
This speed--up  grows sublinearly.
We examine it in view of Amdahl's stating that it should verify
\begin{equation}
\frac{T_{1}}{T_{n_b}}=\frac{1}{(1-x)+\frac{x}{n_b}}\,,\label{amdahl}
\end{equation}
where $0<x<1$ measures the proportion of correctly parallelised tasks during the parallel computation.
For both model constructions, the inverse of the speed--up is an affine function
of the inverse of the number of processors $\frac{1}{n_b}$ in good approximation (Fig.~\ref{fspar}).
This is in agreement with equation~(\ref{amdahl}).
In the case of the data sampled at $\frac{G}{H}=2.4$, the slope of the affine fit indicates that $x\simeq 0.6$,
while we have $x\simeq 0.77$ for artificial data.
This is less than what would be required for high performance computing,
but this is nevertheless a good acceleration for computations at our scale, where computations are performed on a desktop computer.
In our case the speed up is limited by load imbalances between the different optimisations
that run for distinct number of steps.

\bibliographystyle{apalike}
\bibliography{ref_wake}

\end{document}